\documentclass[twocolumn]{aastex63}

\usepackage{xcolor}
\usepackage{enumitem}
\defcitealias{patel17a}{P17}

\accepted{to ApJ}

\shorttitle{Orbits of Magellanic Satellites}
\shortauthors{Patel et al.}

\graphicspath{{./}{figures/}}

\begin{document}

\title{The Orbital Histories of Magellanic Satellites Using Gaia DR2 Proper Motions}

\correspondingauthor{Ekta Patel}
\email{ektapatel@berkeley.edu}

\author{Ekta Patel}
\affiliation{Department of Astronomy, University of California, Berkeley, 501 Campbell Hall, Berkeley, CA, 94720, USA} 
\affiliation{Miller Institute for Basic Research in Science,468 Donner Lab,Berkeley, CA 94720, USA}

\author{Nitya Kallivayalil}
\affiliation{Department of Astronomy, University of Virginia, 530 McCormick Road, Charlottesville, VA 22904, USA}

\author{Nicolas Garavito-Camargo}
\affiliation{Steward Observatory, University of Arizona, 933 North Cherry Avenue, Tucson, AZ 85721, USA}

\author{Gurtina Besla}
\affiliation{Steward Observatory, University of Arizona, 933 North Cherry Avenue, Tucson, AZ 85721, USA}

\author{Daniel R. Weisz}
\affiliation{Department of Astronomy, University of California, Berkeley, 501 Campbell Hall, Berkeley, CA, 94720, USA}

\author{Roeland P. van der Marel}
\affiliation{Space Telescope Science Institute, 3700 San Martin Drive, Baltimore, MD 21218, USA}
\affiliation{Center for Astrophysical Sciences, Department of Physics \& Astronomy, Johns Hopkins University, Baltimore, MD 21218, USA}

\author{Michael Boylan-Kolchin}
\affiliation{Department of Astronomy, The University of Texas at Austin, 2515 Speedway, Stop C1400, Austin, TX 78712-1205, USA}

\author{Marcel S. Pawlowski}
\affiliation{Leibniz-Institut f\"{u}r Astrophysik Potsdam (AIP), An der Sternwarte 16, D-14482 Potsdam, Germany}

\author{Facundo A. G\'{o}mez}
\affiliation{Instituto de Investigaci\'on Multidisciplinar en Ciencia y Tecnolog\'ia, Universidad de La Serena, Ra\'ul Bitr\'an 1305, La Serena, Chile}
\affiliation{Departamento de Astronom\'ia, Universidad de La Serena, Av. Juan Cisternas 1200 Norte, La Serena, Chile}



\begin{abstract} 
With the release of Gaia DR2, it is now possible to measure the proper motions (PMs) of the lowest mass, ultra-faint satellite galaxies in the Milky Way's (MW) halo for the first time. Many of these faint satellites are posited to have been accreted as satellites of the Magellanic Clouds (MCs). Using their 6-dimensional phase space information, we calculate the orbital histories of 13 ultra-faint satellites and five classical dwarf spheroidals in a combined MW+LMC+SMC potential to determine which galaxies are dynamically associated with the MCs. These 18 galaxies are separated into four classes: i.) long-term Magellanic satellites that have been bound to the MCs for at least the last two consecutive orbits around the MCs (Carina 2, Carina 3, Horologium 1, Hydrus 1); ii.) Magellanic satellites that were recently captured by the MCs $<$ 1 Gyr ago (Reticulum 2, Phoenix 2); iii.) MW satellites that have interacted with the MCs (Sculptor 1, Tucana 3, Segue 1); and iv.) MW satellites (Aquarius 2, Canes Venatici 2, Crater 2, Draco 1, Draco 2, Hydra 2, Carina, Fornax, Ursa Minor). Results are reported for a range of MW and LMC masses. Contrary to previous work, we find no dynamical association between Carina, Fornax, and the MCs. Finally, we determine that the addition of the SMC's gravitational potential affects the longevity of satellites as members of the Magellanic system (long-term versus recently captured), but it does not change the total number of Magellanic satellites.
\end{abstract}


\vspace{1cm}
\section{Introduction} 
\label{sec:intro}

In the hierarchical cold dark matter paradigm, dark matter halos of order $10^{11}\, M_{\sun}$ commonly contain tens of their own subhalos with sufficient gravitational potential to host luminous galaxies. The Large Magellanic Cloud (LMC) and M33 are the only two galaxies in the Local Group with halo masses in the $10^{11}\, M_{\sun}$ regime and they also happen to be the most massive satellites of the MW and M31, respectively. As such, the LMC and M33 are expected to have entered the halos of the MW and M31 with a group of their own satellite galaxies \citep[i.e. satellites of satellite galaxies; see][]{donghia08}. Recent studies have quantified predictions for the populations of satellites expected around the LMC and M33, finding that each should host approximately 5-10 ultra-faint dwarf galaxies (UFDs) with $\rm M_* \approx 10^2-10^5\, M_{\sun}$ \citep[e.g.][]{sales11,sales13,dooley17b,patel18b,jahn19} at minimum. 

Nearly 30 new dwarf galaxies have recently been discovered in the vicinity of the Magellanic Clouds \citep[MCs;][]{bechtol15,dwagner15,koposov15b,martin15,laevens15a,kim15a,kim15b,dwagner16, torrealba16a,torrealba16b,torrealba18,koposov18,homma18}. Furthermore, the timely second data release from the Gaia mission \citep{gaiadr2a} has enabled proper motion (PM) measurements for these ultra-faint satellites \citep{simon18,fritz18,kallivayalil18,pace19,massari18}, now making it possible to study their 3D kinematics and orbital histories in unprecedented detail. With this new data from Gaia DR2, several authors have aimed to identify the subset of known UFDs and classical dwarf spheroidals in the MW's halo that were originally satellites of the MCs. 

\citet{kallivayalil18} measured the PMs of 13 UFDs that also had radial velocity measurements using Gaia DR2. They compared the new 3D kinematics of UFDs to the tidal debris of a cosmological analog of the LMC to determine which UFDs have coincident kinematics with the LMC debris, and found that four UFDs (Carina 2, Carina 3, Horologium 1, Hydrus 1) are likely members of the Magellanic system. For UFDs without measured radial velocities at that time, they used the simulation to predict the PMs and radial velocities of expected Magellanic debris, finding that a group of stars in Phoenix 2 have a PM in DR2 consistent with this prediction. \cite{pardy19} and \cite{jahn19} used the orbital poles of UFDs and classical satellites calculated with Gaia DR2 PMs to additionally conclude that Carina and Fornax are also potential Magellanic satellites. 

\citet{erkal19} used Gaia DR2 PMs for 25 UFD satellites and the classical dwarfs to integrate orbits backwards in time, or \emph{rewind orbits}, in a combined MW+LMC potential. By calculating the orbital energy of these 25 galaxies relative to the LMC 5 Gyr ago, they determined that 6 UFDs (Carina 2, Carina 3, Horologium 1, Hydrus 1, Reticulum 2, Phoenix 2) are likely members of the Magellanic system. 

While these analyses have quantified the viability of satellites as members of the Magellanic system, none have accounted for the gravitational influence of the Small Magellanic Cloud (SMC), which is in a binary orbit with the LMC \citep{murai80, besla12}. In some studies \citep[i.e.][]{kallivayalil18,jahn19}, the inclusion of the SMC is inhibited by the simulations in that finding a reasonable cosmological match to the MW+LMC+SMC system is rare \citep[e.g.][]{bk11}. In other cases \citep[i.e.][]{erkal19}, the SMC is omitted as a gravitational mass that exerts non-negligible forces on other galaxies, especially the UFDs. However, this dismisses the competing tidal effects between the interacting MCs, which, in addition to tides from the MW, can perturb the orbits of satellites in a non-negligible way and potentially impact the total number of Magellanic satellites today.

Similarly, existing predictions for the total number of satellites hosted by the LMC and SMC today, in a $\Lambda$CDM paradigm, also omit the dynamical significance of the Clouds' binary dynamics. \cite{dooley17b} quantified the number of satellites expected around both the LMC and SMC under the assumption that each of the Clouds can be treated as an isolated halo. However, this assumption implies that the SMC continued to accrete substructures up until $z=0$, whereas if it were captured $>$ 5 Gyr ago by the LMC, its mass growth may have been truncated at the time of capture and some of those SMC satellites might have been destroyed by the LMC. Thus, while the predictions in \cite{dooley17b} are helpful benchmarks, they may overestimate the number and longevity of Magellanic satellites. 

\citet{jethwa16} do consider the combined gravitational influence of the MW, LMC, and SMC to calculate the probabilities that the \emph{Dark Energy Survey} UFDs belong to the LMC and SMC. This work came before PMs were available, yet they conclude that seven UFDs have a high probability (p $>$ 0.7) for being satellites of the LMC.

The goal of this work is to use Gaia DR2 PMs to calculate the orbital histories of all potentially associated Magellanic satellites, selected based on their membership to the MW's Vast Polar Structure \citep{pawlowski12}, and thereby determine which satellites have a high probability of entering the MW's halo as a group with the MCs. We further distinguish between the Magellanic satellites that have made only one passage around the LMC and those that evidence long-lived companionship. Our analysis explicitly includes the combined gravitational influence of the MW, LMC, and SMC for the first time. We also account for dynamical friction from both the MW and LMC, as well as the binary orbital history of the LMC-SMC and its subsequent effect on candidate Magellanic satellites. 

This paper is organized as follows. \S \ref{sec:data} includes justification for our sample selection and the observational data adopted for these galaxies. \S \ref{sec:model} outlines the analytic orbital model and all model parameters for the MW, LMC, and SMC. It also discusses the orbits of the MCs. In \S \ref{sec:analysis}, we analyze the orbital histories of all 18 candidate Magellanic satellites under the gravitational influence of the MW, MW+LMC, and MW+LMC+SMC. We also calculate the statistical significance of each candidate satellite's orbital histories accounting for the errors in PMs, line-of-sight velocities, and distances. Using these results, we define selection criteria to identify true Magellanic satellites. \S \ref{sec:discussion} includes a comparison to recent literature, a discussion on the mass of the MW and LMC, and how the inclusion of the SMC affects the results. Finally, in \S \ref{sec:discussion} we also demonstrate how smaller PM measurement uncertainties can affect a satellite's membership to the Magellanic system. \S \ref{sec:conclusions} provides a summary of our conclusions.

\section{Data}
\label{sec:data}
Here we briefly describe the selection of satellite galaxies included in our sample and the data used in this study.

\subsection{Sample Selection}
Since \citet{lyndenbell76} it has been suggested that several of the MW's classical dwarf satellites reside in a spatially coherent plane. More recent work has extended this plane to include several stellar streams and globular clusters. This is now referred to as the MW's `Vast Polar Stucture' \citep[VPOS;][]{pawlowski12}. Our goal is to identify the orbital histories of satellites that are dynamical companions to the LMC and SMC today. Since the VPOS is coincident with the orbital plane of the MCs, high probability members of the VPOS comprise our initial sample of possible Magellanic satellites. 

In \citet{fritz18}\footnote{We selected all satellites that have $p(\rm inVPOS) \geq 0.5$ in Table 4 of \citet{fritz18}.}, the following UFDs were identified as having $\geq 50$\% probability of being members of the VPOS: Crater 2, Carina 2, Carina 3, Hydrus 1, Horologium 1, Reticulum 2, Tucana 3, Segue 1, Aquarius 2, Canes Venatici 2. We also use the same criteria to choose the subset of classical satellites that lie in the VPOS: Carina, Draco, Fornax, Sculptor, and Ursa Minor. \citet{pawlowski19} independently analyzed the disk of classical satellites in light of Gaia DR2 PMs and find that Leo II is also consistent with the VPOS but has a high orbital pole uncertainty given its large distance, so we omit Leo II from our sample.

In \citet{kallivayalil18}, it was found that Hydra 2, Draco 2 and Phoenix 2 may also be associated with the MCs. Thus, we additionally include these three UFDs in our sample. For Phoenix 2, \citet{kallivayalil18} were able to measure a PM, but there was no measured radial velocity at the time. There is now a radial velocity measurement \citep{fritz19} as well as an independent PM measurement for Phoenix 2 \citep{pace19}, allowing for a full exploration of its orbital history \citep[see also][]{erkal19}. 

The total sample of candidate Magellanic satellites analyzed in this work is therefore comprised of 13 UFD satellites ($\rm M_* \approx 10^2-10^5\, M_{\sun}$) and 5 classical dwarf spheroidal satellites ($\rm M_* \approx 10^5-10^7\, M_{\sun}$). Their properties are listed in Table \ref{tab:pm_rv} and Table \ref{tab:size_lumninosity_table}. In the sections that follow, we will discuss the methods used to measure PMs and our selection of PM measurements for satellites where multiple measurements have been published.

\subsection{Proper Motions of the Candidate Magellanic Satellites}
Several groups measured PMs for MW dwarf galaxies with Gaia DR2. Given the difficulty in identifying member stars for these relatively sparse dwarf galaxy systems from the larger MW foreground, some works took the approach of cross-matching publicly available spectroscopic member catalogs with DR2 \citep{fritz18, simon18}, while others added photometric members under the assumption that member stars move coherently, forming a clump in PM space, and utilizing the position in the color-magnitude diagram \citep{gaiadr2b, kallivayalil18, massari18, pace19}.

We start with the values from \citet{fritz18}, who presented PMs for all dwarf galaxies in the MW vicinity based on cross-matching confirmed spectroscopic member stars for these dwarfs with Gaia DR2. They also presented the covariances of their reported errors. For dwarfs where additional photometric members are identified, we use PMs and reported errors from the measurement using more member stars, as well as the corresponding covariances \citep[specifically from][]{gaiadr2b, massari18, pace19}. We add a systematic error floor of 0.035 mas to all reported errors as in \cite{fritz18}. Table \ref{tab:pm_rv} lists the PMs, line-of-sight velocities, and distance moduli for all satellites in our sample, including references to the original measurements.

The LMC and SMC PMs and measurement errors are taken from \citet{k13} and \citet{zivick18}, respectively. The LMC measurement is based on multiple epochs of HST data for 22 fields across the galaxy, separated by a 3$-$7 year baseline, and centered on an inertial reference frame made up of background quasars. The long time baselines with HST lead to random errors of only 1$-$2\% per field. The SMC measurement is based on 35 HST fields, also centered on background quasars, and spanning a 3 year baseline, as well as an additional 8 Gaia DR1 stars. The PM measurements of both galaxies are consistent with the Gaia DR2 measurements \citep{gaiadr2b}.

Galactocentric quantities are calculated using the same Cartesian coordinate system (X, Y, Z) as in \citet{k13}. In this system, the origin is at the Galactic center, the X-axis points in the direction from the Sun to the Galactic center, the Y-axis points in the direction of the Sun's Galactic rotation, and the Z-axis points toward the Galactic north pole. The position and velocity of the dwarfs in this frame can be derived from the observed sky positions, distances, line-of-sight velocities, and PMs. Errors in the Galactocentric quantities are calculated by doing 1000 Monte Carlo drawings over the errors in the measured PMs (including reported covariances), radial velocities and distance moduli. The Local Standard of Rest velocity at the solar circle from \citet{mcmillan11} and solar peculiar velocity from \citet{schonrich10} are used in the transformation from sky coordinates to Galactocentric coordinates (see caption for Table~\ref{tab:GCs}).

Table \ref{tab:GCs} provides the Cartesian Galactocentric quantities for each satellite galaxy in our sample. The errors on each position and velocity component represent the standard deviation on that quantity derived from 1000 Monte Carlo samples.

\begin{deluxetable*}{lcccccccc}
\label{tab:pm_rv}
\tabletypesize{\footnotesize}
\tablecaption{Properties of the Candidate Magellanic Satellites}
\tablehead{\colhead{Name} & \colhead{$m-M$} & \colhead{RA} & \colhead{Dec.} & \colhead{$\rm V_{LOS}$}  & \colhead{$\mu_{\alpha^*}$} & $\mu_{\delta}$ [mas/yr]& $C_{\mu_{\alpha},\mu_{\delta}}$  & Notes \\
   & & [deg] & [deg] & [km/s] & [mas/yr]  & [mas/yr]  & [mas/yr]  & }
\startdata
\multicolumn{9}{c}{UFDs}  \\ \hline
Aquarius 2	& 20.16 $\pm$	0.07 &	338.5 &	-9.3 &	-71.1 $\pm$	2.5	& -0.252 $\pm$	0.526 &	0.011	0.448 &	0.131 & DM: [12]; PM: [2]; RV: [12]\\ 
Canes Ventici 2	& 21.02 $\pm$	0.06 &	194.3 &	34.3 &	-128.9 $\pm$	1.2 & 	-0.342 $\pm$	0.232 &	-0.473 $\pm$	0.169 & 	-0.006 & DM: [13]; PM: [2]; RV: [32]\\
Carina 2	& 17.79 $\pm$	0.05 &	114.1 & 	-58.0 &	477.2 $\pm$	1.2 & 	1.79 $\pm$	0.06 &	0.01$\pm$	0.05& 	0.03 & DM: [14]; PM: [3]; RV: [33]\\
Carina 3	& 17.22 $\pm$	0.1 & 	114.6 &	-57.9 &	284.6 $\pm$	3.4 & 	3.046 $\pm$	0.119 &	1.565$\pm$	0.135 & 	0.066 & DM: [14]; PM: [2]; RV: [33]\\
Crater 2	& 20.25 $\pm$	0.1 & 	177.3 &	-18.4 &	87.5 $\pm$	0.4 & 	-0.184 $\pm$	0.061 &	-0.106 $\pm$	0.031 &	-0.041 & DM: [15]; PM: [2]; RV: [34]\\ 
Draco 2	& 16.66 $\pm$	0.04 & 	238.2 &	64.6 &	-347.6 $\pm$	1.8 &	1.242 $\pm$	0.276 &	0.845$\pm$	0.285 &	-0.591 & DM: [16]; PM: [2]; RV: [35]\\
Horologium 1	& 19.6 $\pm$	0.2 & 	43.9 &	-54.1 &	112.8 $\pm$	2.6 &	0.891$\pm$	0.088 & 	-0.55$\pm$	0.08 &	0.294 & DM: [17,18]; PM: [2]; RV: [36]\\
Hydrus 1	& 17.2 $\pm$ 0.04 &	37.4 &	-79.3 &	80.4 $\pm$	0.6 &	3.733$\pm$	0.038 &	-1.605$\pm$	0.036 &	0.264 & DM: [20]; PM: [2]; RV: [20]\\
Hydra 2	& 20.89 $\pm$	0.12 &	185.4 &	-32.0 &	303.1 $\pm$	1.4 &	-0.416$\pm$	0.519 &	0.134$\pm$	0.422 &	-0.427 & DM: [19]; PM: [2]; RV: [37]\\
Phoenix 2	& 19.6 $\pm$	0.2 & 	355	& -54.4 & 	-42	$\pm$6 &	0.49 $\pm$	0.11 &	-1.03 $\pm$	0.12&	-0.48 & DM: [21]; PM: [4]; RV: [11]\\
Reticulum 2	& 17.5 $\pm$ 0.1 &	53.9 &	-54.0 & 	62.8 $\pm$	0.5 &	2.33$\pm$	0.07 &	-1.33$\pm$	0.08 &	0.06 & DM: [21]; PM: [3]; RV: [38] \\
Segue 1	& 16.8 $\pm$ 0.2 &	151.8 & 16.1 &	208.5 $\pm$	0.9 &	-1.697$\pm$	0.195 &	-3.501$\pm$	0.175&	-0.087 & DM: [22]; PM: [2]; RV: [39]\\
Tucana 3	& 16.8 $\pm$	0.1 &	359.1 &	-59.6 &	-102.3 $\pm$	2 &	 -0.025$\pm$0.034 	 & 		-1.661 $\pm$ 0.035	&-0.401 & DM: [21]; PM: [2]; RV: [40,41] \\ \hline 
\multicolumn{9}{c}{classical dwarfs}  \\ \hline
Carina 1 &	20.0$\pm$0.08 &	100.4 & -51.0	&229.1$\pm$	0.1 & 0.495$\pm$0.015 &	0.143$\pm$0.014 & -0.08 & DM: [23,24]; PM: [1]; RV: [42] \\ 
Draco 1 &	19.49$\pm$0.17 &	260.1 & 57.9 &	-291.0$\pm$0.1	& -0.019$\pm$0.009 & -0.145$\pm$0.01 &	-0.08 & DM: [25,26]; PM: [1]; RV: [43] \\
Fornax 1 &	20.72$\pm$0.04 &	40.0 &	-34.4 & 55.3 $\pm$0.3	& 0.376$\pm$0.003 &	-0.413$\pm$0.003	&-0.09 & DM: [27]; PM: [1]; RV: [42,44] \\
Sculptor 1 &	19.64$\pm$0.13 &	15.0 &	-33.7 &	111.4$\pm$0.1 &	0.082$\pm$0.005 &-0.131$\pm$0.004 &	0.23 & DM: [28,29]; PM: [1]; RV: [42,44] \\
Ursa Minor 1 & 19.4$\pm$0.11 &	227.3 &	67.2 &	-246.9$\pm$0.1 & -0.182$\pm$0.01 & 0.074$\pm$0.008 &-0.34 & DM: [30,31]; PM: [1]; RV: [45] \\
LMC & 18.50$\pm$0.1 & 78.76 & -69.19 &  262.2 $\pm$3.4& -1.910$\pm$0.020 & 0.229$\pm$ 0.047 & -- &DM:[5]; PM:[6], RV:[7] \\ 
SMC & 18.99$\pm$0.1 & 13.18 & -72.83  & 145.6$\pm$0.6 & -0.83 $\pm$ 0.02 & -1.21 $\pm$ 0.01 & -- & DM:[8]; PM:[9], RV:[10]\\
\enddata
\tablecomments{Column 1: distance modulus, Column 2 and 3: R.A. and Dec., Column 4: line-of-sight velocity, Column 5 and 6: PMs in the R.A. and Dec. directions (without the additional systematic error included), Column 7: the covariance between the two PM components, Column 8: original reference for each measurement. References: [1] \citet{gaiadr2b}; [2] \citet{fritz18}; [3] \citet{massari18}; [4] \citet{pace19}; [5] \citet{freedman01}; [6] \citet{k13}; [7] \citet{vdm02}; [8] \citet{cioni00}; [9] \citet{zivick18}; [10] \citet{harris06}; [11] \citet{fritz19}; [12] \citet{torrealba16b}; [13] \citet{greco08}; [14] \citet{torrealba18}; [15] \citet{joo18}; [16] \citet{longeard18}; [17] \citet{koposov15}; [18] \citet{bechtol15}; [19] \citet{vivas16}; [20] \citet{koposov18}; [21] \citet{mutlupakdil18}; [22] \citet{belokurov07}; [23] \citet{coppola15}; [24] \citet{vivas13}; [25] \citet{bonanos04}; [26] \citet{kinemuchi08}; [27] \citet{rizzi07}; [28] \citet{martinezvazquez16}; [29] \citet{pietrzynski08}; [30] \citet{carrera02}; [31] \citet{bellazzini02}; [32] \citet{simon07}; [33] \citet{li18b}; [34] \citet{caldwell17}; [35] \citet{martin16}; [36] \citet{koposov15b}; [37] \citet{kirby15}; [38] \citet{simon15}; [39] \citet{simon11}; [40] \citet{simon17}; [41] \citet{li18a}; [42] \citet{walker09b}; [43] \cite{walker15}; [44] \citet{battaglia12}; [45] \citet{kirby10}
}
\end{deluxetable*}
    
\begin{deluxetable*}{ccccccc}
\label{tab:GCs}
\tablecaption{Galactocentric Properties of Candidate Magellanic Satellites}
\tabletypesize{\footnotesize}
\tablehead{ \colhead{}& \colhead{X} &  \colhead{Y}&  \colhead{Z}&  \colhead{$\rm V_x$} & \colhead{$\rm V_y$} &  \colhead{$\rm V_z$} \\
        & [kpc] & [kpc] & [kpc] & [km s$^{-1}$] & [km s$^{-1}$] & [km s$^{-1}$] }
\startdata
Aqu2 & 28.71$\pm$1.23 & 53.16$\pm$1.77 & -85.98$\pm$2.87 & 91.31$\pm$239.21 &  250.76$\pm$212.3 & 130.49$\pm$166.0  \\ 
CanVen2 & -16.37$\pm$0.22 & 18.58$\pm$0.51 & 158.67$\pm$4.32 & -0.66$\pm$162.9 &  -203.05$\pm$150.42 & -70.09$\pm$16.93  \\ 
Car2 & -8.3$\pm$0.0 & -34.54$\pm$0.8 & -10.65$\pm$0.25 & 134.12$\pm$11.0 &  -287.58$\pm$4.14 & 134.95$\pm$13.02  \\ 
Car3 & -8.29$\pm$0.0 & -26.6$\pm$1.24 & -8.06$\pm$0.37 & -10.7$\pm$18.9 &  -151.85$\pm$8.41 & 356.05$\pm$25.9  \\ 
Cra2 & 10.3$\pm$0.88 & -81.23$\pm$3.86 & 75.13$\pm$3.57 & -34.4$\pm$35.2 &  115.88$\pm$21.41 & 2.83$\pm$19.96  \\ 
Dra2 & -10.57$\pm$0.04 & 15.58$\pm$0.28 & 14.61$\pm$0.26 & 22.54$\pm$22.16 &  100.31$\pm$22.35 & -341.04$\pm$25.48  \\ 
Hor1 & -7.16$\pm$0.1 & -48.01$\pm$4.36 & -67.91$\pm$6.16 & -20.24$\pm$30.24 &  -150.18$\pm$45.34 & 152.34$\pm$32.09  \\ 
Hyi1 & 1.87$\pm$0.19 & -19.59$\pm$0.36 & -16.48$\pm$0.3 & -144.15$\pm$6.58 &  -178.7$\pm$8.73 & 288.26$\pm$8.57  \\ 
Hya2 & 47.82$\pm$3.06 & -117.14$\pm$6.39 & 76.34$\pm$4.17 & -165.16$\pm$302.26 &  -92.01$\pm$257.22 & 208.27$\pm$275.52  \\ 
Phx2 & 25.47$\pm$3.14 & -24.81$\pm$2.31 & -71.85$\pm$6.69 & -67.68$\pm$48.82 &  -165.47$\pm$54.59 & 162.72$\pm$31.4  \\ 
Ret2 & -9.63$\pm$0.06 & -20.38$\pm$0.96 & -24.14$\pm$1.14 & 19.92$\pm$12.38 &  -96.74$\pm$17.42 & 218.24$\pm$14.63  \\ 
Seg1 & -19.38$\pm$0.98 & -9.47$\pm$0.84 & 17.67$\pm$1.57 & -98.19$\pm$18.34 &  -205.06$\pm$38.14 & -35.49$\pm$22.9  \\ 
Tuc3 & 0.79$\pm$0.41 & -8.95$\pm$0.4 & -19.03$\pm$0.85 & 23.48$\pm$5.94 &  146.27$\pm$8.05 & 185.68$\pm$5.69  \\  \hline 
Car1 & -24.72$\pm$0.6 & -94.62$\pm$3.48 & -39.26$\pm$1.44 & -36.84$\pm$18.42 &  -50.55$\pm$8.51 & 149.23$\pm$20.28  \\ 
Dra1 & -4.15$\pm$0.32 & 64.88$\pm$5.0 & 45.01$\pm$3.47 & 54.29$\pm$13.85 &  4.15$\pm$8.25 & -151.78$\pm$11.73  \\ 
Fnx1 & -39.58$\pm$0.57 & -48.15$\pm$0.87 & -126.93$\pm$2.3 & 38.14$\pm$22.76 &  -107.56$\pm$21.25 & 76.0$\pm$9.72  \\ 
Scu1 & -5.22$\pm$0.19 & -9.59$\pm$0.6 & -84.12$\pm$5.26 & 16.93$\pm$12.7 &  175.89$\pm$16.04 & -96.14$\pm$1.87  \\ 
UMin1 & -22.16$\pm$0.71 & 52.0$\pm$2.68 & 53.46$\pm$2.75 & -4.26$\pm$10.75 &  46.77$\pm$10.69 & -148.2$\pm$10.61  \\  
LMC & -1.06$\pm0.33$ & -41.05$\pm$1.89 & -27.83$\pm$1.28 & -57.60 $\pm$7.99 & -225.96$\pm$12.60 & 221.16$\pm$16.68\\
SMC & 15.05$\pm$1.07 & -38.10$\pm$1.75 &  -44.18$\pm$2.03 & 17.66$\pm$3.84  & -178.60$\pm$15.89 & 174.36 $\pm$12.47 \\\hline     
\enddata
\tablecomments{All quantities are calculated directly from the values compiled in Table \ref{tab:pm_rv}. Solar reflex motion is taken from \citet{mcmillan11} where $\rm V_{c,peak} (8.29 \, kpc) \approx 239\, km \, s^{-1}$. We adopt the solar peculiar velocity from \citet{schonrich10} who find $\rm (U,V,W)_{\odot}=(11.1^{+0.69}_{-0.75}, 12.24^{+0.47}_{-0.47}, 7.25^{+0.37}_{-0.36})\, km \, s^{-1}$. Note the standard errors on each component represent the standard deviation from one iteration of the Monte Carlo scheme (i.e. 1000 random samples). The horizontal line indicates the division between ultra-faint galaxies and the classical satellite galaxies. Galaxies will appear in this order in tables moving forward.} 
\end{deluxetable*}

\section{Analytic Orbital Models}
\label{sec:model}
In this section, we briefly describe the method used to calculate orbital histories for all satellites in our sample using the Galactocentric positions and velocities provided in Table \ref{tab:GCs} as initial conditions. This method follows the general strategies outlined in \citet{k13}, \citet{gomez15}, and further modified in \citet[][hereafter \citetalias{patel17a}]{patel17a}. 

\subsection{Galaxy Potentials}
To numerically integrate orbits backwards in time, the gravitational potentials of the MW, LMC, SMC, and all satellites are modeled as extended mass distributions. The following subsections outline the specific parameters of each galactic potential.

\subsubsection{Milky Way Potential}
\label{subsubsec:mw}
Two MW dark matter halo potentials are considered throughout this analysis to account for both a light and heavy MW scenario, identical to the MW models in \citetalias{patel17a}. The light MW mass model will be referred to as MW1 and has a virial mass\footnote{We adopt the \citet{brynorman98} definition of virial mass using $\Omega_{\rm m} =0.27$, $h=0.7$, and $\Delta_{\rm vir}=359$.} of $10^{12} \, M_{\sun}$ and virial radius of 261 kpc. The heavy MW mass potential, MW2, has a virial mass of $1.5 \times 10^{12} \, M_{\sun}$ and virial radius of 299 kpc. 

Each MW potential is a composite of an Navarro-Frenk-White (NFW) halo \citep{nfw96}, a Miyamoto-Nagai disk \citep{mn75}, and a Hernquist bulge \citep{hernquist90}. The NFW dark matter halo is adiabatically contracted owing to the presence of the disk using the \texttt{CONTRA} code \citep{contra}. The density profile of the MW's halo is truncated at the virial radius of each model. Beyond the virial radius, the potential of the MW is treated as a point mass as in \citetalias{patel17a}.

The MW's disk mass in each model was chosen to provide the best match to the observed rotation curve from \citet{mcmillan11}, such that the peak velocity reaches $V_c \approx 239 \rm \, km \, s^{-1}$ at the solar radius. Fig. 1 in \citetalias{patel17a} illustrates the rotation curves of our adopted MW models. All MW halo, disk, and bulge parameters for each model are listed in Table \ref{table:mwparams}.

\begin{table}
\centering

\begin{tabular}{lp{1.5cm}p{1.5cm}}\hline\hline 
& MW1 & MW2 \\ \hline
Mvir [10$^{10}$ $M_{\odot}$] & 100 & 150 \\ 
Rvir [kpc] & 261 & 299  \\
c$_{\rm vir}$ & 9.86 & 9.56  \\ 
M$_{\rm d}$ [10$^{10}$ $M_{\odot}$] & 6.5 & 5.5  \\ 
R$_{\rm d}$ [kpc]& 3.5 & 3.5  \\
z$_{\rm d}$ [kpc] &0.53 & 0.53 \\ 
M$_{\rm b}$ [10$^{10}$ $M_{\odot}$] & 1 & 1   \\ 
R$_{\rm b}$ [kpc] & 0.7 & 0.7  \\ \hline
\end{tabular}
\caption{Model parameters for each MW mass model. These are identical to the MW models in \citetalias{patel17a}. From top to bottom the rows list: 1) virial mass following the \citet{brynorman98} definition, 2) virial radius calculated with Eq. A1 from \citet{vdm12ii}, 3) virial concentration, 4) stellar disk mass, 5) stellar disk radial scale length, 6) stellar disk scale height, 7) bulge mass, 8) bulge scale length. \label{table:mwparams}}
\end{table}

\subsubsection{LMC and SMC Potentials}
\label{subsubsec:lmc}
The LMC potential is modelled using two components, namely a Hernquist halo and a Miyamoto-Nagai disk. We consider three total masses for the LMC at infall: $0.8, 1.8, 2.5 \times 10^{11}\, M_{\sun}$, which will be referred to as LMC1, LMC2, and LMC3, respectively. The mass of the LMC's disk is held fixed at its present day stellar mass $M_d = 3\times 10^9 \, M_{\sun}$ \citep{vdm02} for all three models and the Hernquist halo scale radius is varied to match the rotation velocity of $V_c \approx 92$ km s$^{-1}$ at 8.7 kpc \citep{vdmnk14}. All LMC model parameters are listed in Table \ref{table:lmcparams}. 

As in \citet{garavitoc19}, the majority of this work will focus on the intermediate mass LMC2, our fiducial LMC model. This mass is consistent with recent models of the Magellanic system and with the halo mass estimates from abundance matching \citep{besla12, besla13, besla16}. However, we will discuss the effects of a lower (LMC1) and higher (LMC3) LMC mass model throughout this analysis. 

The SMC is modeled as a Hernquist halo since its baryonic content is much less massive than the LMC's, owing to repeated encounters with the LMC \citep{besla12}. Such encounters also imply the halo of the SMC is truncated today. The Hernquist halo scale radius ($\rm r_H$) is determined by matching the mass profile to the dynamical mass within 3 kpc of the center of the SMC, $\rm M(3\, kpc) \approx 2\times10^9 \, M_{\sun}$ \citep{harris06}. \citet{diteodoro19} find that a dynamical mass within 4 kpc of $\rm M(4\, kpc) \approx 1-1.5\times10^9 \, M_{\sun}$, is required to reproduce the SMC's rotation curve. Our models are therefore representative of the SMC's current properties. We consider two different SMC models with halo masses of $5\times10^9 M_{\sun}$ (SMC1) and $3\times10^{10}\, M_{\sun}$ (SMC2), respectively. For such halo masses, the SMC's baryon fraction is 5\% \citep[excluding the gas content of the Magellanic Stream;][]{besla15}. The model parameters for the SMC potentials are listed in Table \ref{table:lmcparams}.

\begin{table}
\centering
\begin{tabular}{llllll}\hline\hline 
& LMC 1& LMC2$^a$ & LMC3 & SMC1 & SMC2\\ \hline
$M_{\rm H}$ [10$^{10}$ $M_{\odot}$] & 8 & 18 & 25 & 0.5 & 3 \\
$R_{\rm vir}$ [kpc] & 113 & 148 & 165 & 45 & 81 \\
$\rm r_H$ [kpc] & 12.5 & 23.1 & 28.8 & 2.5 & 8.6\\
M$_{\rm d}$ [10$^{9}$ $M_{\odot}$] & 3 & 3 & 3  & -- & --\\ 
R$_{\rm d}$ [kpc]& 1.7 & 1.7 & 1.7 & -- & -- \\
z$_{\rm d}$ [kpc] &0.27 & 0.27 & 0.27 & -- & --\\ \hline
\end{tabular}
\caption{Model parameters for the LMC and SMC potentials. The LMC is a two component disk+halo potential and the SMC is only modeled as a Hernquist sphere. From top to bottom the rows list: 1) Hernquist halo mass, 2) virial radius, 3) Hernquist scale radius, 4) stellar disk mass, 5) stellar disk radial scale length, 6) stellar disk scale height.\\
$^a$ indicates the fiducial LMC model
\label{table:lmcparams}}
\end{table}

\subsubsection{Classical Satellites}
All classical satellites considered in this work have stellar masses in the range $M_* = 10^5-10^7\, M_{\sun}$. As such, the dark matter halos of all classical satellites fainter than the MCs are modeled as Plummer spheres \citep{plummer11} with a total halo mass of $10^{10}\, M_{\sun}$ \citep[see][]{bullockbk17}. The Plummer scale radius for each classical satellite is determined by computing the radius at which the halo mass enclosed within the Plummer profile matches the dynamical mass inferred at the half-light radius. The half-light radii and stellar velocity dispersions used to compute dynamical masses, which are derived from the \citet{walker09c} dynamical mass estimator, are compiled in Table \ref{tab:size_lumninosity_table}. The resulting Plummer scale radii ($\rm r_P$) for the classical satellites are as follows: Carina (4.0 kpc), Draco (2.1 kpc), Fornax (4.4 kpc),  Sculptor (2.7 kpc), Ursa Minor (3.9 kpc). 

\begin{deluxetable}{ccccc}
\label{tab:size_lumninosity_table}
\tablecaption{Absolute Magnitude, Size, and Velocity Dispersion of Candidate Magellanic Satellites}
\tabletypesize{\footnotesize}
\tablehead{ \colhead{}& \colhead{$M_V$} &  \colhead{$R_{1/2}$}&  \colhead{$\sigma$} &\colhead{Refs.$^a$} \\
        & & [pc] & [km s$^{-1}$] & }
\startdata
Aqu2 & $-4.36\pm0.14$ & $160\pm26$ & $5.4^{+3.4}_{-0.9}$ & [12,12,12]\\
CanVen2 & $-5.17\pm0.32$ & $71\pm11$ & $4.6\pm1.0$& [46,46,32]\\
Car2 & $-4.50\pm0.10$ & $92\pm8$ & $3.4^{1.2}_{-0.8}$ & [14,14,33]\\
Car3 & $-2.4\pm0.20$ & $30\pm8$ & $5.6^{+4.3}_{-2.1}$ & [14,14,33]\\
Cra2 & $-8.20\pm0.10$ & $1066\pm86$ & $2.7\pm0.3$ & [47,47,34]\\
Dra2 & $-0.80^{+0.40}_{-1.00}$  &  $19^{+4}_{-3}$ & $<5.9$ & [16,16,16]\\
Hor1 & $-3.76\pm0.56$ & $40^{+1}_{-9}$ & $4.9^{+2.8}_{-0.9}$ & [46,46, 36]\\
Hyi1 & $-4.71\pm0.08$ & $53\pm4$ & $2.7^{+0.5}_{-0.4}$ & [20,20,20]\\
Hya2 & $-4.86\pm0.37$ & $67\pm13$ & $<3.6$ & [46,46, 37]\\
Phx2 & $-2.70\pm0.40$ & $37\pm8$ & -- & [21,21,--]\\
Ret2 & $-3.88\pm0.38$ & $51\pm3$ & $3.3\pm0.7$ & [46,46,38]\\
Seg1 & $-1.30\pm0.73$ & $24\pm4$ & $3.7^{+1.4}_{-1.1}$ &[46,46,39] \\
Tuc3 & $-1.49\pm0.20$ & $37\pm9$ & $<1.2$ & [21,21, 40]\\ \hline
Car1 & $-9.45\pm0.05$ & $349\pm4$ & $6.6\pm1.2$ & [46,46,48]\\
Dra1 & $-8.88\pm0.05$ &$219\pm2$ & $9.1\pm1.2$ & [46,46,42]\\
Fnx1 & $-13.34\pm0.14$ & $787\pm9$ & $11.7\pm0.9$ & [46,49,42]\\
Scu1 & $-10.82\pm0.14$ & $308\pm1$ & $9.2\pm1.1$ & [46,46,42]\\ 
UMin1 & $-9.03\pm0.05$ & $383\pm2$ & $9.5\pm1.2$ & [46,46,50]\\
\enddata
\tablecomments{$^a$ Reference numbers correspond to the reference list provided at the end of Table \ref{tab:pm_rv}. Additional references include: [46] \citet{munoz18}; [47] \citet{torrealba16a}; [48] \citet{walker08}; [49] \citet{battaglia06}; [50] \citet{walker09c}}
\end{deluxetable}

\subsubsection{Ultra-faint Satellites}
UFDs, as their name suggests, are much fainter and have significantly lower stellar masses than the classical dwarfs ($M_* = 10^2-10^5$ vs. $M_* = 10^5-10^7$). UFDs are also significantly smaller in size with half-light radii typically $\lesssim 200\, \rm pc$ and  stellar velocity dispersions $\lesssim 5 \, \rm km\, s^{-1}$. Table \ref{tab:size_lumninosity_table} lists the absolute magnitudes, half-light radii, and stellar velocity disperions for UFD galaxies in our sample. 

Given their low stellar masses, UFDs are also modeled as Plummer spheres but with a total halo mass of $10^9\, M_{\sun}$ \citep{jeon17}. Applying the \citet{walker09c} mass estimator to the measured velocity dispersion and half light radii of Aqu2, Ret2, and Hor1, and searching for the radius at which the mass enclosed within a Plummer profile is equivalent to the dynamical mass yields scale radii of 1.2, 0.8, and 0.6 kpc for these satellites respectively. Thus, all UFDs are assigned the same Plummer scale radius of 1 kpc for simplicity since velocity dispersion values are unavailable or uncertain for some UFDs in our sample (see Table \ref{tab:size_lumninosity_table}). Since Crater 2 (Cra2) is a known outlier on the size-luminosity relation with a size similar to Fornax and the SMC but a luminosity that is consistent with the UFD satellites \citep[][see Table \ref{tab:size_lumninosity_table}]{torrealba16a}, we adopt a more appropriate Plummer scale radius of 9 kpc for Cra2.

\subsection{Dynamical Friction and Numerical Integration Scheme}
\label{sec:DFall}
\subsubsection{Acceleration from the MW}
\label{subsec:df}
We present orbital solutions for each candidate Magellanic satellite galaxy in three different scenarios. First, we calculate the orbit of each satellite in the presence of the MW only. Then we consider the combined MW and LMC potential. Finally, we add the SMC and calculate orbits in the full MW+LMC+SMC potential. The progression of adding one galactic potential at a time allows us to disentangle the influence of each additional massive body. In every scenario, each galaxy experiences the gravitational influence of every other galaxy (up to $N_{\rm gal}=4$). Though its gravitational potential is static, the MW's center of mass is not held fixed and therefore moves in response to the LMC's close passage as in \citet{gomez15} and \citetalias{patel17a}. 

Satellites passing through the halo of the MW experience dynamical friction (DF) as approximated by the Chandrasekhar formula \citep{chandrasekhar}:

\begin{equation}\textbf{F}_{df}= \rm - \frac{4\pi G^2 M_{sat}^2 ln\Lambda \rho(r)}{v^2} \left[ erf(X) - \frac{2X}{\sqrt\pi} exp(-X^2)\right] \frac{\textbf{v}}{v}.\label{eq:df} \end{equation} 

Here, $\rho(r)$ is the density of the MW's adiabatically contracted dark matter halo at a distance $r$ from the Galactic center. $X=v/\sqrt{2\sigma}$ where $\sigma$ is the one-dimensional galaxy velocity dispersion for an NFW halo derived in \citet{zentner03}. DF also depends on the total mass of the satellite $\rm M_{sat}$ as well as its total velocity $v$. The Coulomb logarithm (ln$\rm\Lambda$) is calibrated with respect to each class of satellite (i.e. massive satellites like the LMC, classical satellites including the SMC, and UFDs). For all MW-LMC acceleration calculations, we adopt the Coulomb parametrization in \citet{vdm12iii}:

\begin{equation}\rm ln \Lambda = max[L, ln(r/Ca_s)^{\alpha}], \label{eq:dfvdm} \end{equation} 

\noindent where L=0, C=1.22, $\alpha$=1.0. These values are constants that parametrize the best-fitting match for the orbit of a 1:10 host-satellite mass ratio from N-body simulations. $a_s$ is the scale radius of the satellite, which is the Hernquist scale length ($\rm r_{H}$) or Plummer scale length ($\rm r_P$) depending on the satellite's potential. 

We adopt the Coulomb logarithm from \cite{hashimoto03} for the DF approximation used for the SMC, the classical satellites, and the UFDs as they move through the MW's halo:

\begin{equation}\rm ln \Lambda = \frac{r}{1.4\ a_s},\label{eq:dfhashi} \end{equation} 

\noindent where $a_s$ is once again the satellite scale radius and $r$ is the distance of the satellite from the MW's Galactic Center. The total acceleration felt by all satellites owing to the MW only is then:

 \begin{equation} \rm \ddot r_{sat, MW} = \frac{d\Phi_{bulge, MW}}{d\textbf{r}} + \frac{d\Phi_{disk,MW}}{d\textbf{r}} + \frac{d\Phi_{halo,MW}}{d\textbf{r}} + \frac{\textbf{F}_{df}}{M_{sat}}
 \label{eq:totalsatforces_mw}
 \end{equation}
 
\noindent and the total acceleration felt by the MW as a result of each satellite is:
\begin{equation} \rm \ddot r_{MW} =  \frac{d\Phi_{sat}}{d\textbf{r}}. \label{eq:mwforce}\end{equation}

Note that in the case of the LMC, the MW will experience two acceleration forces since the LMC is modelled as a disk plus halo potential (i.e. $\rm \Phi_{LMC} = \Phi_{disk} +  \Phi_{halo}$).

\subsubsection{Acceleration from the LMC}
 \begin{figure*}
    \centering
    \includegraphics[scale=0.55, trim=0mm 5mm 0mm 0mm]{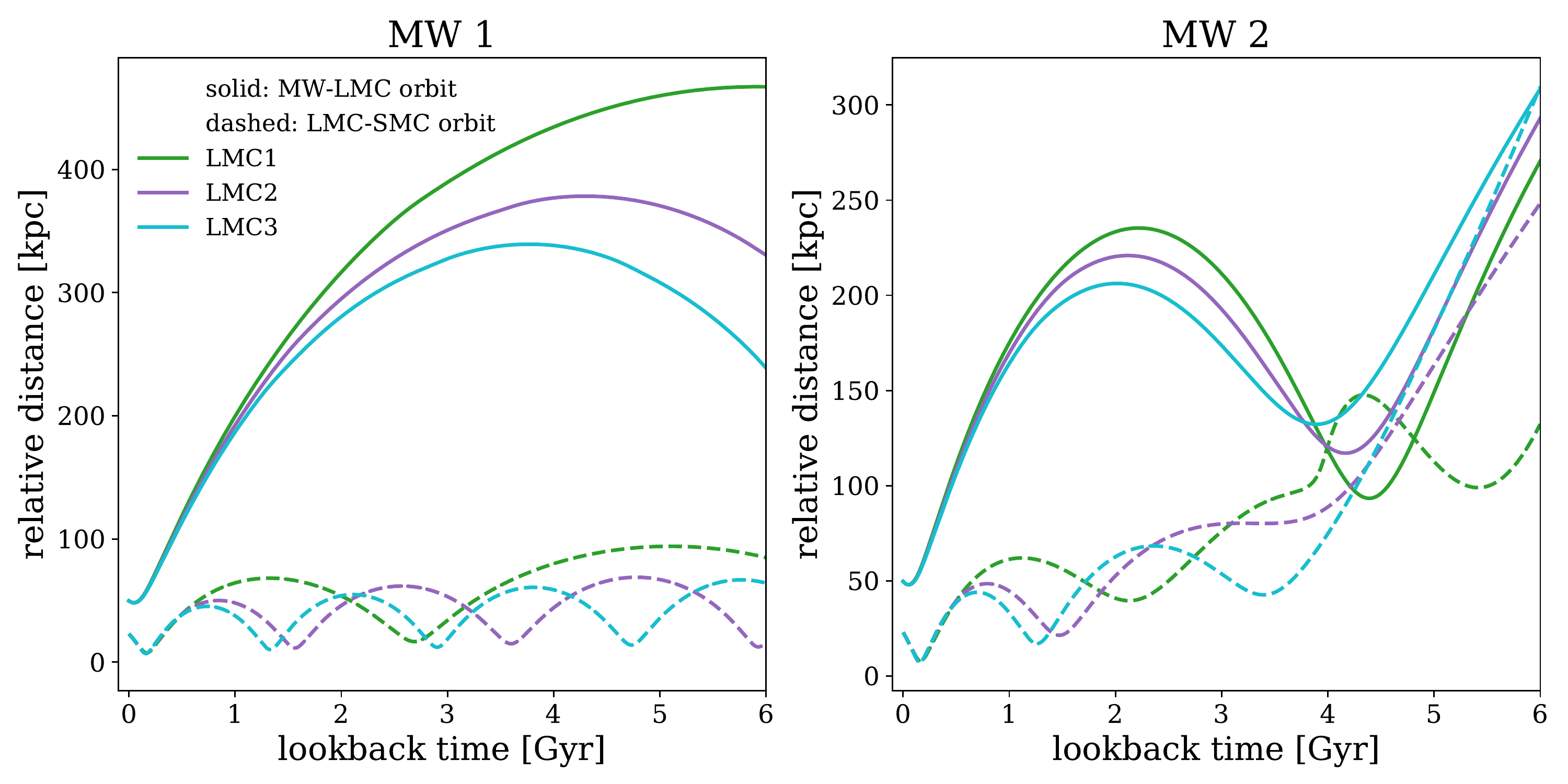}
    \caption{Direct orbits for the LMC (solid lines) relative to the MW and the SMC relative to the LMC (dashed lines). The left panel shows all orbits calculated for the low mass MW1 model, while the right panel illustrates orbits in the high mass MW2 model. All orbits are calculated in the combined MW+LMC+SMC potential. In MW1, the LMC is on a wide period orbit and only completes one pericentric passsage in the last 6 Gyr regardless of LMC mass. The SMC makes multiple passes around the LMC in MW1. In MW2 the LMC completes two pericentric passages in the last 6 Gyr. The binary orbit of the LMC-SMC is disrupted at times greater than 3 Gyr ago in a high mass MW model, shortly after the system makes a close encounter with the MW.} 
    \label{fig:MW_LMC_SMC}
\end{figure*}

\begin{table}
    \centering
    \begin{tabular}{ccc}\hline \hline 
     
     {\bf MW1}  & $\rm r_{inner}$ [kpc] & $\rm r_{outer}$ [kpc]  \\ \hline
    LMC1 & 15.7 & 41.9  \\ 
    LMC2$^a$ & 16.9 & 60.8  \\ 
    LMC3 & 17.3 & 74.2  \\ \hline \hline
    {\bf MW2} & $\rm r_{inner}$ [kpc] & $\rm r_{outer}$ [kpc] \\ \hline 
    LMC1 & 16.3 & 46.1  \\ 
    LMC2$^a$ & 17.6 & 68.2  \\ 
    LMC3 & 18.0 & 84.2  \\ \hline
    \end{tabular}
    \caption{Distance from the center of the LMC where the density of the MW and LMC are equal today. DF due to galaxies passing through the LMC's halo is implemented when satellites pass within $\rm r_{outer}$.\\
    $^a$ indicates the fiducal LMC model}
    \label{tab:rt}
\end{table}

Since the LMC is 8-25 times more massive than the classical satellites and 80-250 times more massive than the UFDs in our models, it too will exert a drag force that slows the orbital motion of satellites that pass through regions where the LMC's halo DM density is in excess of the ambient MW halo. As we calculate orbits backwards in time, DF translates to an acceleration force. This is accounted for using the same dynamical friction approximation adopted in \citet{bekki05} and \citet{b07} to account for the effect of the SMC passing through the LMC's halo. 

\begin{equation}
    \textbf{F}_{df,\rm LMC}= \rm 0.428 ln \Lambda\frac{G M_{sat}^2}{r^2} \label{eq:df_mcs}
\end{equation}

Here $r$ is now the distance between the satellite and the center of the LMC and $\rm ln \Lambda =0.3$ \citep[instead of $\rm ln \Lambda =0.2$ as in][]{b07}. This value for $\rm ln \Lambda$ was chosen by finding the best analytic match to the LMC-SMC orbit from N-body simulations, prior to accretion by the MW \citep{besla10, besla12}. Between SMC1 and SMC2, we find that SMC1 provides the better fit to this simulated orbit and will use it as the fiducial SMC model throughout this analysis. 

The DF approximation given in Equation \ref{eq:df_mcs} is applied to all candidate Magellanic satellites in addition to the SMC when they fall within the region of the LMC's halo where its density dominates over the MW's. This radius is determined by finding the distance at which the density profile of the MW (as described in \S  \ref{subsubsec:mw}) is equivalent to the LMC's density profile (as described in \S \ref{subsubsec:lmc}). In doing so, we find two distances at which these quantities are equivalent, denoted as the \emph{inner} and \emph{outer} radius. These radii act as pseudo-truncation radii, thus DF owing to the LMC is only active when candidate Magellanic satellites or the SMC pass within the \emph{outer} radius ($\rm r_{outer}$) as listed in Table \ref{tab:rt}. The total acceleration felt by all satellites due to the LMC is summarized as:

 \begin{equation} \rm \ddot r_{sat, LMC} = \frac{d\Phi_{disk, LMC}}{d\textbf{r}} + \frac{d\Phi_{halo, LMC}}{d\textbf{r}} + \frac{\textbf{F}_{df, LMC}}{M_{sat}}.
 \label{eq:totalsatforces_lmc}
 \end{equation}
 
The LMC in turn experiences the acceleration of each satellite as in Eq. \ref{eq:mwforce} (replacing the MW subscript with the LMC). We have also checked whether any DF forces should be included for the UFD satellites as they pass through the halo of the SMC using the same prescription as in Equation \ref{eq:dfhashi}, however, our tests showed that this effect is negligible so we have omitted it from our model. 

\subsection{Orbits of the Magellanic Clouds}
As in \citetalias{patel17a}, the symplectic leapfrog integration method from \citet{gadget} is used to numerically integrate the equations of motions backwards in time. Orbits are only calculated for the last 6 Gyr as the mass evolution of the MW and mass loss due to tides are not included in our framework. Furthermore, the Appendix of \citetalias{patel17a} showed that a 6 Gyr integration period provides a good match to corresponding orbits of satellite galaxies from a cosmological simulation, but that at earlier times, cosmological orbits begin to deviate from the analytic results. Secondly, \citet{santistevan20} found that MW-mass galaxies have acquired about 80\% of their mass by 6 Gyr ago, so integrating further than this would require a more complex orbital model that handles the mass evolution of the MW. Finally, integrating for a shorter orbital period would not be ideal as it often takes $\sim$5-7 Gyr for satellites to complete multiple orbits around the MW (and the MCs). 

 Orbits resulting from the positions and velocities in Table \ref{tab:GCs} will be referred to as \emph{direct orbits}, i.e., the orbits calculated from the Galactocentric quantities derived directly from the transformation of average proper motion, line-of-sight velocity, and distance modulus to Cartesian coordinates centered on the MW as described in \S \ref{sec:data}. These orbits do not represent the measurement errors on the observational quantities.

Fig. \ref{fig:MW_LMC_SMC} shows the orbit of the LMC relative to the MW for all three LMC models in both MW mass potentials. In MW1, all LMC mass models are on a first infall, long period orbit with a recent pericenter occurring $\sim$50 Myr ago. In the more massive MW2 potential, all LMC models complete two pericentric passages. The first occurs at $\sim$4 Gyr ago at a distance of 100-150 kpc and the second occurs at $\sim$50 Myr ago at approximately 50 kpc. 

Fig. \ref{fig:MW_LMC_SMC} also shows the orbit of SMC1 (dashed lines) relative to the LMC in all three LMC models. The left panel shows orbits in the MW1 potential and the right panel shows the same orbits calculated in the MW2 potential. For both MW masses, the time and distance at the most recent LMC-SMC pericentric passage are  consistent with results from \citet{zivick18}, who find an impact parameter of $7.5\pm2.5$ kpc at $147\pm33$ Myr ago. 

The SMC completes multiple pericentric passages about the LMC for MW1, whereas the binary LMC-SMC orbit is disrupted at times earlier than 3 Gyr ago for MW2 \citep[see also][]{bekki05, k13,zivick18}. \cite{k13} found that the latter solution is quite implausible and that a MW mass of $\lesssim 1.5 \times 10^{12}\, M_{\sun}$ is preferred to form a long-lived LMC-SMC binary.  Given the extensive work that has been carried out on the orbit of the LMC-SMC system, we count the SMC as a satellite of the LMC moving forward.

\section{Analysis of Orbital Histories}
\label{sec:analysis}

In this section, we present the direct orbital histories for all candidate Magellanic satellites. In \S \ref{subsec:stats}, we present the statistical significance of these orbital solutions by calculating 1000 orbital histories for each candidate satellite using the fiducial LMC model and both MW masses. This analysis samples the 1$\sigma$ error space of the PMs, line-of-sight velocities, and distance moduli. Finally, in \S \ref{subsec:identification} we identify which candidate Magellanic satellites exhibit orbital histories that confirm they are dynamically associated members of the Magellanic system accounting for both the latest PM measurements and the acceleration of the SMC for the first time.

\subsection{Orbits of the Classical Satellites}
\begin{figure*}[ht!]
    \centering
    \includegraphics[scale=0.52]{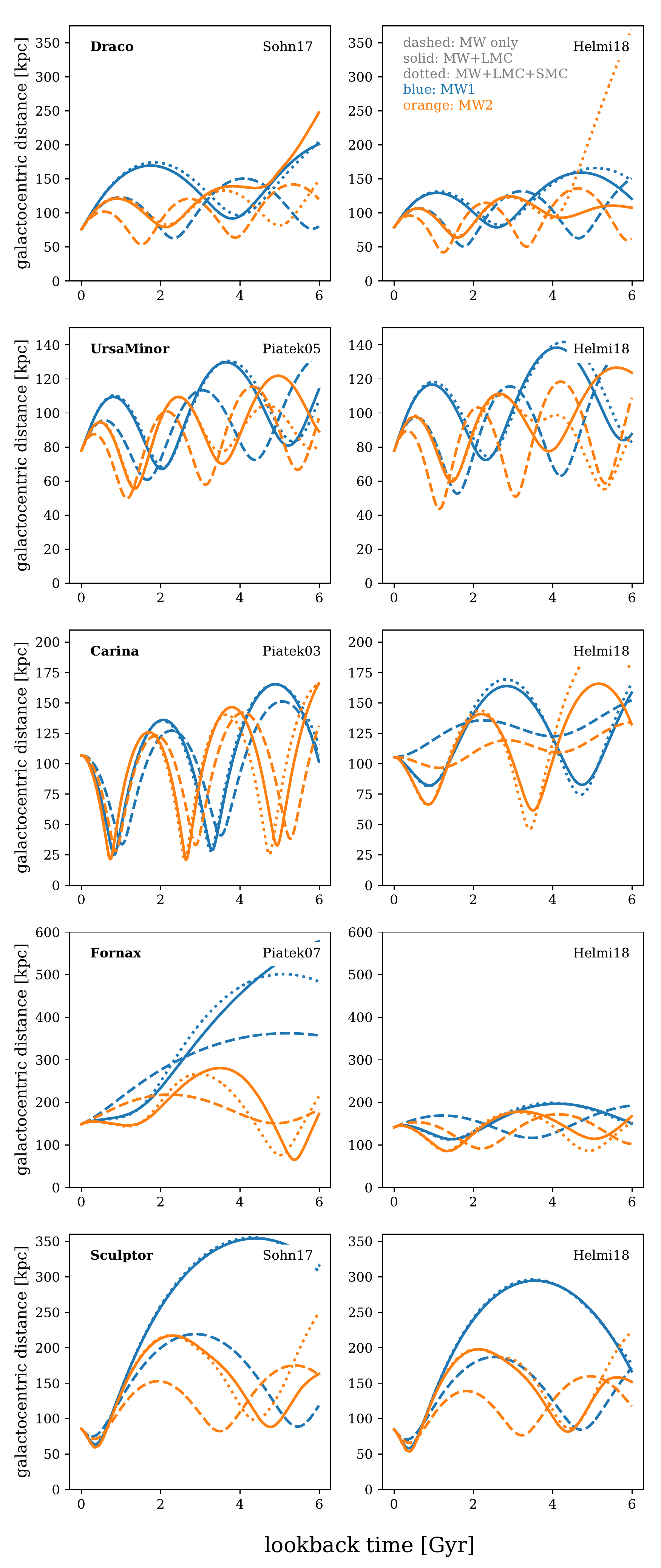}
    \caption{Direct orbits for the classical satellite galaxies included in our sample. Blue and orange lines indicate orbits calculated in MW1 and MW2, respectively. The left column shows direct orbits calculated using pre-Gaia DR2 PMs. The right column shows direct orbits calculated with Gaia DR2 PMs (see Tables \ref{tab:pm_rv} and \ref{tab:GCs}.) All classical satellites are noticeably impacted by the addition of the LMC, regardless of the PM measurement used. Carina and Fornax exhibit significant changes in their orbits as a result of PM differences. The LMC impacts the dynamics of Carina, Fornax and Sculptor most strongly, while the SMC does little to change the dynamics of the classical satellites.
    }
    \label{fig:classical_vpos}
\end{figure*}

Fig. \ref{fig:classical_vpos} shows direct orbits for the classical satellite galaxies in our sample. Note that all distances are shown relative to the Galactic Center. All blue lines correspond to MW1, while all orange lines correspond to MW2. This color scheme will remain fixed in all subsequent figures of orbital histories. The fiducial LMC mass model (LMC2) is adopted in all cases. We discuss how orbits evolve when the mass of the LMC is lower (LMC1) and higher (LMC3) in \S \ref{subsec:identification}. Each satellite's orbital history is calculated in three potentials: the MW only (dashed lines), the MW+LMC (solid lines), and the MW+LMC+SMC (dotted lines). 

Since all of the classical dwarf spheroidal galaxies have PM measurements pre-dating Gaia DR2, the left column of Fig. \ref{fig:classical_vpos} shows the direct orbits using the most recent pre-Gaia DR2 PM for each classical dwarf galaxy. These  come from \citet{sohn17} for Draco and Sculptor, \citet{piatek05} for Ursa Minor, \citet{piatek03} for Carina, and \citet{piatek07} for Fornax, as denoted in the the top right of each panel. The right column shows the direct orbits using the Gaia DR2 PMs from \citet{gaiadr2b}. 

In the cases of Carina, Fornax, and Ursa Minor the Gaia DR2 and pre-existing PMs are consistent with each other at the $2\sigma$ level of the old measurement, although the latter have large error bars (100-200 $\mu$as/yr). The Gaia DR2 PMs for these galaxies reach much higher precision (3-15 $\mu$as/yr). Draco and Sculptor's previous PM measurements were made using HST and a baseline of nearly 10 years \citep{sohn17}, thus the most recent PM measurements reach similar precision (5-20 $\mu$as/yr).

Overall, Fig. \ref{fig:classical_vpos} shows that all of the classical satellites are noticeably impacted by the gravitational influence of the LMC (dashed vs. solid lines). This effect manifests in different ways for each individual classical satellite such that the inclusion of the LMC can change the length of the orbital period, increase or decrease the distance at pericenter (apocenter), as well as alter the timing of pericenter (apocenter). However, the addition of the SMC (dotted lines) has little effect on the orbital properties of the classical satellites. This is not surprising since the adopted mass of the SMC is only 50\% of the mass used for the classical satellite galaxies.

Fig. \ref{fig:classical_vpos} shows that Carina's orbit is similar for the MW+LMC+SMC potential using both PMs. Adopting the Gaia DR2 PM leads to an orbital period that is larger by a factor of $\sim$1.3. Draco's direct orbits are also consistent between the previous and Gaia DR2 PMs. The orbits have similar periods and the most notable difference is a decrease in the distances achieved at apocenter by $\sim$25 kpc using the Gaia DR2 PMs. There is little to no difference between the orbits calculated with previous and Gaia DR2 PMs for Ursa Minor and Sculptor. 

Fornax shows the most significant differences between the previously measured and Gaia DR2 PMs. In the MW+LMC+SMC potential, Fornax's Gaia DR2 orbit indicates that it has completed nearly two orbital passages in the last 6 Gyr with the most recent passage around the MW occurring at $\sim$1.5 Gyr ago at a distance of 90-120 kpc, which is much closer than the orbit calculated using the \citet{piatek07} PMs. Carina and Fornax have recently been posited as satellites of the MCs \citep{pardy19, jahn19}. In \S \ref{subsec:stats} and  \S \ref{subsec:identification}, we will explore the statistical likelihood of this based on their orbital histories relative to the LMC.

\subsection{Orbits of the Ultra-faint Satellites}
\begin{figure*}[ht!]
    \centering
    \includegraphics[scale=0.58, trim=0mm 5mm 0mm 0mm]{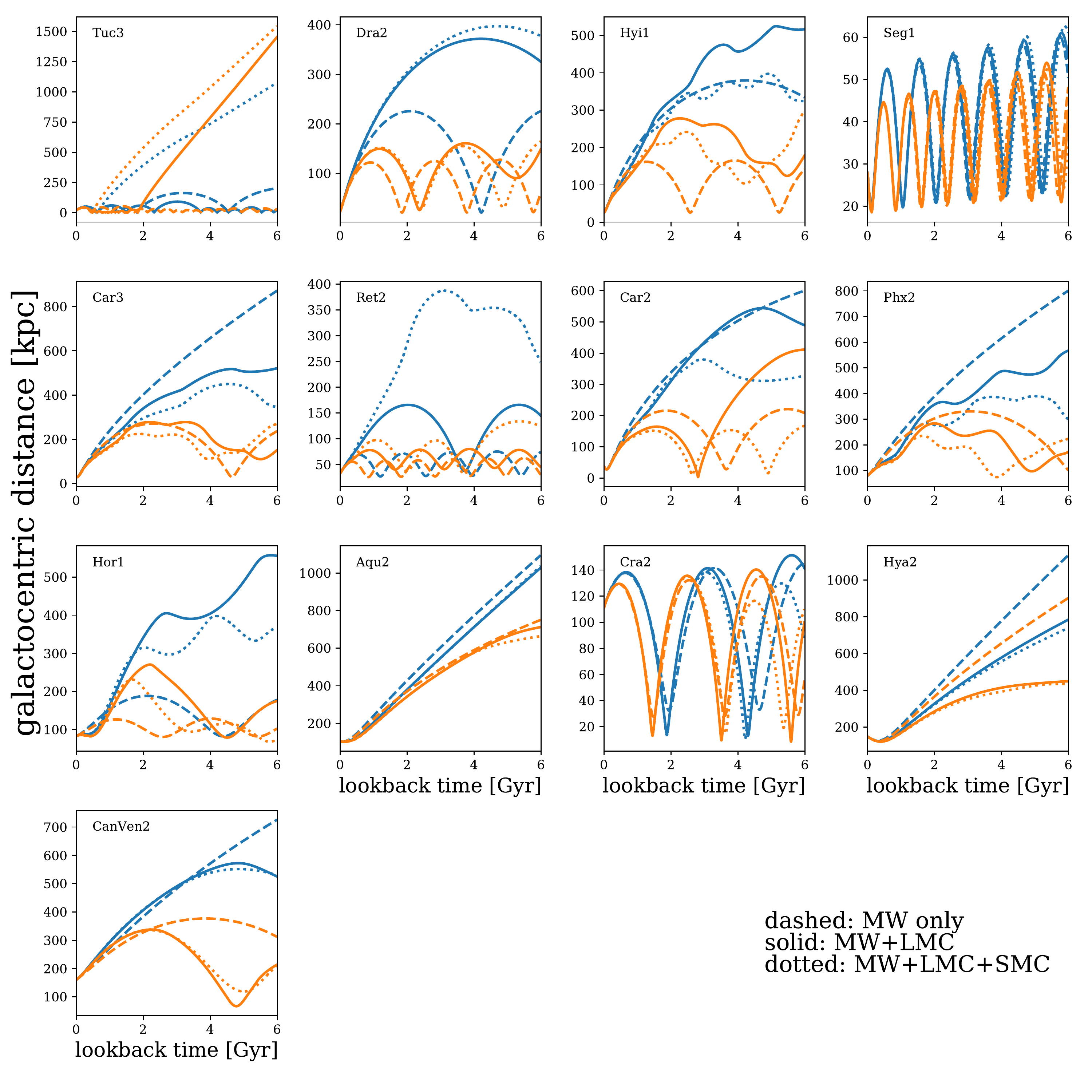}
    \caption{Direct orbits for all UFD satellite galaxies included in our sample. The blue lines indicate orbits calculated in MW1 while the orange lines represent MW2. Orbits are shown in the MW only potential (dashed lines), the MW+LMC potential (solid lines), and the MW+LMC+SMC potential (dotted lines) using galactocentric quantities derived from Gaia DR2 PMs (see Tables \ref{tab:pm_rv} and \ref{tab:GCs}). All satellites, with the exception of Seg1, Aqu2, Hya2, and CanVen2, are notably perturbed by the inclusion of the LMC. The addition of the SMC further perturbs the orbits of Hyi1, Car3, Ret2, Car2, Phx2, Tuc3 and Hor1. Of these, Tuc3 and Ret2 are the most highly affected, illustrating that the SMC can change the long-term dynamics of specific satellite's orbits.}
    \label{fig:UFD_orbits}
\end{figure*}

Fig. \ref{fig:UFD_orbits} shows the direct orbits as a function of lookback time for all 13 UFD satellites in our sample. All colors and line styles represent the same model parameters as in Fig. \ref{fig:classical_vpos}. 

For every satellite with the exception of Seg1, there are noticeable differences in the resulting orbital histories when satellites experience only the MW's gravity (dashed lines) versus the combined MW+LMC potential (solid lines). These differences manifest as changes in the orbital period, distance at pericenter and apocenter, as well as the timing of these critical orbital parameters. The inclusion of the LMC does not affect each satellite's orbit in the same way. For example, including the influence of the LMC decreases the orbital period of Car2 by $\sim$1 Gyr (for MW2) and increases the orbital period of Dra2 (for MW1 and MW2) by $\sim$0.3 Gyr. 

For Cra2, the impact of the LMC is different, such that it decreases the distance achieved at pericenter from $\sim30$ kpc to $\sim$10 kpc, making it well-aligned with previous conclusions that Cra2 may have suffered from extreme tidal stripping \citep{torrealba16a, sanders18, fattahi18, fu19, erkal19}. Hyi1, Car3, Car2, Phx2, and Hor1 also exhibit noticeable perturbations when the LMC potential is included. These satellites have all previously been claimed to be Magellanic satellites by other authors \citep[][see \S \ref{subsec:lit_comparison}]{kallivayalil18, erkal19, jahn19}. 

When the SMC's potential is additionally included (dotted lines), the orbits of the satellites are further perturbed \citep[see][]{jethwa16}. This is particularly interesting in the case of Ret2 where the orbital solution in the combined MW+LMC+SMC for the low mass MW (MW1) shows deviations of hundreds of kiloparsecs from the orbit in the MW+LMC potential, suggesting it may be more perturbed by the SMC than the LMC. Tuc3 is another case where the SMC changes the long-term dynamics of a satellite even though the timing and distance at the most recent pericenter with respect to the MW and with to the LMC remain the same. Carefully determining which of the Clouds plays a more significant role in these satellite's orbits requires futher attention and is beyond the scope of this work.

\subsection{Statistical Significance of Orbital Histories}
\label{subsec:stats}
As the direct orbits only represent one set of orbital solutions, we tabulate the average orbital properties across 1000 orbital calculations in the combined MW+LMC+SMC potential for each candidate Magellanic satellite. These orbits use Galactocentric positions and velocities derived from the Monte Carlo scheme discussed in \S \ref{sec:data} as initial conditions. Average orbital properties and corresponding standard errors are calculated with respect to the LMC and are listed in Table \ref{tab:orbparams_MW1} (MW1) and Table \ref{tab:orbparams_MW2} (MW2). 

In each table Columns 1-8 list the fraction of 1000 orbits where the satellite reaches pericenter and apocenter ($\rm f_{peri}$, $\rm f_{apo}$), the fraction of orbits where the distance at pericenter is less than $\rm r_{outer}$ ($\rm f_{r_{outer}}$, see \S \ref{subsubsec:lmc} and Table \ref{tab:rt}), the distance of the most recent pericenter ($\rm r_{peri}$) and apocenter ($\rm r_{apo}$), and the time at which these occur on average ($\rm t_{peri}$, $\rm t_{apo}$). The second half of each table (Columns 9-16) lists the same quantities for the second pericenter and apocenter as a function of lookback time. Tables listing the orbital properties calculated with respect to LMC1 and LMC3 are provided in Appendix \ref{sec:appendixA} and Appendix \ref{sec:appendixB}. Orbital properties calculated with respect to the MW are provided in Appendix \ref{sec:appendixC}.

\subsection{Identifying Magellanic Satellites}
\label{subsec:identification}

\begin{figure*}[ht!]
\centering
\includegraphics[scale=0.75, trim=0mm 5mm 0mm 2mm]{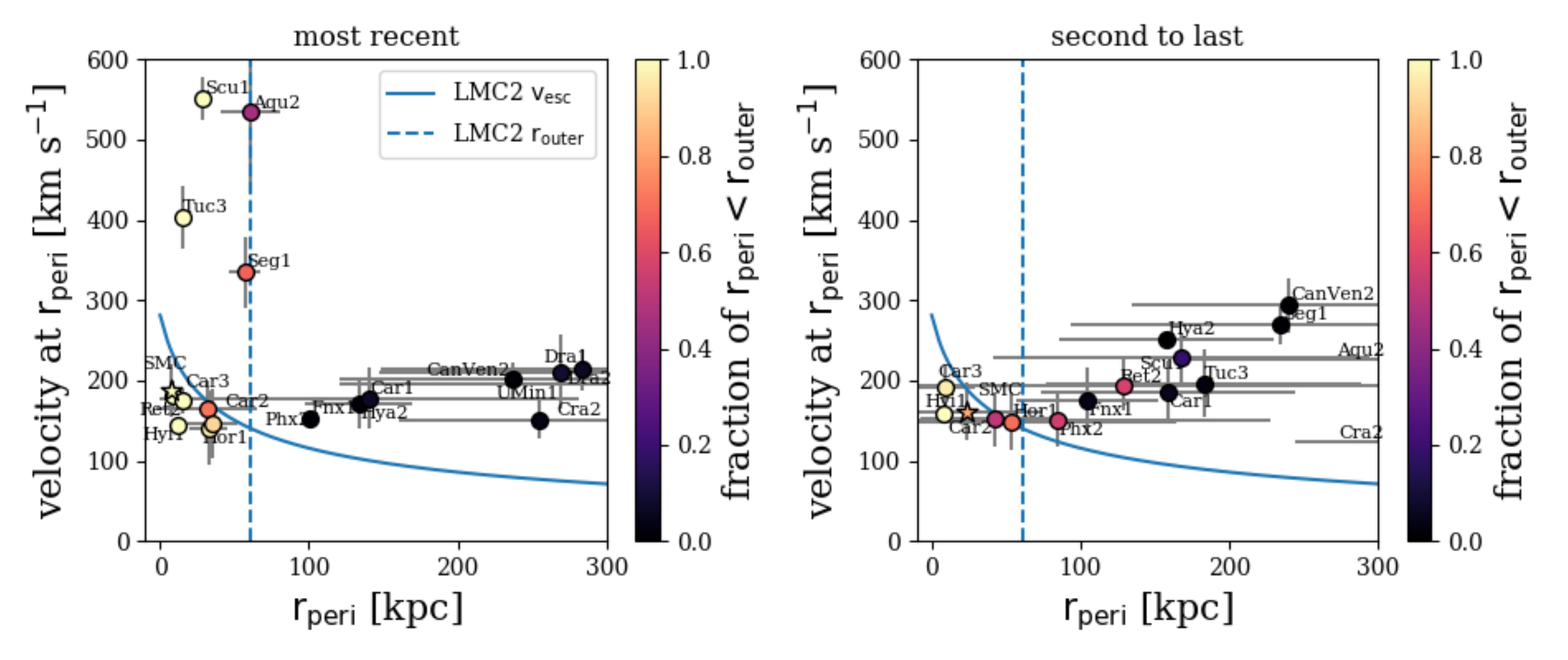}
\caption{Distance at pericenter ($\rm r_{peri}$) versus the velocity at pericenter for the fraction of 1000 orbits where $\rm r_{peri}< \rm r_{outer}$ ($\rm f_{r_{outer},1}$, indicated by the colorbar). All quantities are with respect to the LMC for the most recent passage (left) and the second to last passage around the LMC (right). These orbital parameters are calculated for MW1 and the fiducial LMC model (LMC2). The blue dashed line is $\rm r_{outer}$ for MW1 and LMC2. The solid blue curve represents the escape velocity curve for LMC2. Seg1, Tuc3, and Scu1 are all MW satellites that have recent encounters with the LMC. Ret2 and Phx2 are recently captured Magellanic satellites, while Car2, Car3, Hor1, and Hyi1 are long-term Magellanic satellites.}
\label{fig:orbparams}
\end{figure*}

\begin{figure*}[ht!]
\centering
\includegraphics[scale=0.75, trim=0mm 5mm 0mm 5mm]{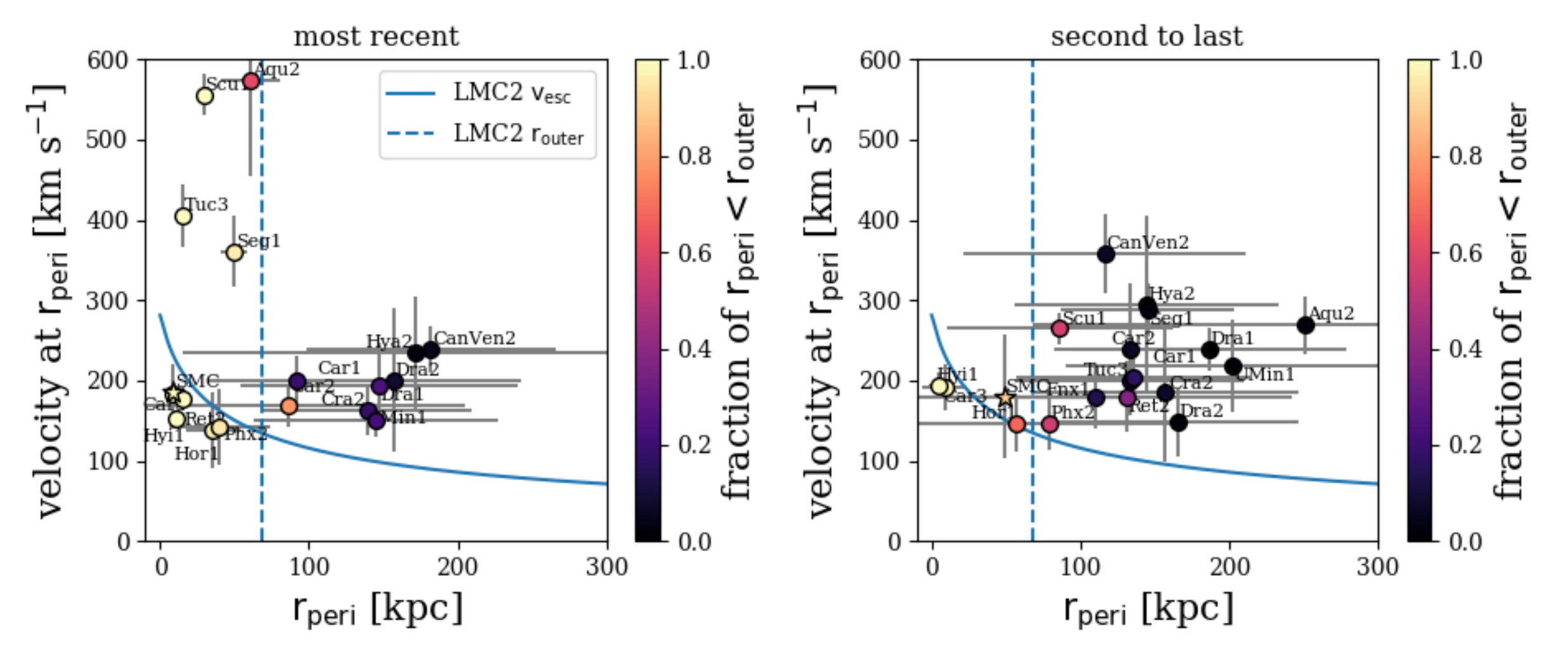}
\caption{Same as Fig. \ref{fig:orbparams} except orbital parameters are calculated for MW2 and the fiducial LMC model (LMC2). Seg1, Tuc3, and Scu1 are all MW satellites that recently passed nearby the MCs at velocities well above the escape speed of the LMC. Ret2 and Phx2 are recently captured Magellanic satellites, while Car3, Hor1, and Hyi1 are long-term Magellanic satellites. In MW2, Car2 no longer qualifies as a satellite of the MCs.}
\label{fig:orbparams2}
\end{figure*}

To determine which of the candidate Magelllanic satellites are true dynamical companions, we examine the orbital properties calculated relative to the fiducial LMC model. The left panel of Fig. \ref{fig:orbparams} illustrates $\rm r_{peri}$ versus the velocity at $\rm r_{peri}$ for the most recent pericentric passage in MW1. The average velocity and standard deviation is computed using only the subset of orbital solutions where $\rm r_{peri} < r_{outer}$, denoted as $\rm f_{r_{router}}$ and indicated by the colorbar. The dashed blue line represents $\rm r_{outer}$ and the solid blue curve is the escape velocity of the fiducial LMC model. Using the properties shown in Fig. \ref{fig:orbparams}, three criteria are defined to determine membership to the Magellanic system. 

\textbf{Criterion 1}: First, we limit the sample of candidate satellites to only those galaxies whose orbits are dominated by the gravitational potential of the LMC rather than the MW's for a high percentage of orbits. This is accomplished by selecting satellites with $\rm f_{r_{router}} > 0.5$, indicating that more than 50\% of the PM error space allows for a closest approach within $\rm r_{outer}$. By doing so, the following galaxies remain: Seg1, Tuc3, Scu1, Car2, Car3, Hor1, Hyi1, Ret2, and Phx2. 

\textbf{Criterion 2}: Next, we examine which of the remaining candidate satellites have velocities that are comparable to or less than the escape velocity of the LMC. All candidate satellites whose velocities at $\rm r_{peri}$ fall below the blue solid curve ($\rm v_{esc}$ of the LMC) in Fig. \ref{fig:orbparams} remain. Seg1, Tuc3, and Scu1 have significantly higher velocities than the LMC's escape speed. These galaxies are likely MW satellites that orbit within 50 kpc of the Galactic Center and consequently pass nearby the LMC. Additionally, all three satellites are especially unlikely to be companions of the LMC as they are on retrograde orbits compared to other satellites in the VPOS, including the MCs \citep{sohn17,fritz18}. We will refer to these galaxies as \emph{MW satellites that recently interacted with the MCs}.

Our results for Tuc3's orbit are well-aligned with recent literature wherein models of the formation of Tuc3's stellar stream require a recent, close encounter with the LMC \citep{erkal18, li18b, simon18}. This further suggests that Segue 1 may also have faint tidal debris resulting from a close passage with the LMC.

\textbf{Criterion 3}: Of the satellites that remain (Car2, Car3, Hor1, Hyi1, Phx2, Ret2), all six are to the left of the dashed blue line and below the solid blue line, indicating that they are bound to the LMC. Each satellite completes a recent passage around the LMC in the last 0.5 Gyr. To further separate these satellites into those that only recently passed around the LMC once versus those that may have completed multiple tightly bound orbits around the LMC, the right side of Fig. \ref{fig:orbparams} illustrates the same quantities for the second to last pericentric passage. By applying Criterion 1 and  Criterion 2 to the orbital properties at the second to last pericentric passage, Car2, Car3, Hor1, and Hyi1 remain. These satellites are therefore designated \emph{long-term Magellanic satellites} since they complete two bound orbits on average around the LMC in the last 2.5-3 Gyr. The SMC, a long-term satellite, is also included in Fig. \ref{fig:orbparams} for reference. Ret2 and Phx2 are classified as \emph{recently captured Magellanic satellites} since they only complete one bound orbit around the MCs in the last 1 Gyr. 

\begin{figure*}[ht!]
    \centering
    \includegraphics[scale=0.75, trim=10mm 0mm 0mm 3mm]{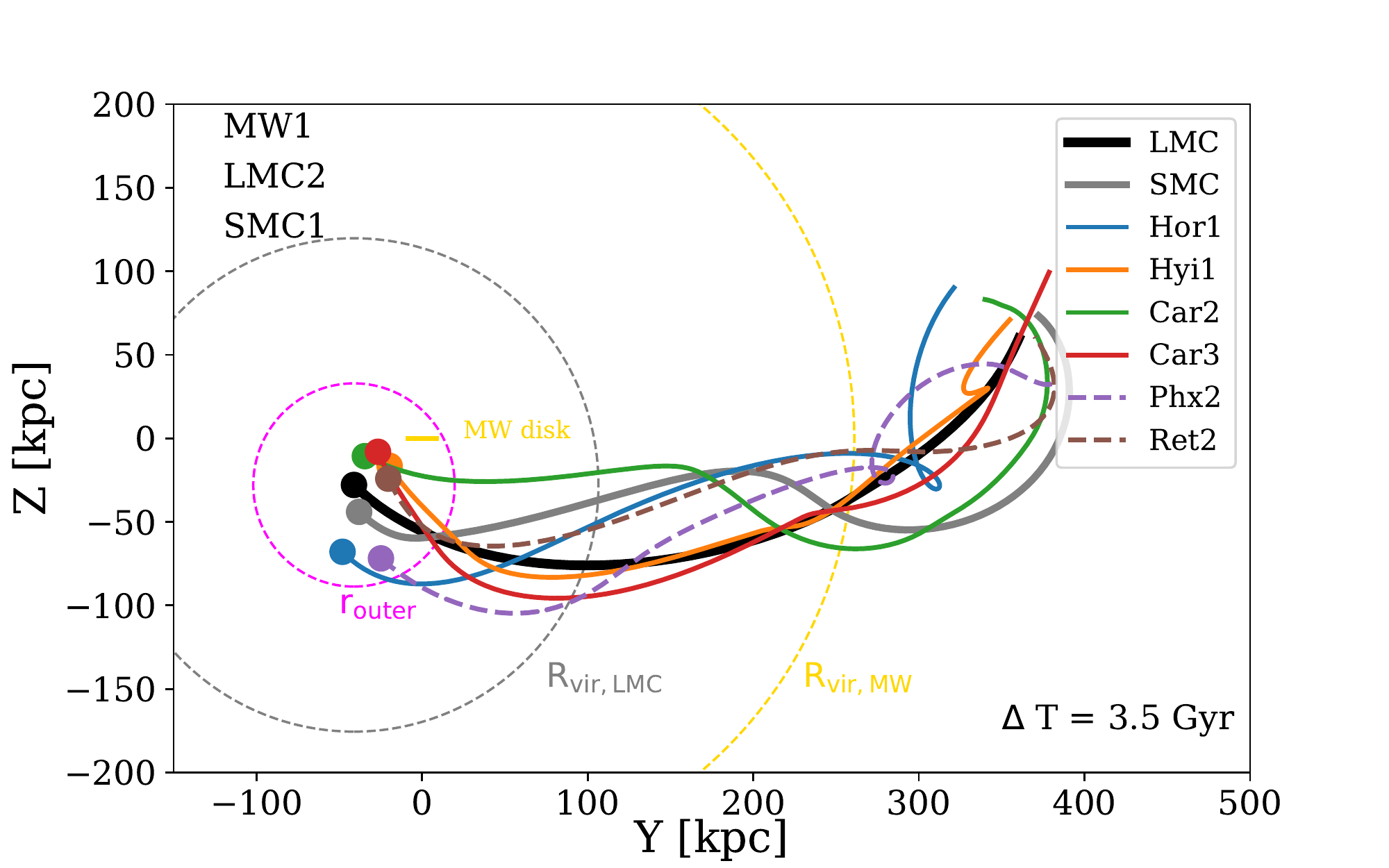}
    \caption{Direct orbits of all Magellanic satellites for the last 3.5 Gyr projected in the YZ-galactocentric plane. Recently captured Magellanic satellites (Ret2, Phx2) are illustrated with dashed lines and long-term Magellanic satellites (Car2, Car3, Hor1, Hyi1) are plotted with solid lines for MW1 using the fiducial LMC model. The disk of the MW lies along the z-axis. The orbit of the LMC (SMC) is illustrated in black (gray). The filled circles represent the positions of all satellites today. The magenta dashed circle indicates $\rm r_{outer}$ of the LMC and the gray dashed circle is the virial radius of the LMC. The gold dashed circle is the virial radius of the MW. The orbits of all Magellanic satellites follow the orbital path of the LMC. }
    \label{fig:LMCsats_YZplane}
\end{figure*}

For MW1 and the fiducial LMC, the orbits of these six Magellanic satellites are shown in Fig. \ref{fig:LMCsats_YZplane} along with the orbits of the LMC and the SMC for the last 3.5 Gyr. Orbits are plotted in the YZ-plane relative to the MW's Galactic Center. The disk of the MW lies along the z-axis. The orbits of all Magellanic satellites clearly follow the orbital path of the LMC/SMC. A 3D animation showing the orbits of all 18 candidate Magellanic satellites using the MW1 and LMC2 models is available at \url{https://bit.ly/35wH5Tr}.  

Fig. \ref{fig:orbparams2} is the same as Fig. \ref{fig:orbparams} but for MW2 in the fiducial LMC model. Applying Criterion 1 and Criterion 2 to the left panel of Fig. \ref{fig:orbparams2}, we conclude that Seg1, Tuc3, and Scu1 are still MW satellites that make a close passage around the LMC in the last 1 Gyr. Applying Criterion 3 to Fig. \ref{fig:orbparams2}, Ret2 and Phx2 are recently captured Magellanic satellites, while Car3, Hor1, Hyi1, and Phx2 are all long-term Magellanic satellite. Car2 now falls outside of the selection criteria due to an increase in $\rm r_{peri}$ by $\sim$50 kpc in MW2. This is likely attributed to the difference in the LMC's orbital history for MW1 and MW2. Satellites are less likely to remain members of the Magellanic system as the MCs pass around the MW two times in the last 6 Gyr \citep[i.e more severe tidal stripping owing to the MW in MW2 may yield fewer Magellanic satellites; see][]{sales11}. Note that the SMC is unbound from the LMC early in MW2 as a high mass MW cannot sustain a long-lived LMC-SMC binary \citep[see][]{k13}.

Table \ref{tab:summary_sats_frac} provides a summary of candidate satellites separated into the classes identified above for all MW and LMC mass combinations. Analogous figures for LMC1 and LMC3 are provided in Appendix \ref{sec:appendixA} and Appendix \ref{sec:appendixB}, respectively. 

The following galaxies are ruled out as Magellanic satellites: Car1, Dra1, UMin1, Fnx1, Cra2, CanVen2, Dra2, Hya2. While Aqu2, Tuc3, Seg1, and Scu1 can have close encounters with the LMC in specific MW-LMC mass combinations (see Table \ref{tab:summary_sats_frac}), we stress that they are not dynamically associated members of the Magellanic system. In \S \ref{subsec:lit_comparison}, we compare these results to other recent studies and discuss how differing sets of selection criteria for identifying Magellanic satellites can lead to alternative conclusions.

\begin{deluxetable}{c|c|c}
\tabletypesize{\footnotesize}
\label{tab:summary_sats_frac}
\tablecaption{Identification of Magellanic Satellites and Recent Encounters in a MW+LMC+SMC Potential}
\tablehead{\colhead{} & \colhead{MW1} & \colhead{MW2}}
\startdata \hline
\multicolumn{3}{c}{MW satellites, recent interaction with MCs}  \\ \hline
LMC1 & Tuc3, Scu1 & Tuc3, Scu1 \\
LMC2 & Seg1, Tuc3, Scu1 & Seg1, Tuc3, Scu1 \\ 
LMC3 & Aqu2, Seg1, Tuc3, Scu1 & Aqu2, Seg1, Tuc3, Scu1\\ \hline
\multicolumn{3}{c}{recently captured Magellanic satellites}  \\ \hline
LMC1 & Ret2 & Ret2  \\
LMC2 & Ret2, Phx2 & Ret2, Phx2 \\
LMC3 & Ret2 &  Ret2 \\ \hline
\multicolumn{3}{c}{long-term Magellanic satellites}\\ \hline
LMC1 & Car3, Hyi1 & Car3, Hyi1 \\
LMC2 & Car2, Car3, Hor1, Hyi1 & Car3, Hor1, Hyi1 \\
LMC3 & Car2, Car3, Hor1, Hyi1, Phx2 & Car3, Hor1, Hyi1, Phx2 \\
\enddata
\tablecomments{MW satellites, recent interaction with MCs: orbits where $\rm f_{r_{outer},1} > 0.5$ and velocity at most recent $\rm r_{peri}>$  $\rm v_{esc,LMC}$. Magellanic satellites, bound late: also have $\rm f_{r_{outer},1} > 0.5$ and velocity at most recent pericenter  $> \rm v_{esc,LMC}$ (i.e. at least one bound orbit around the LMC). Magellanic satellites, bound early: Magellanic satellites that additionally satisfy the same set of criteria also for the second to last pericentric passage (i.e. at least two bound orbits around the LMC).}
\end{deluxetable}

\begin{splitdeluxetable*}{lcccccccBlccccccc}
\tabletypesize{\footnotesize}
\label{tab:orbparams_MW1}
\tablecaption{Orbital properties with respect to LMC2 in MW1.}
\tablehead{\colhead{Name} & \colhead{$\rm f_{peri,1}$ } & \colhead{$\rm f_{r_{outer},1}$} & \colhead{$\rm r_{peri,1}$ [kpc]} & \colhead{$\rm t_{peri,1}$ [Gyr]} & \colhead{$\rm f_{apo,1}$} & \colhead{$\rm r_{apo,1}$ [kpc]} & \colhead{$\rm t_{apo,1}$ [Gyr]} & \colhead{Name} & \colhead{$\rm f_{peri,2}$} &  \colhead{$\rm f_{r_{outer},2}$} & \colhead{$\rm r_{peri,2}$ [kpc]} & \colhead{$\rm t_{peri,2}$ [Gyr]} & \colhead{$\rm f_{apo,2}$} & \colhead{$\rm r_{apo,2}$ [kpc]} & \colhead{$\rm t_{apo,2}$ [Gyr]}\\ \hline
\multicolumn{8}{c}{most recent} & \multicolumn{8}{c}{second to last}}
\startdata
Aqu2 & 1.0 & 0.46 & 61.07$\pm$20.0 & 0.16$\pm$0.06 & 0.21 & 472.19$\pm$160.67 & 3.07$\pm$1.14 & Aqu2 & 1.0 &  0.01 & 343.33$\pm$181.56 & 4.23$\pm$1.11 & 0.08 & 406.01$\pm$149.01 & 4.82$\pm$0.78 \\ 
CanVen2 & 0.24 & 0.01 & 236.77$\pm$115.85 & 3.38$\pm$1.23 & 0.31 & 346.53$\pm$115.15 & 1.88$\pm$1.33 & CanVen2 & 0.24 &  0.0 & 240.01$\pm$105.2 & 4.51$\pm$0.97 & 0.08 & 321.98$\pm$109.63 & 4.58$\pm$1.18 \\ 
Car2 & 0.81 & 0.71 & 31.74$\pm$33.71 & 1.24$\pm$0.48 & 0.84 & 77.38$\pm$81.01 & 0.66$\pm$0.85 & Car2 & 0.81 &  0.51 & 41.96$\pm$49.18 & 4.03$\pm$1.0 & 0.75 & 123.1$\pm$94.32 & 3.08$\pm$1.03 \\ 
Car3 & 1.0 & 1.0 & 8.86$\pm$3.08 & 0.18$\pm$0.05 & 0.99 & 58.8$\pm$42.4 & 1.0$\pm$0.5 & Car3 & 1.0 &  0.96 & 9.42$\pm$19.26 & 1.75$\pm$0.81 & 0.92 & 56.49$\pm$23.13 & 2.45$\pm$0.86 \\ 
Cra2 & 1.0 & 0.04 & 254.33$\pm$93.5 & 2.6$\pm$0.82 & 1.0 & 348.69$\pm$60.96 & 1.47$\pm$0.33 & Cra2 & 1.0 &  0.0 & 431.89$\pm$187.86 & 4.97$\pm$0.69 & 0.88 & 426.21$\pm$159.8 & 4.56$\pm$0.65 \\ 
Dra2 & 0.35 & 0.03 & 315.3$\pm$195.24 & 4.41$\pm$1.22 & 0.63 & 535.71$\pm$151.29 & 3.13$\pm$1.09 & Dra2 & 0.35 &  0.0 & 409.33$\pm$163.4 & 5.1$\pm$0.59 & 0.08 & 433.24$\pm$152.79 & 4.95$\pm$0.72 \\ 
Hor1 & 0.98 & 0.97 & 32.59$\pm$12.49 & 0.27$\pm$0.32 & 0.92 & 140.61$\pm$176.8 & 1.56$\pm$1.42 & Hor1 & 0.98 &  0.7 & 53.75$\pm$110.3 & 2.09$\pm$1.37 & 0.69 & 73.3$\pm$88.87 & 2.59$\pm$1.21 \\ 
Hyi1 & 1.0 & 1.0 & 11.81$\pm$2.43 & 0.27$\pm$0.04 & 1.0 & 30.3$\pm$2.95 & 0.77$\pm$0.08 & Hyi1 & 1.0 &  1.0 & 8.4$\pm$2.49 & 1.24$\pm$0.14 & 1.0 & 28.95$\pm$4.95 & 1.68$\pm$0.21 \\ 
Hya2 & 0.26 & 0.02 & 133.48$\pm$36.1 & 0.86$\pm$1.06 & 0.15 & 219.52$\pm$120.46 & 1.02$\pm$1.29 & Hya2 & 0.26 &  0.0 & 157.59$\pm$71.85 & 5.05$\pm$0.86 & 0.06 & 387.1$\pm$165.66 & 4.14$\pm$1.05 \\ 
Phx2 & 0.97 & 0.91 & 34.75$\pm$16.5 & 0.43$\pm$0.4 & 0.89 & 181.56$\pm$186.47 & 2.2$\pm$1.3 & Phx2 & 0.97 &  0.56 & 84.1$\pm$143.76 & 2.96$\pm$1.26 & 0.57 & 103.64$\pm$120.62 & 3.67$\pm$1.1 \\ 
Ret2 & 1.0 & 1.0 & 15.76$\pm$2.92 & 0.12$\pm$0.02 & 0.92 & 199.74$\pm$217.83 & 1.91$\pm$1.42 & Ret2 & 1.0 &  0.57 & 128.98$\pm$199.41 & 2.46$\pm$1.39 & 0.75 & 165.14$\pm$206.47 & 3.07$\pm$1.31 \\ 
Seg1 & 0.99 & 0.67 & 56.84$\pm$10.85 & 0.32$\pm$0.11 & 0.99 & 70.17$\pm$7.28 & 0.12$\pm$0.04 & Seg1 & 0.99 &  0.0 & 234.4$\pm$140.59 & 1.85$\pm$1.1 & 0.98 & 261.92$\pm$151.59 & 1.5$\pm$1.03 \\ 
Tuc3 & 1.0 & 1.0 & 14.82$\pm$3.25 & 0.08$\pm$0.01 & 0.72 & 219.89$\pm$143.61 & 1.26$\pm$1.05 & Tuc3 & 1.0 &  0.03 & 182.68$\pm$106.25 & 1.41$\pm$1.17 & 0.55 & 298.44$\pm$142.43 & 2.24$\pm$1.37 \\ 
Car1 & 0.68 & 0.08 & 140.66$\pm$140.59 & 1.73$\pm$1.33 & 0.93 & 280.58$\pm$234.89 & 1.94$\pm$1.81 & Car1 & 0.68 &  0.05 & 158.4$\pm$85.28 & 4.57$\pm$0.97 & 0.52 & 272.18$\pm$123.27 & 3.96$\pm$1.01 \\ 
Dra1 & 0.78 & 0.07 & 283.83$\pm$135.04 & 4.27$\pm$0.97 & 0.99 & 417.1$\pm$86.54 & 2.43$\pm$0.46 & Dra1 & 0.78 &  0.0 & 304.81$\pm$0.0 & 4.58$\pm$0.0 & 0.1 & 431.22$\pm$221.96 & 5.56$\pm$0.41 \\ 
Fnx1 & 1.0 & 0.0 & 100.1$\pm$4.52 & 0.14$\pm$0.07 & 0.84 & 366.87$\pm$259.49 & 2.88$\pm$2.08 & Fnx1 & 1.0 &  0.04 & 104.49$\pm$47.78 & 1.7$\pm$0.96 & 0.37 & 259.48$\pm$88.69 & 4.07$\pm$0.81 \\ 
Scu1 & 1.0 & 1.0 & 28.91$\pm$4.88 & 0.11$\pm$0.01 & 0.83 & 338.49$\pm$83.57 & 2.1$\pm$0.75 & Scu1 & 1.0 &  0.18 & 167.35$\pm$126.52 & 4.35$\pm$0.85 & 0.2 & 256.8$\pm$61.37 & 5.27$\pm$0.57 \\ 
UMin1 & 0.87 & 0.09 & 269.27$\pm$122.35 & 3.78$\pm$1.02 & 0.99 & 384.19$\pm$63.84 & 2.13$\pm$0.37 & UMin1 & 0.87 &  0.0 & 439.41$\pm$237.89 & 4.87$\pm$1.28 & 0.32 & 492.95$\pm$201.59 & 5.5$\pm$0.39 \\ 
\enddata
\tablecomments{Columns 1-8 refer to the most recent occurrence of a pericenter and apocenter. Columns 9-16 refer to the second to last instance where these minima and maxima occur. $\rm f_{peri,i}$ ($\rm f_{apo,i}$) is the fraction of 1000 orbits where a pericenter (apocenter) is recovered\footnote{Every unique orbital solution does not result in the same number of apocenters and pericenters as a function of lookback time given the large PM uncertainties. Furthermore, some satellites on first infall never reach an apocenter within the the last 6 Gyr.}. $\rm f_{r_{outer},i}$ is the fraction of 1000 orbits with $\rm r_{peri} < r_{\rm outer}$ (see \S \ref{subsec:df}).}
\end{splitdeluxetable*}

\begin{splitdeluxetable*}{lcccccccBlccccccc}
\tabletypesize{\footnotesize}
\label{tab:orbparams_MW2}
\tablecaption{Orbital properties with respect to LMC2 in MW2.}
\tablehead{\colhead{Name} & \colhead{$\rm f_{peri,1}$ } & \colhead{$\rm f_{r_{outer},1}$} & \colhead{$\rm r_{peri,1}$ [kpc]} & \colhead{$\rm t_{peri,1}$ [Gyr]} & \colhead{$\rm f_{apo,1}$} & \colhead{$\rm r_{apo,1}$ [kpc]} & \colhead{$\rm t_{apo,1}$ [Gyr]} & \colhead{Name} & \colhead{$\rm f_{peri,2}$} &  \colhead{$\rm f_{r_{outer},2}$} & \colhead{$\rm r_{peri,2}$ [kpc]} & \colhead{$\rm t_{peri,2}$ [Gyr]} & \colhead{$\rm f_{apo,2}$} & \colhead{$\rm r_{apo,2}$ [kpc]} & \colhead{$\rm t_{apo,2}$ [Gyr]}\\ \hline
\multicolumn{8}{c}{most recent} & \multicolumn{8}{c}{second to last}}
\startdata
Aqu2 & 1.0 & 0.61 & 60.95$\pm$19.87 & 0.16$\pm$0.06 & 0.36 & 394.88$\pm$171.75 & 2.1$\pm$0.85 & Aqu2 & 0.34 &  0.02 & 251.1$\pm$183.45 & 3.46$\pm$0.95 & 0.2 & 293.41$\pm$120.06 & 4.18$\pm$0.92 \\ 
CanVen2 & 0.49 & 0.05 & 181.92$\pm$83.48 & 2.33$\pm$1.05 & 0.53 & 296.2$\pm$102.69 & 1.08$\pm$1.13 & CanVen2 & 0.2 &  0.06 & 116.18$\pm$95.01 & 4.72$\pm$0.83 & 0.35 & 379.18$\pm$183.57 & 4.07$\pm$0.85 \\ 
Car2 & 0.99 & 0.78 & 85.67$\pm$118.73 & 1.54$\pm$1.42 & 1.0 & 121.68$\pm$139.86 & 0.93$\pm$1.14 & Car2 & 0.83 &  0.06 & 133.35$\pm$50.7 & 4.02$\pm$0.84 & 0.96 & 259.38$\pm$97.27 & 3.21$\pm$1.0 \\ 
Car3 & 1.0 & 1.0 & 8.19$\pm$3.1 & 0.18$\pm$0.05 & 1.0 & 52.04$\pm$36.9 & 0.86$\pm$0.38 & Car3 & 1.0 &  0.99 & 8.8$\pm$15.23 & 1.52$\pm$0.61 & 0.97 & 61.21$\pm$44.92 & 2.32$\pm$0.85 \\ 
Cra2 & 1.0 & 0.18 & 139.11$\pm$70.12 & 2.31$\pm$0.61 & 1.0 & 270.27$\pm$29.85 & 0.96$\pm$0.13 & Cra2 & 0.82 &  0.08 & 157.11$\pm$88.94 & 4.52$\pm$0.91 & 0.92 & 245.66$\pm$108.93 & 3.61$\pm$0.74 \\ 
Dra2 & 0.96 & 0.09 & 156.95$\pm$85.75 & 3.25$\pm$1.02 & 0.99 & 332.3$\pm$107.4 & 1.56$\pm$0.56 & Dra2 & 0.39 &  0.02 & 165.29$\pm$80.87 & 4.74$\pm$0.78 & 0.65 & 282.88$\pm$104.43 & 4.3$\pm$0.95 \\ 
Hor1 & 0.97 & 0.97 & 35.68$\pm$17.9 & 0.22$\pm$0.38 & 0.97 & 125.39$\pm$134.31 & 1.36$\pm$1.33 & Hor1 & 0.86 &  0.68 & 56.87$\pm$78.5 & 1.88$\pm$1.42 & 0.77 & 98.65$\pm$99.62 & 2.51$\pm$1.31 \\ 
Hyi1 & 1.0 & 1.0 & 11.1$\pm$2.51 & 0.27$\pm$0.03 & 1.0 & 35.12$\pm$7.94 & 0.82$\pm$0.16 & Hyi1 & 1.0 &  1.0 & 4.42$\pm$5.9 & 1.31$\pm$0.27 & 1.0 & 38.32$\pm$13.42 & 1.84$\pm$0.33 \\ 
Hya2 & 0.33 & 0.03 & 171.24$\pm$155.53 & 1.27$\pm$1.6 & 0.24 & 304.57$\pm$215.14 & 1.56$\pm$1.7 & Hya2 & 0.09 &  0.01 & 144.2$\pm$88.82 & 4.45$\pm$0.85 & 0.13 & 330.55$\pm$129.64 & 3.42$\pm$1.02 \\ 
Phx2 & 0.97 & 0.95 & 40.15$\pm$34.26 & 0.4$\pm$0.54 & 0.95 & 165.11$\pm$145.54 & 1.88$\pm$1.24 & Phx2 & 0.84 &  0.56 & 78.44$\pm$90.98 & 2.61$\pm$1.23 & 0.68 & 138.36$\pm$117.13 & 3.51$\pm$1.2 \\ 
Ret2 & 1.0 & 1.0 & 15.53$\pm$3.01 & 0.13$\pm$0.02 & 1.0 & 210.87$\pm$122.79 & 1.78$\pm$0.86 & Ret2 & 0.98 &  0.38 & 131.35$\pm$110.18 & 2.68$\pm$1.24 & 0.81 & 219.62$\pm$138.69 & 3.26$\pm$1.09 \\ 
Seg1 & 1.0 & 0.96 & 49.9$\pm$8.74 & 0.29$\pm$0.08 & 1.0 & 69.35$\pm$31.27 & 0.1$\pm$0.17 & Seg1 & 1.0 &  0.03 & 145.19$\pm$57.96 & 1.31$\pm$0.55 & 1.0 & 176.33$\pm$53.66 & 0.94$\pm$0.37 \\ 
Tuc3 & 1.0 & 1.0 & 15.15$\pm$3.24 & 0.07$\pm$0.01 & 0.85 & 171.86$\pm$112.99 & 0.99$\pm$0.84 & Tuc3 & 0.83 &  0.1 & 133.46$\pm$76.63 & 1.27$\pm$0.97 & 0.74 & 224.04$\pm$106.54 & 2.02$\pm$1.21 \\ 
Car1 & 0.96 & 0.19 & 91.19$\pm$59.3 & 1.38$\pm$0.93 & 1.0 & 145.07$\pm$99.05 & 0.77$\pm$0.76 & Car1 & 0.75 &  0.17 & 135.55$\pm$78.63 & 3.57$\pm$1.02 & 0.85 & 195.61$\pm$99.61 & 2.8$\pm$0.99 \\ 
Dra1 & 0.89 & 0.24 & 147.27$\pm$93.11 & 3.77$\pm$1.05 & 0.96 & 282.86$\pm$47.45 & 1.78$\pm$0.67 & Dra1 & 0.2 &  0.0 & 186.27$\pm$91.85 & 5.29$\pm$0.57 & 0.5 & 308.2$\pm$152.04 & 4.68$\pm$0.64 \\ 
Fnx1 & 1.0 & 0.0 & 100.73$\pm$3.88 & 0.12$\pm$0.03 & 1.0 & 209.65$\pm$157.71 & 1.39$\pm$1.26 & Fnx1 & 0.91 &  0.14 & 110.39$\pm$53.27 & 2.03$\pm$1.53 & 0.74 & 215.8$\pm$83.4 & 2.88$\pm$0.68 \\ 
Scu1 & 1.0 & 1.0 & 29.2$\pm$4.97 & 0.11$\pm$0.01 & 1.0 & 232.14$\pm$57.28 & 1.09$\pm$0.32 & Scu1 & 1.0 &  0.56 & 85.93$\pm$76.06 & 2.53$\pm$0.5 & 0.99 & 271.3$\pm$115.62 & 3.89$\pm$0.7 \\ 
UMin1 & 0.94 & 0.22 & 145.29$\pm$82.16 & 3.42$\pm$0.96 & 0.98 & 265.65$\pm$36.64 & 1.55$\pm$0.55 & UMin1 & 0.36 &  0.02 & 202.23$\pm$112.33 & 5.23$\pm$0.54 & 0.63 & 287.09$\pm$151.25 & 4.42$\pm$0.6 \\ 
\enddata
\end{splitdeluxetable*}

\section{Discussion}
\label{sec:discussion}

\subsection{Comparison to Recent Literature}
\label{subsec:lit_comparison}

Here we will discuss our results in the context of a few recent studies on Magellanic satellites that are the most relevant to this analysis. \cite{jethwa16} derived probabilities for membership to the Magellanic system for 14 UFDs discovered in DES. Though PMs were not available at the time, they integrated orbits in a combined MW+LMC+SMC potential, including dynamical friction and tidal shredding. Satellites were initially radially distributed in a way that is consistent with cosmological simulations. \citet{jethwa16} found that seven UFDs have a high probability (p $>$ 0.7) of being LMC satellites based on their positions ($<$ 50 kpc from the LMC), and that of the four UFDs with measured velocities, these values are consistent with the LMC. The only overlapping satellites between our work and \cite{jethwa16} are Ret2 and Hor1, and both studies find that these satellites are highly likely satellites of the LMC. 

\citet{kallivayalil18} compared the measured 3D kinematics of UFDs with the position and velocities of an LMC analog's debris in a high-resolution simulation and concluded that Car2, Car3, Hor1, and Hyi1 are all satellites that entered the MW's halo with the MCs. They also found that Ret2 is not consistent with the kinematics of simulated LMC debris in all three velocity components, but its orbital pole is consistent with that of the debris, hinting at potential association. 

\citet{kallivayalil18} indicated that Hya2 and Dra2 require more detailed orbital modeling, which we undertake in this paper.  We find no association between these two galaxies and the MCs using our selection criteria. Our conclusions for satellite membership of four UFDs (Car2, Car3, Hyi1, Hor1) are consistent with \citet{kallivayalil18}. Furthermore, the identification of recently captured Magellanic satellites in this work confirms the conclusion from \citet{kallivayalil18} that Ret2 is a tentative Magellanic satellite. 

\citet{pardy19} used the Auriga simulations to count the abundance of satellites around LMC analogs and found that the LMC is expected to host $\sim$3 satellites with $M_* \geq 10^5 \, M_{\sun}$ and within two times $\rm R_{200}$ of the LMC. The SMC counts towards this prediction and they additionally claimed that Carina and Fornax are also satellites of the LMC given the coherence between their orbital poles on the sky compared to that of the LMC. However, we do not find that Carina and Fornax are associated to the MCs using our satellite criteria, which accounts for the orbital histories of the galaxies in addition to their current kinematics and dynamics.  

\citet{jahn19} used the subset of simulations from the FIRE suite hosting an LMC analog to calculate the expected abundance of LMC satellite galaxies and found that it can host 5-10 satellites with $M_* \geq 10^4 \, M_{\sun}$ within its virial radius. Like \cite{pardy19}, they also used the recently measured PMs of UFDs and classical dwarfs to additionally quantify which of these galaxies have 3D angular momenta vectors that are consistent with the MCs, similar to the analysis of \citet{sales11, sales17}. These authors concluded that given their current angular momenta, Car2, Car3, Hor1, Hyi1, Carina, and Fornax are all satellites of the LMC, in addition to the SMC. Using the satellite selection criteria defined in this work, we find good agreement with both \cite{jahn19} and \cite{pardy19} in the ultra-faint regime, but we do not find that Carina and Fornax are dynamically associated Magellanic satellites even though their orbital poles are aligned today. 

\cite{erkal19} calculated the orbital energy of satellites 5 Gyr ago to determine whether they were energetically bound to the LMC. This process is repeated for 10,000 Monte Carlo realizations to derive a probability for being an LMC satellite as a function of LMC mass. In doing so, they found that Car2, Car3, Hor1, Hyi1, Phx2, and Ret2 are highly probable satellites of the LMC in addition to the SMC and that an LMC mass of $1.5\times10^{11}\, M_{\sun}$ is required for all to be bound simultaneously. In general, the results from \citet{erkal19} are in good agreement with our fiducial LMC model. Two main differences include that we find Car2 is not an LMC satellite in a high mass MW model (MW2; see \S \ref{subsec:mass_disc}) and that Ret2 is only recently captured by the MCs in our categorization (i.e. it was not bound to the LMC 5 Gyr ago). 

It is worth noting that each of the aforementioned analyses uses different criteria to select satellites that may be of Magellanic origin. We stress that even in our own analysis an alternative set of selection criteria may lead to different conclusions. For example, if we chose Magellanic satellites that satisfy $\rm r_{peri} < R_{vir,LMC}$ instead of $\rm r_{peri} < r_{outer}$ and remove the escape velocity criteria in \S \ref{subsec:identification}, a greater number of Magellanic satellites are identified. In particular, Carina and Fornax would be LMC satellites under these criteria, in line with the conclusions in \citet{pardy19, jahn19}. These modified criteria would also falsely count the MW satellites that only recently interacted with the MCs once as Magellanic satellites.

\subsection{Masses of the LMC and the MW}
\label{subsec:mass_disc}
The identification of Magellanic satellites discussed in \S \ref{subsec:identification} and summarized in Table \ref{tab:summary_sats_frac} is sensitive to both the mass of the LMC and the mass of the MW. For fixed LMC mass, but variable MW mass, results are usually the same.  But, for fixed MW mass, and variable LMC mass, there are some notable differences.

 For a fixed MW1 mass model, higher LMC masses tend towards more satellites classified as long-term Magellanic satellites since the LMC's gravity overcomes the MW's as the MW-LMC mass ratio decreases. For example, in LMC1 only Car3 and Hyi1 are long-term Magellanic satellites. LMC2 adds Car2 and Hor1, and furthermore for LMC3, Phx2 is additionally a long-term Magellanic satellite. For all LMC mass models in MW1, Ret2 is always a recently captured Magellanic satellite. This suggests that Ret2 requires an even more massive LMC (i.e. $>2.5 \times 10^{11}\, M_{\sun}$) for it to be bound to the MCs even though \citet{erkal19} find that Ret2 needs the LMC's mass to be $\geq 9.5\times 10^{10}\, M_{\sun}$ for it to be energetically bound.

In a similar fashion, increasing the LMC's mass leads to more MW satellites having recent interactions with the MCs. For LMC1, only Tuc3 and Scu1 pass Criteria 1 (see \S \ref{subsec:identification}). For the fiducial LMC2 model, Seg1 is additionally a MW satellite that interacts with the MCs recently. Finally, for LMC3, Aqu2 also follows suit. Like Scu1, Tuc3, and Seg1, Aqu2 is also on a retrograde orbit relative to the LMC and other satellites in the VPOS.

For a fixed MW2 mass model, all results are the same as MW1 with the exception of Car2 for LMC2 and LMC3. This demonstrates that the mass of the LMC drives the classification, not the mass of the MW. Car2 is never a Magellanic satellites in MW2 as its distance at pericenter increases to values beyond $\rm r_{outer}$. This is in contrast to \citet{erkal19} who find that Car2 requires a relatively low mass LMC ($M=2\times10^{10}\, M_{\sun}$) for it to be bound. However, there are several differences between our orbital model and that of \cite{erkal19} that may account for this discrepancy, including: 1) the gravitational influence of the SMC, 2) the addition of a disk potential for the LMC, 3) modelling satellites as extended objects, and 4) implementing DF from both the MW and the LMC.

A low mass MW (MW1) and massive LMC (LMC2, LMC3) are the most favorable for producing the highest total number of MW satellites with recent interactions with the MCs (4 galaxies at maximum) and Magellanic satellites (6 galaxies at maximum). This is due in large part to the LMC being on first infall and only making one passage around the MW recently, resulting in less tidal stripping of satellites. Secondly, a more massive LMC brings a greater number of satellites with it, as expected from hierarchical $\Lambda$CDM.

\subsection{Inclusion of the SMC Potential}
\label{subsec:smc_inclusion}

\begin{deluxetable}{c|c|c}
\tabletypesize{\footnotesize}
\label{tab:summary_sats_frac_noSMC}
\tablecaption{Identification of Magellanic Satellites and Recent Encounters in a MW+LMC Potential (no SMC)}
\tablehead{\colhead{} & \colhead{MW1} & \colhead{MW2}}
\startdata \hline
\multicolumn{3}{c}{MW satellites, recent interaction with the MCs}  \\ \hline
LMC1 & Tuc3, Scu1 & Tuc3, Scu1 \\
LMC2 & Seg1, Tuc3, Scu1 & Aqu2, Seg1, Tuc3, Scu1 \\ 
LMC3 & Aqu2, Seg1, Tuc3, Scu1 & Aqu2, Seg1, Tuc3, Scu1\\ \hline
\multicolumn{3}{c}{recently captured Magellanic satellites}  \\ \hline
LMC1 & Ret2 & Hyi1, Ret2  \\
LMC2 & Hor1, Ret2, Phx2 & Hor1, Ret2, Phx2 \\
LMC3 & Phx2, Ret2 &  Phx2, Ret2 \\ \hline
\multicolumn{3}{c}{long-term Magellanic satellites}\\ \hline
LMC1 & Car3, Hyi1 & Car3 \\
LMC2 & Car2, Car3, Hyi1 & Car3, Hyi1 \\
LMC3 & Car2, Car3, Hor1, Hyi1 & Car3, Hor1, Hyi1 \\
\enddata
\end{deluxetable}

To understand how the inclusion of the SMC impacts our analysis of Magellanic satellites, we recalculate the orbital properties for all 18 galaxies in a MW+LMC gravitational potential, neglecting the SMC. Using these properties, we re-classify galaxies into the categories defined in \S \ref{subsec:identification} and present the results for all six MW-LMC mass combinations in Table \ref{tab:summary_sats_frac_noSMC}. 

When orbital properties are computed in a MW+LMC potential, we find nearly the same results for galaxies in the `MW satellites, recent interaction with the MCs' category. The only difference is that Aqu2 also interacts with the MCs in the MW2-LMC2 mass combination even though it is not identified as such when orbits are calculated in the MW+LMC+SMC potential. This suggests that the SMC may even perturb galaxies on first infall, retrograde orbits like Aqu2.

Overall the total number of Magellanic satellite remains the same for the MW+LMC potential compared to the MW+LMC+SMC potential, and the same six satellites are always placed in the `long-term' and `recently captured' categories: Car2, Car3, Hor1, Hyi, Ret2, Phx2. However, more UFDs are classified into the `recently captured' category in the MW-LMC potential.

The SMC can cause some generic changes to the average distance and timing of pericenter and apocenter. As a result, Ret2 and Phx2 are \emph{always} recently captured by the MCs regardless of the MW and/or the LMC's mass with no SMC. This is due to an increase in the distance at the second pericenter for both satellites, likely caused by the decreased mass of the combined MCs when the SMC is not included.

Similarly, Hyi1 and Hor1 are also occasionally recently captured Magellanic satellites, whereas they are always long-term Magellanic satellites for the combined MW+LMC+SMC potential. We conclude that the SMC's gravitational influence changes the predicted longevity of satellites as Magellanic satellites, increasing the number of satellites that entered the MW's halo with the MCs by one if the MCs are on first infall (i.e. the MW1 model). These results are consistent with \cite{jethwa16} who find that the inclusion of the SMC only impacts one of the UFDs they study.

We note that other MW satellites, such as the Sagittarius dSph, may also have had interactions with the MCs \citep[e.g.][]{zhao98}, potentially perturbing the orbits of the MCs and any satellites associated with them. However, investigating the influence of Sagittarius requires high resolution N-body simulations that account for the mass loss satellites experience as they repeatedly pass around the MW as well as the mass evolution of the MW and LMC, so we defer this to future work.

\subsection{Effect of Reducing Proper Motion Uncertainty}
PM uncertainties will decrease as the time baselines between Gaia data releases increases. Future PM measurements with HST+JWST will also yield higher precision PMs for many of the galaxies included in our sample. For example, JWST ERS 1334 will yield an improved PM for Dra2 and HST GO-14734 will obtain first-epoch imaging for eight galaxies in our sample that can be followed up with JWST to obtain improved PMs. 
 
 Given these future prospects, we recalculate the orbital properties of Ret2 and Phx2 after reducing the uncertainty in the Gaia DR2 PMs to 25\% of their current values\footnote{For Gaia, this roughly corresponds to a 7 year baseline between DR1 and the final data release, so it is possible to reach this precision in the next decade.} to determine how smaller PM uncertainties affect the identification of Magellanic satellites. We also set the PM covariance term to zero for this exercise\footnote{We have checked that setting the PM covariance to zero with the current PM values does not significantly affect the average and standard errors on orbital properties reported in \S \ref{subsec:stats} for a fair comparison.}. We focus on Ret2 and Phx2 because they are the only recently captured Magellanic satellites (see \S  \ref{subsec:identification}). 

\begin{figure*}[ht!]
\centering
\includegraphics[scale=0.75, trim=0mm 5mm 0mm 0mm]{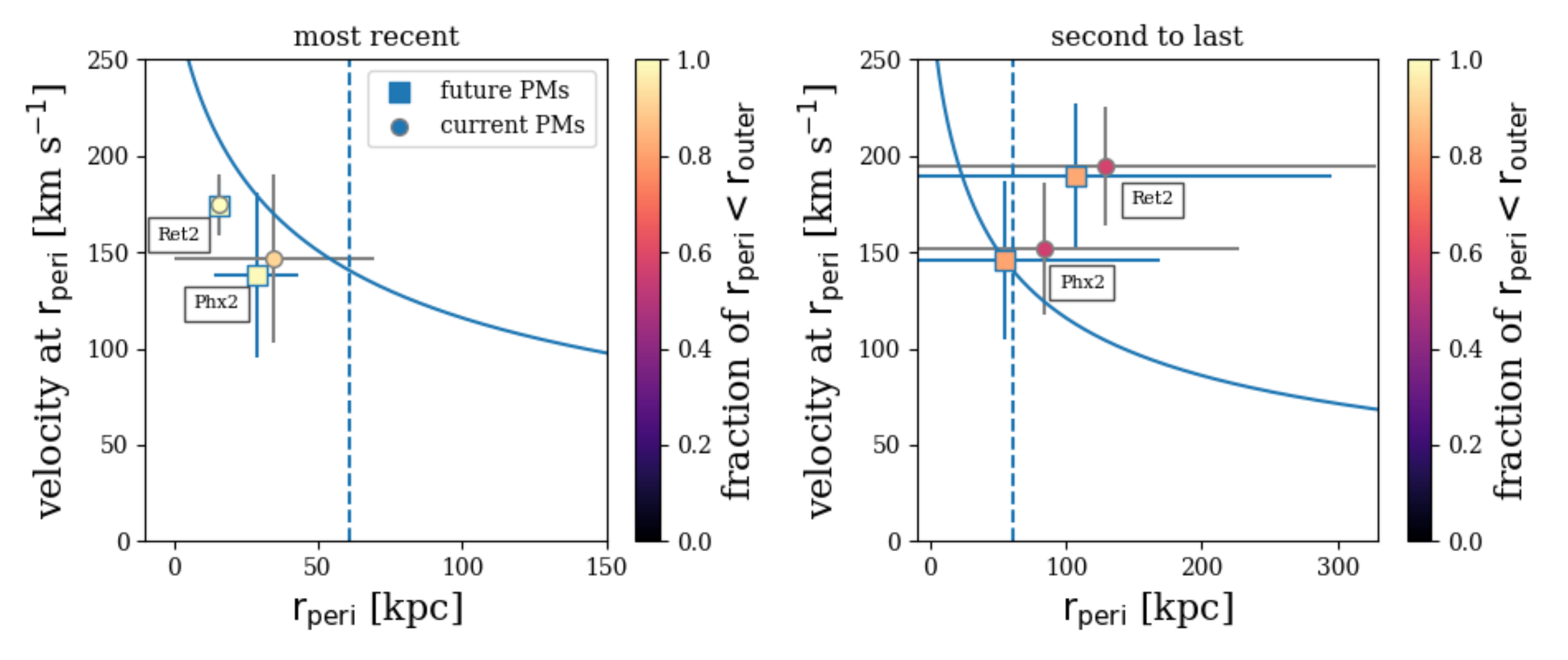}
\caption{Distance and velocity at pericenter for Ret2 and Phx2 where filled circles and errors are identical to the data from Fig. \ref{fig:orbparams}. Filled squares (future PMs) illustrate the same properties calculated when the measured uncertainty in PM components has been reduced to 25\% of their current values. Smaller PM uncertainties decrease the distance and velocity at the most recent pericenter for Phx2 and yield similar average results, though with smaller error bars for Ret2. For the second pericentric passage, the distance at pericenter is reduced by $\sim$20 kpc and the fraction of orbits where $\rm r_{peri} < r_{outer}$ rises from 0.6 to 0.8 for both Ret2 and Phx2. A more precise PM measurement makes Phx2 a long-term Magellanic satellite, while Ret2 is still recently captured by the MCs.}
\label{fig:PMreduction}
\end{figure*}

Fig. \ref{fig:PMreduction} shows the resulting orbital properties for Ret2 and Phx2 when the PM uncertainties are reduced (filled squares) while keeping the most likely PM values fixed\footnote{In reality, the most likely value for both PM components will also shift by $\sim 1 \sigma$ on average, further increasing the chances that satellites will be re-classified from one category to another.}. The original values for the same properties are also plotted (filled circles) for reference. The orbital properties at the most recent pericenter (left panel) remain similar to the original results reported in Fig. \ref{fig:orbparams}. At the second pericentric passage (right panel), more significant changes in $\rm r_{peri}$ and the fraction of orbits where $\rm r_{peri} < r_{outer}$ are noticeable. There is a similar effect on both Ret2 and Phx2 in the right panel such that the average value of $\rm r_{peri}$ decreases by $\sim$20 kpc and the fraction of satisfactory orbits increases to nearly 0.8 \citep[see also Section 3 of][]{erkal19}. 

With smaller PM uncertainties, Phx2 becomes a long-term Magellanic satellite, while Ret2 remains a recently captured Magellanic satellite. However, it is yet to be determined whether this truly suggests Ret2 was only recently captured by the MCs or if this is an artifact of large uncertainties on orbital parameters even with smaller PM uncertainties. More precise PM measurements are therefore necessary to confirm or invalidate their short-lived nature as Magellanic satellites.

\section{Conclusions}
\label{sec:conclusions}
We have used Gaia DR2 PMs to calculate the orbital histories of 13 UFD galaxies and 5 classical dwarf spheroidals within the VPOS to identify which galaxies are the most likely to be Magellanic satellites. These orbits are computed in a static MW+LMC+SMC potential where all galaxies, including the MW, are free to move in response to the gravitational influence of each other. Dynamical friction from the MW and LMC are also included where the latter is calibrated to a realistic SMC orbit from N-body simulations.

We also calculate orbits in a MW only and MW+LMC potential for comparison. Orbits are calculated for both a low mass MW1 ($\rm M_{vir}=10^{12}\, M_{\sun}$) and high mass MW2 ($\rm M_{vir}=1.5\times10^{12}\,M_{\sun}$) potential as well as three different LMC mass models ($\rm M_{vir}=0.8,1.8,2.5 \times 10^{11}\, M_{\sun}$). Our findings are summarized below:

\begin{enumerate}[leftmargin=*]
\item Direct orbital histories for all 18 galaxies in our sample using the fiducial LMC model (LMC2) are presented in Figs. \ref{fig:classical_vpos} and \ref{fig:UFD_orbits}. These orbits represent one orbital solution calculated from the average PM, line-of-sight velocity, and distance modulus converted to Galactocentric quantities. For the classical dwarfs, direct orbits using both previously measured PMs and Gaia DR2 PMs are calculated. We find consistency for all satellites except Fornax, which completes multiple passages around the MW at closer distances than predicted by previous PMs. The orbits of all five classical satellites are noticeably impacted by the inclusion of the LMC. These differences manifest as changes in the orbital period, distance at pericenter and apocenter, as well as the timing of these critical points. The SMC has a less significant effect on the orbits of classical dwarfs.

\item The gravitational influence of the LMC and SMC each impact the direct orbits of the UFD satellites. The SMC has a more noticeable effect compared to the orbits of the classical dwarfs, such that it too can alter the timing and distances at pericenter and apocenter. The LMC most significantly perturbs the direct orbits of the following UFD satellites: Car2, Car3, Hor1, Hyi1, Ret2, Tuc3, and Phx2. The addition of SMC in particular highly affects the orbits and long-term dynamics of Tuc3 and Ret2.

\item By evaluating the statistical significance of orbital properties calculated relative to the LMC in a combined LMC+SMC+MW potential, we separate galaxies into the following classes: 1) long-term Magellanic satellites, 2) recently captured Magellanic satellites, 3) MW satellites that have recently interacted with the MCs, and 4) MW satellites. For the fiducial LMC model (LMC2), Car2, Car3, Hor1, and Hyi1 are identified as long-term satellites, while Ret2 and Phx2 are recently captured Magellanic satellites (see Table \ref{tab:summary_sats_frac}). 

\item The masses of the MW and LMC play key roles in the classification of Magellanic satellites. In a low mass MW (MW1), the LMC is on first infall only completing one recent passage around the MW, whereas for a high mass MW, the LMC completes two pericentric passages in the last 6 Gyr. The binarity of the LMC-SMC orbit is also short-lived for MW2. As a result, the highest number of Magellanic satellites are identified for a low mass MW (MW1) and high mass LMC (LMC2, LMC3) combination. Results are similar between MW1 and MW2 except that Car2 is not a Magellanic satellite for a high mass MW (MW2). 

\item In Table \ref{tab:summary_sats_frac_noSMC}, we tabulate the impact of the SMC's gravitational influence on the orbital histories of galaxies in our sample. By calculating the statistical significance of orbital properties in a MW+LMC potential (no SMC), we find the same total number of Magellanic satellites but with a larger fraction that are recently captured by the MCs. This suggests that the SMC impacts the implied longevity of Magellanic satellites. Ret2 and Phx2 are exclusively identified as recently captured Magellanic satellites in the MW+LMC potential, whereas they can be long-term Magellanic satellites in the MW+LMC+SMC potential. Hor1 and Hyi1 are also categorized as recently captured Magellanic satellites in certain MW-LMC mass combinations, but are always long-term Magellanic satellites in the MW+LMC+SMC scenario. 

\item PMs will become more precise as upcoming measurements from Gaia, HST, and JWST are taken with longer time baselines between epochs. We tested whether reducing the PM measurement errors of Ret2 and Phx2 to 25\% of their current values provides more narrow constraints on their orbital histories. Ret2 is still always recently captured by the MCs, but Phx2 can be a long-term Magellanic satellite with smaller PM uncertainties. However, improved PMs are necessary to determine whether our results for recently captured Magellanic satellites are truly short-lived members of the Magellanic system or if this is an artifact of large orbital uncertainties.
\end{enumerate}

Our findings that a total of 3-6 of the 18 galaxies analyzed in this work are identified as Magellanic satellites are consistent with the low end of cosmological expectations \citep[e.g.][]{sales13, sales17, deason15, dooley17b, jahn19}. The recent findings of \citet{nadler19} are most applicable to our analysis as they account for the survey footprints in which our sample of UFDs were discovered. These authors use an observational selection function combined with theoretical models to determine that $4.7\pm1.8$ satellites observed with DES and PS1 are LMC-associated satellites. In \citet{nadler19} \emph{LMC-associated} refers to surviving satellites residing within the LMC's virial radius at the time the LMC falls into the MW's halo (which they find is  $\leq 2$ Gyr ago). While our definition of Magellanic satellites differs from \citet{nadler19}, the consistency between our results is promising.

Varying criteria have recently been used to identify Magellanic satellites, leading to a wide range of conclusions for the same galaxies. A common definition for identifying satellites around complex systems such as the MCs is therefore necessary both in cosmological studies and in studies like this that use astrometry and orbital histories to determine membership. If UFDs are detected around M33 in the near future, as predicted in \cite{patel18b}, and PMs are obtained in the decade to follow, this will be crucial for determining whether these satellites are dynamically associated to M33 or M31. 

Chemical abundance measurements and forthcoming star formation histories (SFHs) will both play key roles in identifying observational trends that complement the orbital histories of Magellanic satellites presented in this work. Chemical abundance ratios provide one opportunity for uncovering the formation histories of UFDs, and these signatures may provide an independent method of separating Magellanic UFDs from MW UFDs. Detailed chemical abundance analyses have been carried out for four of our Magellanic satellites \citep{ji16, nagasawa18, ji19}. However, more analysis is necessary to conclusively state whether there are obvious differences between Magellanic and MW UFDs.

SFHs are only available for two of the UFD satellites in our sample \citep{weisz14, brown14}. Upcoming SFHs of Magellanic satellites derived from deep HST imaging (HST program GO-14734; P.I. - N. Kallivayalil) will specifically illuminate differences between SFHs of the UFDs that are of Magellanic origin and those that are purely satellites of the MW (Sacchi et al., in prep.).

\section*{Acknowledgements}
{EP was supported by the National Science Foundation through the Graduate Research Fellowship Program funded by Grant Award No. DGE-1746060 and is currently supported by the Miller Institute for Basic Research, University of California Berkeley. NK is supported by NSF CAREER award 1455260. GB acknowledges support from NSF Grant AST-1714979. NGC is supported by NASA 17-ATP17-0006 and HST AR 15004. DRW acknowledges fellowship support from the Alfred P. Sloan Foundation and the Alexander von Humboldt Foundation. This support was provided by NASA through grant numbers HST-GO-15476 and JWST-DD-ERS-1334 from the Space Telescope Science Institute, which is operated by AURA, Inc., under NASA contract NAS5-26555. MBK acknowledges support from NSF CAREER award AST-1752913, NSF grant AST-1910346, NASA grant NNX17AG29G, and HST-AR-14282, HST-AR-14554, HST-AR-15006, HST-GO-14191, and HST-GO-15658 from STScI. FAG acknowledges financial support from CONICYT through the project FONDECYT Regular Nr. 1181264, and funding from the Max Planck Society through a Partner Group grant. EP would like to thank Rachel Smullen for stimulating discussions that have improved the quality of this work. The authors would also like to thank the referee for their helpful comments.} 

\software{\texttt{astropy} \citep{astropy}, \texttt{matplotlib} \citep{matplotlib}, \texttt{numpy} \citep{numpy}, and \texttt{scipy} \citep{scipy}.}

\bibliography{thesisrefs}{}

\begin{thebibliography}{}
\expandafter\ifx\csname natexlab\endcsname\relax\def\natexlab#1{#1}\fi
\providecommand{\url}[1]{\href{#1}{#1}}
\providecommand{\dodoi}[1]{doi:~\href{http://doi.org/#1}{\nolinkurl{#1}}}
\providecommand{\doeprint}[1]{\href{http://ascl.net/#1}{\nolinkurl{http://ascl.net/#1}}}
\providecommand{\doarXiv}[1]{\href{https://arxiv.org/abs/#1}{\nolinkurl{https://arxiv.org/abs/#1}}}

\bibitem[{{Battaglia} {et~al.}(2012){Battaglia}, {Irwin}, {Tolstoy}, {de Boer},
  \& {Mateo}}]{battaglia12}
{Battaglia}, G., {Irwin}, M., {Tolstoy}, E., {de Boer}, T., \& {Mateo}, M.
  2012, \apjl, 761, L31, \dodoi{10.1088/2041-8205/761/2/L31}

\bibitem[{{Battaglia} {et~al.}(2006){Battaglia}, {Tolstoy}, {Helmi}, {Irwin},
  {Letarte}, {Jablonka}, {Hill}, {Venn}, {Shetrone}, {Arimoto}, {Primas},
  {Kaufer}, {Francois}, {Szeifert}, {Abel}, \& {Sadakane}}]{battaglia06}
{Battaglia}, G., {Tolstoy}, E., {Helmi}, A., {et~al.} 2006, \aap, 459, 423,
  \dodoi{10.1051/0004-6361:20065720}

\bibitem[{{Bechtol} {et~al.}(2015){Bechtol}, {Drlica-Wagner}, {Balbinot},
  {Pieres}, {Simon}, {Yanny}, {Santiago}, {Wechsler}, {Frieman}, {Walker},
  {Williams}, {Rozo}, {Rykoff}, {Queiroz}, {Luque}, {Benoit-L{\'e}vy},
  {Tucker}, {Sevilla}, {Gruendl}, {da Costa}, {Fausti Neto}, {Maia}, {Abbott},
  {Allam}, {Armstrong}, {Bauer}, {Bernstein}, {Bernstein}, {Bertin}, {Brooks},
  {Buckley-Geer}, {Burke}, {Carnero Rosell}, {Castander}, {Covarrubias},
  {D'Andrea}, {DePoy}, {Desai}, {Diehl}, {Eifler}, {Estrada}, {Evrard},
  {Fernandez}, {Finley}, {Flaugher}, {Gaztanaga}, {Gerdes}, {Girardi},
  {Gladders}, {Gruen}, {Gutierrez}, {Hao}, {Honscheid}, {Jain}, {James},
  {Kent}, {Kron}, {Kuehn}, {Kuropatkin}, {Lahav}, {Li}, {Lin}, {Makler},
  {March}, {Marshall}, {Martini}, {Merritt}, {Miller}, {Miquel}, {Mohr},
  {Neilsen}, {Nichol}, {Nord}, {Ogando}, {Peoples}, {Petravick}, {Plazas},
  {Romer}, {Roodman}, {Sako}, {Sanchez}, {Scarpine}, {Schubnell}, {Smith},
  {Soares-Santos}, {Sobreira}, {Suchyta}, {Swanson}, {Tarle}, {Thaler},
  {Thomas}, {Wester}, {Zuntz}, \& {DES Collaboration}}]{bechtol15}
{Bechtol}, K., {Drlica-Wagner}, A., {Balbinot}, E., {et~al.} 2015, \apj, 807,
  50, \dodoi{10.1088/0004-637X/807/1/50}

\bibitem[{{Bekki} \& {Chiba}(2005)}]{bekki05}
{Bekki}, K., \& {Chiba}, M. 2005, \mnras, 356, 680,
  \dodoi{10.1111/j.1365-2966.2004.08510.x}

\bibitem[{{Bellazzini} {et~al.}(2002){Bellazzini}, {Ferraro}, {Origlia},
  {Pancino}, {Monaco}, \& {Oliva}}]{bellazzini02}
{Bellazzini}, M., {Ferraro}, F.~R., {Origlia}, L., {et~al.} 2002, \aj, 124,
  3222, \dodoi{10.1086/344794}

\bibitem[{{Belokurov} {et~al.}(2007){Belokurov}, {Zucker}, {Evans}, {Kleyna},
  {Koposov}, {Hodgkin}, {Irwin}, {Gilmore}, {Wilkinson}, {Fellhauer},
  {Bramich}, {Hewett}, {Vidrih}, {De Jong}, {Smith}, {Rix}, {Bell}, {Wyse},
  {Newberg}, {Mayeur}, {Yanny}, {Rockosi}, {Gnedin}, {Schneider}, {Beers},
  {Barentine}, {Brewington}, {Brinkmann}, {Harvanek}, {Kleinman}, {Krzesinski},
  {Long}, {Nitta}, \& {Snedden}}]{belokurov07}
{Belokurov}, V., {Zucker}, D.~B., {Evans}, N.~W., {et~al.} 2007, \apj, 654,
  897, \dodoi{10.1086/509718}

\bibitem[{{Besla}(2015)}]{besla15}
{Besla}, G. 2015, ArXiv e-prints.
\newblock \doarXiv{1511.03346}

\bibitem[{{Besla} {et~al.}(2013){Besla}, {Hernquist}, \& {Loeb}}]{besla13}
{Besla}, G., {Hernquist}, L., \& {Loeb}, A. 2013, \mnras, 428, 2342,
  \dodoi{10.1093/mnras/sts192}

\bibitem[{{Besla} {et~al.}(2007){Besla}, {Kallivayalil}, {Hernquist},
  {Robertson}, {Cox}, {van der Marel}, \& {Alcock}}]{b07}
{Besla}, G., {Kallivayalil}, N., {Hernquist}, L., {et~al.} 2007, \apj, 668,
  949, \dodoi{10.1086/521385}

\bibitem[{Besla {et~al.}(2010)Besla, Kallivayalil, Hernquist, van~der Marel,
  Cox, \& Kere{\v s}}]{besla10}
Besla, G., Kallivayalil, N., Hernquist, L., {et~al.} 2010, The Astrophysical
  Journal Letters, 721, L97.
\newblock \url{http://stacks.iop.org/2041-8205/721/i=2/a=L97}

\bibitem[{{Besla} {et~al.}(2012){Besla}, {Kallivayalil}, {Hernquist}, {van der
  Marel}, {Cox}, \& {Kere{\v s}}}]{besla12}
{Besla}, G., {Kallivayalil}, N., {Hernquist}, L., {et~al.} 2012, \mnras, 421,
  2109, \dodoi{10.1111/j.1365-2966.2012.20466.x}

\bibitem[{{Besla} {et~al.}(2016){Besla}, {Mart{\'{\i}}nez-Delgado}, {van der
  Marel}, {Beletsky}, {Seibert}, {Schlafly}, {Grebel}, \& {Neyer}}]{besla16}
{Besla}, G., {Mart{\'{\i}}nez-Delgado}, D., {van der Marel}, R.~P., {et~al.}
  2016, \apj, 825, 20, \dodoi{10.3847/0004-637X/825/1/20}

\bibitem[{{Bonanos} {et~al.}(2004){Bonanos}, {Stanek}, {Szentgyorgyi},
  {Sasselov}, \& {Bakos}}]{bonanos04}
{Bonanos}, A.~Z., {Stanek}, K.~Z., {Szentgyorgyi}, A.~H., {Sasselov}, D.~D., \&
  {Bakos}, G.~{\'A}. 2004, \aj, 127, 861, \dodoi{10.1086/381073}

\bibitem[{{Boylan-Kolchin} {et~al.}(2011){Boylan-Kolchin}, {Besla}, \&
  {Hernquist}}]{bk11}
{Boylan-Kolchin}, M., {Besla}, G., \& {Hernquist}, L. 2011, \mnras, 414, 1560,
  \dodoi{10.1111/j.1365-2966.2011.18495.x}

\bibitem[{{Brown} {et~al.}(2014){Brown}, {Tumlinson}, {Geha}, {Kirby}, {Vand
  enBerg}, {Kalirai}, {Simon}, {Avila}, {Munoz}, {Guhathakurta}, {Renzini},
  {Ferguson}, {Vargas}, \& {Gennaro}}]{brown14}
{Brown}, T.~M., {Tumlinson}, J., {Geha}, M., {et~al.} 2014, \memsai, 85, 493.
\newblock \doarXiv{1310.0824}

\bibitem[{{Bryan} \& {Norman}(1998)}]{brynorman98}
{Bryan}, G.~L., \& {Norman}, M.~L. 1998, \apj, 495, 80, \dodoi{10.1086/305262}

\bibitem[{{Bullock} \& {Boylan-Kolchin}(2017)}]{bullockbk17}
{Bullock}, J.~S., \& {Boylan-Kolchin}, M. 2017, \araa, 55, 343,
  \dodoi{10.1146/annurev-astro-091916-055313}

\bibitem[{{Caldwell} {et~al.}(2017){Caldwell}, {Walker}, {Mateo}, {Olszewski},
  {Koposov}, {Belokurov}, {Torrealba}, {Geringer-Sameth}, \&
  {Johnson}}]{caldwell17}
{Caldwell}, N., {Walker}, M.~G., {Mateo}, M., {et~al.} 2017, \apj, 839, 20,
  \dodoi{10.3847/1538-4357/aa688e}

\bibitem[{{Carrera} {et~al.}(2002){Carrera}, {Aparicio},
  {Mart{\'\i}nez-Delgado}, \& {Alonso-Garc{\'\i}a}}]{carrera02}
{Carrera}, R., {Aparicio}, A., {Mart{\'\i}nez-Delgado}, D., \&
  {Alonso-Garc{\'\i}a}, J. 2002, \aj, 123, 3199, \dodoi{10.1086/340702}

\bibitem[{{Chandrasekhar}(1943)}]{chandrasekhar}
{Chandrasekhar}, S. 1943, \apj, 97, 255, \dodoi{10.1086/144517}

\bibitem[{{Cioni} {et~al.}(2000){Cioni}, {van der Marel}, {Loup}, \&
  {Habing}}]{cioni00}
{Cioni}, M. R.~L., {van der Marel}, R.~P., {Loup}, C., \& {Habing}, H.~J. 2000,
  \aap, 359, 601.
\newblock \doarXiv{astro-ph/0003223}

\bibitem[{{Coppola} {et~al.}(2015){Coppola}, {Marconi}, {Stetson}, {Bono},
  {Braga}, {Ripepi}, {Dall'Ora}, {Musella}, {Buonanno}, {Fabrizio}, {Ferraro},
  {Fiorentino}, {Iannicola}, {Monelli}, {Nonino}, {Th{\'e}venin}, \&
  {Walker}}]{coppola15}
{Coppola}, G., {Marconi}, M., {Stetson}, P.~B., {et~al.} 2015, \apj, 814, 71,
  \dodoi{10.1088/0004-637X/814/1/71}

\bibitem[{{Deason} {et~al.}(2015){Deason}, {Wetzel}, {Garrison-Kimmel}, \&
  {Belokurov}}]{deason15}
{Deason}, A.~J., {Wetzel}, A.~R., {Garrison-Kimmel}, S., \& {Belokurov}, V.
  2015, \mnras, 453, 3568, \dodoi{10.1093/mnras/stv1939}

\bibitem[{{Di Teodoro} {et~al.}(2019){Di Teodoro}, {McClure-Griffiths},
  {Jameson}, {D{\'e}nes}, {Dickey}, {Stanimirovi{\'c}}, {Staveley-Smith},
  {Anderson}, {Bunton}, {Chippendale}, {Lee-Waddell}, {MacLeod}, \&
  {Voronkov}}]{diteodoro19}
{Di Teodoro}, E.~M., {McClure-Griffiths}, N.~M., {Jameson}, K.~E., {et~al.}
  2019, \mnras, 483, 392, \dodoi{10.1093/mnras/sty3095}

\bibitem[{{D'Onghia} \& {Lake}(2008)}]{donghia08}
{D'Onghia}, E., \& {Lake}, G. 2008, \apj, 686, L61, \dodoi{10.1086/592995}

\bibitem[{{Dooley} {et~al.}(2017){Dooley}, {Peter}, {Carlin}, {Frebel},
  {Bechtol}, \& {Willman}}]{dooley17b}
{Dooley}, G.~A., {Peter}, A.~H.~G., {Carlin}, J.~L., {et~al.} 2017, \mnras,
  472, 1060, \dodoi{10.1093/mnras/stx2001}

\bibitem[{{Drlica-Wagner} {et~al.}(2015){Drlica-Wagner}, {Bechtol}, {Rykoff},
  {Luque}, {Queiroz}, {Mao}, {Wechsler}, {Simon}, {Santiago}, {Yanny},
  {Balbinot}, {Dodelson}, {Fausti Neto}, {James}, {Li}, {Maia}, {Marshall},
  {Pieres}, {Stringer}, {Walker}, {Abbott}, {Abdalla}, {Allam},
  {Benoit-L{\'e}vy}, {Bernstein}, {Bertin}, {Brooks}, {Buckley-Geer}, {Burke},
  {Carnero Rosell}, {Carrasco Kind}, {Carretero}, {Crocce}, {da Costa},
  {Desai}, {Diehl}, {Dietrich}, {Doel}, {Eifler}, {Evrard}, {Finley},
  {Flaugher}, {Fosalba}, {Frieman}, {Gaztanaga}, {Gerdes}, {Gruen}, {Gruendl},
  {Gutierrez}, {Honscheid}, {Kuehn}, {Kuropatkin}, {Lahav}, {Martini},
  {Miquel}, {Nord}, {Ogando}, {Plazas}, {Reil}, {Roodman}, {Sako}, {Sanchez},
  {Scarpine}, {Schubnell}, {Sevilla-Noarbe}, {Smith}, {Soares-Santos},
  {Sobreira}, {Suchyta}, {Swanson}, {Tarle}, {Tucker}, {Vikram}, {Wester},
  {Zhang}, {Zuntz}, \& {DES Collaboration}}]{dwagner15}
{Drlica-Wagner}, A., {Bechtol}, K., {Rykoff}, E.~S., {et~al.} 2015, \apj, 813,
  109, \dodoi{10.1088/0004-637X/813/2/109}

\bibitem[{{Drlica-Wagner} {et~al.}(2016){Drlica-Wagner}, {Bechtol}, {Allam},
  {Tucker}, {Gruendl}, {Johnson}, {Walker}, {James}, {Nidever}, {Olsen},
  {Wechsler}, {Cioni}, {Conn}, {Kuehn}, {Li}, {Mao}, {Martin}, {Neilsen},
  {Noel}, {Pieres}, {Simon}, {Stringfellow}, {van der Marel}, \&
  {Yanny}}]{dwagner16}
{Drlica-Wagner}, A., {Bechtol}, K., {Allam}, S., {et~al.} 2016, \apjl, 833, L5,
  \dodoi{10.3847/2041-8205/833/1/L5}

\bibitem[{{Erkal} \& {Belokurov}(2019)}]{erkal19}
{Erkal}, D., \& {Belokurov}, V.~A. 2019, arXiv e-prints, arXiv:1907.09484.
\newblock \doarXiv{1907.09484}

\bibitem[{{Erkal} {et~al.}(2018){Erkal}, {Li}, {Koposov}, {Belokurov},
  {Balbinot}, {Bechtol}, {Buncher}, {Drlica-Wagner}, {Kuehn}, {Marshall},
  {Mart{\'\i}nez-V{\'a}zquez}, {Pace}, {Shipp}, {Simon}, {Stringer}, {Vivas},
  {Wechsler}, {Yanny}, {Abdalla}, {Allam}, {Annis}, {Avila}, {Bertin},
  {Brooks}, {Buckley-Geer}, {Burke}, {Carnero Rosell}, {Carrasco Kind},
  {Carretero}, {D'Andrea}, {da Costa}, {Davis}, {De Vicente}, {Doel}, {Eifler},
  {Evrard}, {Flaugher}, {Frieman}, {Garc{\'\i}a-Bellido}, {Gaztanaga},
  {Gerdes}, {Gruen}, {Gruendl}, {Gschwend}, {Gutierrez}, {Hartley},
  {Hollowood}, {Honscheid}, {James}, {Krause}, {Maia}, {March}, {Menanteau},
  {Miquel}, {Ogando}, {Plazas}, {Sanchez}, {Santiago}, {Scarpine}, {Schindler},
  {Sevilla-Noarbe}, {Smith}, {Smith}, {Soares-Santos}, {Sobreira}, {Suchyta},
  {Swanson}, {Tarle}, {Tucker}, \& {Walker}}]{erkal18}
{Erkal}, D., {Li}, T.~S., {Koposov}, S.~E., {et~al.} 2018, \mnras, 481, 3148,
  \dodoi{10.1093/mnras/sty2518}

\bibitem[{Fattahi {et~al.}(2018)Fattahi, Navarro, Frenk, Oman, Sawala, \&
  Schaller}]{fattahi18}
Fattahi, A., Navarro, J.~F., Frenk, C.~S., {et~al.} 2018, Monthly Notices of
  the Royal Astronomical Society, 476, 3816, \dodoi{10.1093/mnras/sty408}

\bibitem[{{Freedman} {et~al.}(2001){Freedman}, {Madore}, {Gibson}, {Ferrarese},
  {Kelson}, {Sakai}, {Mould}, {Kennicutt}, {Ford}, {Graham}, {Huchra},
  {Hughes}, {Illingworth}, {Macri}, \& {Stetson}}]{freedman01}
{Freedman}, W.~L., {Madore}, B.~F., {Gibson}, B.~K., {et~al.} 2001, \apj, 553,
  47, \dodoi{10.1086/320638}

\bibitem[{{Fritz} {et~al.}(2018){Fritz}, {Battaglia}, {Pawlowski},
  {Kallivayalil}, {van der Marel}, {Sohn}, {Brook}, \& {Besla}}]{fritz18}
{Fritz}, T.~K., {Battaglia}, G., {Pawlowski}, M.~S., {et~al.} 2018, \aap, 619,
  A103, \dodoi{10.1051/0004-6361/201833343}

\bibitem[{{Fritz} {et~al.}(2019){Fritz}, {Carrera}, {Battaglia}, \&
  {Taibi}}]{fritz19}
{Fritz}, T.~K., {Carrera}, R., {Battaglia}, G., \& {Taibi}, S. 2019, \aap, 623,
  A129, \dodoi{10.1051/0004-6361/201833458}

\bibitem[{{Fu} {et~al.}(2019){Fu}, {Simon}, \& {Alarc{\'o}n Jara}}]{fu19}
{Fu}, S.~W., {Simon}, J.~D., \& {Alarc{\'o}n Jara}, A.~G. 2019, \apj, 883, 11,
  \dodoi{10.3847/1538-4357/ab3658}

\bibitem[{{Gaia Collaboration} {et~al.}(2018{\natexlab{a}}){Gaia
  Collaboration}, {Brown}, {Vallenari}, {Prusti}, {de Bruijne}, {Babusiaux},
  {Bailer-Jones}, {Biermann}, {Evans}, {Eyer}, {Jansen}, {Jordi}, {Klioner},
  {Lammers}, {Lindegren}, {Luri}, {Mignard}, {Panem}, {Pourbaix}, {Randich},
  {Sartoretti}, {Siddiqui}, {Soubiran}, {van Leeuwen}, {Walton}, {Arenou},
  {Bastian}, {Cropper}, {Drimmel}, {Katz}, {Lattanzi}, {Bakker}, {Cacciari},
  {Casta{\~n}eda}, {Chaoul}, {Cheek}, {De Angeli}, {Fabricius}, {Guerra},
  {Holl}, {Masana}, {Messineo}, {Mowlavi}, {Nienartowicz}, {Panuzzo},
  {Portell}, {Riello}, {Seabroke}, {Tanga}, {Th{\'e}venin}, {Gracia-Abril},
  {Comoretto}, {Garcia-Reinaldos}, {Teyssier}, {Altmann}, {Andrae}, {Audard},
  {Bellas-Velidis}, {Benson}, {Berthier}, {Blomme}, {Burgess}, {Busso},
  {Carry}, {Cellino}, {Clementini}, {Clotet}, {Creevey}, {Davidson}, {De
  Ridder}, {Delchambre}, {Dell'Oro}, {Ducourant},
  {Fern{\'a}ndez-Hern{\'a}ndez}, {Fouesneau}, {Fr{\'e}mat}, {Galluccio},
  {Garc{\'\i}a-Torres}, {Gonz{\'a}lez-N{\'u}{\~n}ez}, {Gonz{\'a}lez-Vidal},
  {Gosset}, {Guy}, {Halbwachs}, {Hambly}, {Harrison}, {Hern{\'a}ndez},
  {Hestroffer}, {Hodgkin}, {Hutton}, {Jasniewicz}, {Jean-Antoine-Piccolo},
  {Jordan}, {Korn}, {Krone-Martins}, {Lanzafame}, {Lebzelter}, {L{\"o}ffler},
  {Manteiga}, {Marrese}, {Mart{\'\i}n-Fleitas}, {Moitinho}, {Mora}, {Muinonen},
  {Osinde}, {Pancino}, {Pauwels}, {Petit}, {Recio-Blanco}, {Richards},
  {Rimoldini}, {Robin}, {Sarro}, {Siopis}, {Smith}, {Sozzetti}, {S{\"u}veges},
  {Torra}, {van Reeven}, {Abbas}, {Abreu Aramburu}, {Accart}, {Aerts},
  {Altavilla}, {{\'A}lvarez}, {Alvarez}, {Alves}, {Anderson}, {Andrei},
  {Anglada Varela}, {Antiche}, {Antoja}, {Arcay}, {Astraatmadja}, {Bach},
  {Baker}, {Balaguer-N{\'u}{\~n}ez}, {Balm}, {Barache}, {Barata}, {Barbato},
  {Barblan}, {Barklem}, {Barrado}, {Barros}, {Barstow}, {Bartholom{\'e}
  Mu{\~n}oz}, {Bassilana}, {Becciani}, {Bellazzini}, {Berihuete}, {Bertone},
  {Bianchi}, {Bienaym{\'e}}, {Blanco-Cuaresma}, {Boch}, {Boeche}, {Bombrun},
  {Borrachero}, {Bossini}, {Bouquillon}, {Bourda}, {Bragaglia}, {Bramante},
  {Breddels}, {Bressan}, {Brouillet}, {Br{\"u}semeister}, {Brugaletta},
  {Bucciarelli}, {Burlacu}, {Busonero}, {Butkevich}, {Buzzi}, {Caffau},
  {Cancelliere}, {Cannizzaro}, {Cantat-Gaudin}, {Carballo}, {Carlucci},
  {Carrasco}, {Casamiquela}, {Castellani}, {Castro-Ginard}, {Charlot},
  {Chemin}, {Chiavassa}, {Cocozza}, {Costigan}, {Cowell}, {Crifo}, {Crosta},
  {Crowley}, {Cuypers}, {Dafonte}, {Damerdji}, {Dapergolas}, {David}, {David},
  {de Laverny}, {De Luise}, {De March}, {de Martino}, {de Souza}, {de Torres},
  {Debosscher}, {del Pozo}, {Delbo}, {Delgado}, {Delgado}, {Di Matteo},
  {Diakite}, {Diener}, {Distefano}, {Dolding}, {Drazinos}, {Dur{\'a}n},
  {Edvardsson}, {Enke}, {Eriksson}, {Esquej}, {Eynard Bontemps}, {Fabre},
  {Fabrizio}, {Faigler}, {Falc{\~a}o}, {Farr{\`a}s Casas}, {Federici},
  {Fedorets}, {Fernique}, {Figueras}, {Filippi}, {Findeisen}, {Fonti},
  {Fraile}, {Fraser}, {Fr{\'e}zouls}, {Gai}, {Galleti}, {Garabato},
  {Garc{\'\i}a-Sedano}, {Garofalo}, {Garralda}, {Gavel}, {Gavras}, {Gerssen},
  {Geyer}, {Giacobbe}, {Gilmore}, {Girona}, {Giuffrida}, {Glass}, {Gomes},
  {Granvik}, {Gueguen}, {Guerrier}, {Guiraud}, {Guti{\'e}rrez-S{\'a}nchez},
  {Haigron}, {Hatzidimitriou}, {Hauser}, {Haywood}, {Heiter}, {Helmi}, {Heu},
  {Hilger}, {Hobbs}, {Hofmann}, {Holland}, {Huckle}, {Hypki}, {Icardi},
  {Jan{\ss}en}, {Jevardat de Fombelle}, {Jonker}, {Juh{\'a}sz}, {Julbe},
  {Karampelas}, {Kewley}, {Klar}, {Kochoska}, {Kohley}, {Kolenberg},
  {Kontizas}, {Kontizas}, {Koposov}, {Kordopatis}, {Kostrzewa-Rutkowska},
  {Koubsky}, {Lambert}, {Lanza}, {Lasne}, {Lavigne}, {Le Fustec}, {Le
  Poncin-Lafitte}, {Lebreton}, {Leccia}, {Leclerc}, {Lecoeur-Taibi},
  {Lenhardt}, {Leroux}, {Liao}, {Licata}, {Lindstr{\o}m}, {Lister}, {Livanou},
  {Lobel}, {L{\'o}pez}, {Managau}, {Mann}, {Mantelet}, {Marchal}, {Marchant},
  {Marconi}, {Marinoni}, {Marschalk{\'o}}, {Marshall}, {Martino}, {Marton},
  {Mary}, {Massari}, {Matijevi{\v{c}}}, {Mazeh}, {McMillan}, {Messina},
  {Michalik}, {Millar}, {Molina}, {Molinaro}, {Moln{\'a}r}, {Montegriffo},
  {Mor}, {Morbidelli}, {Morel}, {Morris}, {Mulone}, {Muraveva}, {Musella},
  {Nelemans}, {Nicastro}, {Noval}, {O'Mullane}, {Ord{\'e}novic},
  {Ord{\'o}{\~n}ez-Blanco}, {Osborne}, {Pagani}, {Pagano}, {Pailler},
  {Palacin}, {Palaversa}, {Panahi}, {Pawlak}, {Piersimoni}, {Pineau}, {Plachy},
  {Plum}, {Poggio}, {Poujoulet}, {Pr{\v{s}}a}, {Pulone}, {Racero}, {Ragaini},
  {Rambaux}, {Ramos-Lerate}, {Regibo}, {Reyl{\'e}}, {Riclet}, {Ripepi}, {Riva},
  {Rivard}, {Rixon}, {Roegiers}, {Roelens}, {Romero-G{\'o}mez}, {Rowell},
  {Royer}, {Ruiz-Dern}, {Sadowski}, {Sagrist{\`a} Sell{\'e}s}, {Sahlmann},
  {Salgado}, {Salguero}, {Sanna}, {Santana-Ros}, {Sarasso}, {Savietto},
  {Schultheis}, {Sciacca}, {Segol}, {Segovia}, {S{\'e}gransan}, {Shih},
  {Siltala}, {Silva}, {Smart}, {Smith}, {Solano}, {Solitro}, {Sordo}, {Soria
  Nieto}, {Souchay}, {Spagna}, {Spoto}, {Stampa}, {Steele},
  {Steidelm{\"u}ller}, {Stephenson}, {Stoev}, {Suess}, {Surdej}, {Szabados},
  {Szegedi-Elek}, {Tapiador}, {Taris}, {Tauran}, {Taylor}, {Teixeira},
  {Terrett}, {Teyssand ier}, {Thuillot}, {Titarenko}, {Torra Clotet}, {Turon},
  {Ulla}, {Utrilla}, {Uzzi}, {Vaillant}, {Valentini}, {Valette}, {van Elteren},
  {Van Hemelryck}, {van Leeuwen}, {Vaschetto}, {Vecchiato}, {Veljanoski},
  {Viala}, {Vicente}, {Vogt}, {von Essen}, {Voss}, {Votruba}, {Voutsinas},
  {Walmsley}, {Weiler}, {Wertz}, {Wevers}, {Wyrzykowski}, {Yoldas},
  {{\v{Z}}erjal}, {Ziaeepour}, {Zorec}, {Zschocke}, {Zucker}, {Zurbach}, \&
  {Zwitter}}]{gaiadr2a}
{Gaia Collaboration}, {Brown}, A.~G.~A., {Vallenari}, A., {et~al.}
  2018{\natexlab{a}}, \aap, 616, A1, \dodoi{10.1051/0004-6361/201833051}

\bibitem[{{Gaia Collaboration} {et~al.}(2018{\natexlab{b}}){Gaia
  Collaboration}, {Helmi}, {van Leeuwen}, {McMillan}, {Massari}, {Antoja},
  {Robin}, {Lindegren}, {Bastian}, {Arenou}, \& et~al.}]{gaiadr2b}
{Gaia Collaboration}, {Helmi}, A., {van Leeuwen}, F., {et~al.}
  2018{\natexlab{b}}, \aap, 616, A12, \dodoi{10.1051/0004-6361/201832698}

\bibitem[{{Garavito-Camargo} {et~al.}(2019){Garavito-Camargo}, {Besla},
  {Laporte}, {Johnston}, {G{\'o}mez}, \& {Watkins}}]{garavitoc19}
{Garavito-Camargo}, N., {Besla}, G., {Laporte}, C. F.~P., {et~al.} 2019, \apj,
  884, 51, \dodoi{10.3847/1538-4357/ab32eb}

\bibitem[{{Gnedin} {et~al.}(2004){Gnedin}, {Kravtsov}, {Klypin}, \&
  {Nagai}}]{contra}
{Gnedin}, O.~Y., {Kravtsov}, A.~V., {Klypin}, A.~A., \& {Nagai}, D. 2004, \apj,
  616, 16, \dodoi{10.1086/424914}

\bibitem[{{G{\'o}mez} {et~al.}(2015){G{\'o}mez}, {Besla}, {Carpintero},
  {Villalobos}, {O'Shea}, \& {Bell}}]{gomez15}
{G{\'o}mez}, F.~A., {Besla}, G., {Carpintero}, D.~D., {et~al.} 2015, \apj, 802,
  128, \dodoi{10.1088/0004-637X/802/2/128}

\bibitem[{{Greco} {et~al.}(2008){Greco}, {Dall'Ora}, {Clementini}, {Ripepi},
  {Di Fabrizio}, {Kinemuchi}, {Marconi}, {Musella}, {Smith}, {Rodgers},
  {Kuehn}, {Beers}, {Catelan}, \& {Pritzl}}]{greco08}
{Greco}, C., {Dall'Ora}, M., {Clementini}, G., {et~al.} 2008, \apjl, 675, L73,
  \dodoi{10.1086/533585}

\bibitem[{{Harris} \& {Zaritsky}(2006)}]{harris06}
{Harris}, J., \& {Zaritsky}, D. 2006, \aj, 131, 2514, \dodoi{10.1086/500974}

\bibitem[{{Hashimoto} {et~al.}(2003){Hashimoto}, {Funato}, \&
  {Makino}}]{hashimoto03}
{Hashimoto}, Y., {Funato}, Y., \& {Makino}, J. 2003, \apj, 582, 196,
  \dodoi{10.1086/344260}

\bibitem[{{Hernquist}(1990)}]{hernquist90}
{Hernquist}, L. 1990, \apj, 356, 359, \dodoi{10.1086/168845}

\bibitem[{{Homma} {et~al.}(2018){Homma}, {Chiba}, {Okamoto}, {Komiyama},
  {Tanaka}, {Tanaka}, {Ishigaki}, {Hayashi}, {Arimoto}, {Garmilla}, {Lupton},
  {Strauss}, {Miyazaki}, {Wang}, \& {Murayama}}]{homma18}
{Homma}, D., {Chiba}, M., {Okamoto}, S., {et~al.} 2018, \pasj, 70, S18,
  \dodoi{10.1093/pasj/psx050}

\bibitem[{Hunter(2007)}]{matplotlib}
Hunter, J.~D. 2007, Computing in Science Engineering, 9, 90,
  \dodoi{10.1109/MCSE.2007.55}

\bibitem[{{Jahn} {et~al.}(2019){Jahn}, {Sales}, {Wetzel}, {Boylan-Kolchin},
  {Chan}, {El-Badry}, {Lazar}, \& {Bullock}}]{jahn19}
{Jahn}, E.~D., {Sales}, L.~V., {Wetzel}, A., {et~al.} 2019, \mnras, 489, 5348,
  \dodoi{10.1093/mnras/stz2457}

\bibitem[{{Jeon} {et~al.}(2017){Jeon}, {Besla}, \& {Bromm}}]{jeon17}
{Jeon}, M., {Besla}, G., \& {Bromm}, V. 2017, \apj, 848, 85,
  \dodoi{10.3847/1538-4357/aa8c80}

\bibitem[{{Jethwa} {et~al.}(2016){Jethwa}, {Erkal}, \& {Belokurov}}]{jethwa16}
{Jethwa}, P., {Erkal}, D., \& {Belokurov}, V. 2016, \mnras, 461, 2212,
  \dodoi{10.1093/mnras/stw1343}

\bibitem[{{Ji} {et~al.}(2016){Ji}, {Frebel}, {Simon}, \& {Chiti}}]{ji16}
{Ji}, A.~P., {Frebel}, A., {Simon}, J.~D., \& {Chiti}, A. 2016, \apj, 830, 93,
  \dodoi{10.3847/0004-637X/830/2/93}

\bibitem[{{Ji} {et~al.}(2019){Ji}, {Li}, {Simon}, {Marshall}, {Vivas}, {Pace},
  {Bechtol}, {Drlica-Wagner}, {Koposov}, {Hansen}, {Allam}, {Gruendl},
  {Johnson}, {McNanna}, {Noel}, {Tucker}, {Walker}, \& {MagLiteS
  Collaboration}}]{ji19}
{Ji}, A.~P., {Li}, T.~S., {Simon}, J.~D., {et~al.} 2019, arXiv e-prints,
  arXiv:1912.04963.
\newblock \doarXiv{1912.04963}

\bibitem[{Jones {et~al.}(2001--)Jones, Oliphant, Peterson, {et~al.}}]{scipy}
Jones, E., Oliphant, T., Peterson, P., {et~al.} 2001--, {SciPy}: Open source
  scientific tools for {Python}.
\newblock \url{http://www.scipy.org/}

\bibitem[{{Joo} {et~al.}(2018){Joo}, {Kyeong}, {Yang}, {Han}, {Sung}, {Kim},
  {Jeong}, {Ree}, {Rey}, {Jerjen}, {Kim}, {Cha}, \& {Lee}}]{joo18}
{Joo}, S.-J., {Kyeong}, J., {Yang}, S.-C., {et~al.} 2018, \apj, 861, 23,
  \dodoi{10.3847/1538-4357/aac4a3}

\bibitem[{{Kallivayalil} {et~al.}(2013){Kallivayalil}, {van der Marel},
  {Besla}, {Anderson}, \& {Alcock}}]{k13}
{Kallivayalil}, N., {van der Marel}, R.~P., {Besla}, G., {Anderson}, J., \&
  {Alcock}, C. 2013, \apj, 764, 161, \dodoi{10.1088/0004-637X/764/2/161}

\bibitem[{{Kallivayalil} {et~al.}(2018){Kallivayalil}, {Sales}, {Zivick},
  {Fritz}, {Del Pino}, {Sohn}, {Besla}, {van der Marel}, {Navarro}, \&
  {Sacchi}}]{kallivayalil18}
{Kallivayalil}, N., {Sales}, L.~V., {Zivick}, P., {et~al.} 2018, \apj, 867, 19,
  \dodoi{10.3847/1538-4357/aadfee}

\bibitem[{{Kim} \& {Jerjen}(2015)}]{kim15b}
{Kim}, D., \& {Jerjen}, H. 2015, \apjl, 808, L39,
  \dodoi{10.1088/2041-8205/808/2/L39}

\bibitem[{{Kim} {et~al.}(2015){Kim}, {Jerjen}, {Mackey}, {Da Costa}, \&
  {Milone}}]{kim15a}
{Kim}, D., {Jerjen}, H., {Mackey}, D., {Da Costa}, G.~S., \& {Milone}, A.~P.
  2015, \apjl, 804, L44, \dodoi{10.1088/2041-8205/804/2/L44}

\bibitem[{{Kinemuchi} {et~al.}(2008){Kinemuchi}, {Harris}, {Smith},
  {Silbermann}, {Snyder}, {La Cluyz{\'e}}, \& {Clark}}]{kinemuchi08}
{Kinemuchi}, K., {Harris}, H.~C., {Smith}, H.~A., {et~al.} 2008, \aj, 136,
  1921, \dodoi{10.1088/0004-6256/136/5/1921}

\bibitem[{{Kirby} {et~al.}(2015){Kirby}, {Simon}, \& {Cohen}}]{kirby15}
{Kirby}, E.~N., {Simon}, J.~D., \& {Cohen}, J.~G. 2015, \apj, 810, 56,
  \dodoi{10.1088/0004-637X/810/1/56}

\bibitem[{{Kirby} {et~al.}(2010){Kirby}, {Guhathakurta}, {Simon}, {Geha},
  {Rockosi}, {Sneden}, {Cohen}, {Sohn}, {Majewski}, \& {Siegel}}]{kirby10}
{Kirby}, E.~N., {Guhathakurta}, P., {Simon}, J.~D., {et~al.} 2010, \apjs, 191,
  352, \dodoi{10.1088/0067-0049/191/2/352}

\bibitem[{{Koposov} {et~al.}(2015{\natexlab{a}}){Koposov}, {Belokurov},
  {Torrealba}, \& {Evans}}]{koposov15}
{Koposov}, S.~E., {Belokurov}, V., {Torrealba}, G., \& {Evans}, N.~W.
  2015{\natexlab{a}}, \apj, 805, 130, \dodoi{10.1088/0004-637X/805/2/130}

\bibitem[{{Koposov} {et~al.}(2015{\natexlab{b}}){Koposov}, {Casey},
  {Belokurov}, {Lewis}, {Gilmore}, {Worley}, {Hourihane}, {Randich}, {Bensby},
  {Bragaglia}, {Bergemann}, {Carraro}, {Costado}, {Flaccomio}, {Francois},
  {Heiter}, {Hill}, {Jofre}, {Lando}, {Lanzafame}, {de Laverny}, {Monaco},
  {Morbidelli}, {Sbordone}, {Mikolaitis}, \& {Ryde}}]{koposov15b}
{Koposov}, S.~E., {Casey}, A.~R., {Belokurov}, V., {et~al.} 2015{\natexlab{b}},
  \apj, 811, 62, \dodoi{10.1088/0004-637X/811/1/62}

\bibitem[{{Koposov} {et~al.}(2018){Koposov}, {Walker}, {Belokurov}, {Casey},
  {Geringer-Sameth}, {Mackey}, {Da Costa}, {Erkal}, {Jethwa}, {Mateo},
  {Olszewski}, \& {Bailey}}]{koposov18}
{Koposov}, S.~E., {Walker}, M.~G., {Belokurov}, V., {et~al.} 2018, \mnras, 479,
  5343, \dodoi{10.1093/mnras/sty1772}

\bibitem[{{Laevens} {et~al.}(2015){Laevens}, {Martin}, {Ibata}, {Rix},
  {Bernard}, {Bell}, {Sesar}, {Ferguson}, {Schlafly}, {Slater}, {Burgett},
  {Chambers}, {Flewelling}, {Hodapp}, {Kaiser}, {Kudritzki}, {Lupton},
  {Magnier}, {Metcalfe}, {Morgan}, {Price}, {Tonry}, {Wainscoat}, \&
  {Waters}}]{laevens15a}
{Laevens}, B. P.~M., {Martin}, N.~F., {Ibata}, R.~A., {et~al.} 2015, \apjl,
  802, L18, \dodoi{10.1088/2041-8205/802/2/L18}

\bibitem[{{Li} {et~al.}(2018{\natexlab{a}}){Li}, {Simon}, {Kuehn}, {Pace},
  {Erkal}, {Bechtol}, {Yanny}, {Drlica-Wagner}, {Marshall}, {Lidman},
  {Balbinot}, {Carollo}, {Jenkins}, {Mart{\'\i}nez-V{\'a}zquez}, {Shipp},
  {Stringer}, {Vivas}, {Walker}, {Wechsler}, {Abdalla}, {Allam}, {Annis},
  {Avila}, {Bertin}, {Brooks}, {Buckley-Geer}, {Burke}, {Carnero Rosell},
  {Carrasco Kind}, {Carretero}, {Cunha}, {D'Andrea}, {da Costa}, {Davis}, {De
  Vicente}, {Doel}, {Eifler}, {Evrard}, {Flaugher}, {Frieman},
  {Garc{\'\i}a-Bellido}, {Gaztanaga}, {Gerdes}, {Gruen}, {Gruendl}, {Gschwend},
  {Gutierrez}, {Hartley}, {Hollowood}, {Honscheid}, {James}, {Krause}, {Maia},
  {March}, {Menanteau}, {Miquel}, {Plazas}, {Sanchez}, {Santiago}, {Scarpine},
  {Schindler}, {Schubnell}, {Sevilla-Noarbe}, {Smith}, {Smith},
  {Soares-Santos}, {Sobreira}, {Suchyta}, {Swanson}, {Tarle}, {Tucker}, \& {DES
  Collaboration}}]{li18b}
{Li}, T.~S., {Simon}, J.~D., {Kuehn}, K., {et~al.} 2018{\natexlab{a}}, \apj,
  866, 22, \dodoi{10.3847/1538-4357/aadf91}

\bibitem[{{Li} {et~al.}(2018{\natexlab{b}}){Li}, {Simon}, {Pace}, {Torrealba},
  {Kuehn}, {Drlica-Wagner}, {Bechtol}, {Vivas}, {van der Marel}, {Wood},
  {Yanny}, {Belokurov}, {Jethwa}, {Zucker}, {Lewis}, {Kron}, {Nidever},
  {S{\'a}nchez-Conde}, {Ji}, {Conn}, {James}, {Martin}, {Martinez-Delgado},
  {No{\"e}l}, \& {MagLiteS Collaboration}}]{li18a}
{Li}, T.~S., {Simon}, J.~D., {Pace}, A.~B., {et~al.} 2018{\natexlab{b}}, \apj,
  857, 145, \dodoi{10.3847/1538-4357/aab666}

\bibitem[{{Longeard} {et~al.}(2018){Longeard}, {Martin}, {Starkenburg},
  {Ibata}, {Collins}, {Geha}, {Laevens}, {Rich}, {Aguado}, {Arentsen},
  {Carlberg}, {C{\^o}t{\'e}}, {Hill}, {Jablonka}, {Gonz{\'a}lez Hern{\'a}ndez},
  {Navarro}, {S{\'a}nchez-Janssen}, {Tolstoy}, {Venn}, \&
  {Youakim}}]{longeard18}
{Longeard}, N., {Martin}, N., {Starkenburg}, E., {et~al.} 2018, \mnras, 480,
  2609, \dodoi{10.1093/mnras/sty1986}

\bibitem[{{Lynden-Bell}(1976)}]{lyndenbell76}
{Lynden-Bell}, D. 1976, \mnras, 174, 695, \dodoi{10.1093/mnras/174.3.695}

\bibitem[{{Martin} {et~al.}(2015){Martin}, {Nidever}, {Besla}, {Olsen},
  {Walker}, {Vivas}, {Gruendl}, {Kaleida}, {Mu{\~n}oz}, {Blum}, {Saha}, {Conn},
  {Bell}, {Chu}, {Cioni}, {de Boer}, {Gallart}, {Jin}, {Kunder}, {Majewski},
  {Martinez-Delgado}, {Monachesi}, {Monelli}, {Monteagudo}, {No{\"e}l},
  {Olszewski}, {Stringfellow}, {van der Marel}, \& {Zaritsky}}]{martin15}
{Martin}, N.~F., {Nidever}, D.~L., {Besla}, G., {et~al.} 2015, \apjl, 804, L5,
  \dodoi{10.1088/2041-8205/804/1/L5}

\bibitem[{{Martin} {et~al.}(2016){Martin}, {Ibata}, {Lewis}, {McConnachie},
  {Babul}, {Bate}, {Bernard}, {Chapman}, {Collins}, {Conn}, {Crnojevi{\'c}},
  {Fardal}, {Ferguson}, {Irwin}, {Mackey}, {McMonigal}, {Navarro}, \&
  {Rich}}]{martin16}
{Martin}, N.~F., {Ibata}, R.~A., {Lewis}, G.~F., {et~al.} 2016, \apj, 833, 167,
  \dodoi{10.3847/1538-4357/833/2/167}

\bibitem[{{Mart{\'\i}nez-V{\'a}zquez}
  {et~al.}(2016){Mart{\'\i}nez-V{\'a}zquez}, {Stetson}, {Monelli}, {Bernard},
  {Fiorentino}, {Gallart}, {Bono}, {Cassisi}, {Dall'Ora}, {Ferraro},
  {Iannicola}, \& {Walker}}]{martinezvazquez16}
{Mart{\'\i}nez-V{\'a}zquez}, C.~E., {Stetson}, P.~B., {Monelli}, M., {et~al.}
  2016, \mnras, 462, 4349, \dodoi{10.1093/mnras/stw1895}

\bibitem[{{Massari} \& {Helmi}(2018)}]{massari18}
{Massari}, D., \& {Helmi}, A. 2018, \aap, 620, A155,
  \dodoi{10.1051/0004-6361/201833367}

\bibitem[{{McMillan}(2011)}]{mcmillan11}
{McMillan}, P.~J. 2011, \mnras, 418, 1565,
  \dodoi{10.1111/j.1365-2966.2011.19520.x}

\bibitem[{{Miyamoto} \& {Nagai}(1975)}]{mn75}
{Miyamoto}, M., \& {Nagai}, R. 1975, \pasj, 27, 533

\bibitem[{{Mu{\~n}oz} {et~al.}(2018){Mu{\~n}oz}, {C{\^o}t{\'e}}, {Santana},
  {Geha}, {Simon}, {Oyarz{\'u}n}, {Stetson}, \& {Djorgovski}}]{munoz18}
{Mu{\~n}oz}, R.~R., {C{\^o}t{\'e}}, P., {Santana}, F.~A., {et~al.} 2018, \apj,
  860, 66, \dodoi{10.3847/1538-4357/aac16b}

\bibitem[{{Murai} \& {Fujimoto}(1980)}]{murai80}
{Murai}, T., \& {Fujimoto}, M. 1980, \pasj, 32, 581

\bibitem[{{Mutlu-Pakdil} {et~al.}(2018){Mutlu-Pakdil}, {Sand}, {Carlin},
  {Spekkens}, {Caldwell}, {Crnojevi{\'c}}, {Hughes}, {Willman}, \&
  {Zaritsky}}]{mutlupakdil18}
{Mutlu-Pakdil}, B., {Sand}, D.~J., {Carlin}, J.~L., {et~al.} 2018, \apj, 863,
  25, \dodoi{10.3847/1538-4357/aacd0e}

\bibitem[{{Nadler} {et~al.}(2019){Nadler}, {Wechsler}, {Bechtol}, {Mao},
  {Green}, {Drlica-Wagner}, {McNanna}, {Mau}, {Pace}, {Simon}, {Kravtsov},
  {Dodelson}, {Li}, {Riley}, {Wang}, {Abbott}, {Aguena}, {Allam}, {Annis},
  {Avila}, {Bernstein}, {Bertin}, {Brooks}, {Burke}, {Carnero Rosell},
  {Carrasco Kind}, {Carretero}, {Costanzi}, {da Costa}, {De Vicente}, {Desai},
  {Evrard}, {Flaugher}, {Fosalba}, {Frieman}, {Garc{\'\i}a-Bellido},
  {Gaztanaga}, {Gerdes}, {Gruen}, {Gschwend}, {Gutierrez}, {Hartley}, {Hinton},
  {Honscheid}, {Krause}, {Kuehn}, {Kuropatkin}, {Lahav}, {Maia}, {Marshall},
  {Menanteau}, {Miquel}, {Palmese}, {Paz-Chinch{\'o}n}, {Plazas}, {Sanchez},
  {Santiago}, {Scarpine}, {Serrano}, {Smith}, {Soares-Santos}, {Suchyta},
  {Tarle}, {Thomas}, {Varga}, \& {Walker}}]{nadler19}
{Nadler}, E.~O., {Wechsler}, R.~H., {Bechtol}, K., {et~al.} 2019, arXiv
  e-prints, arXiv:1912.03303.
\newblock \doarXiv{1912.03303}

\bibitem[{{Nagasawa} {et~al.}(2018){Nagasawa}, {Marshall}, {Li}, {Hansen},
  {Simon}, {Bernstein}, {Balbinot}, {Drlica-Wagner}, {Pace}, {Strigari},
  {Pellegrino}, {DePoy}, {Suntzeff}, {Bechtol}, {Walker}, {Abbott}, {Abdalla},
  {Allam}, {Annis}, {Benoit-L{\'e}vy}, {Bertin}, {Brooks}, {Carnero Rosell},
  {Carrasco Kind}, {Carretero}, {Cunha}, {D'Andrea}, {da Costa}, {Davis},
  {Desai}, {Doel}, {Eifler}, {Flaugher}, {Fosalba}, {Frieman},
  {Garc{\'\i}a-Bellido}, {Gaztanaga}, {Gerdes}, {Gruen}, {Gruendl}, {Gschwend},
  {Gutierrez}, {Hartley}, {Honscheid}, {James}, {Jeltema}, {Krause}, {Kuehn},
  {Kuhlmann}, {Kuropatkin}, {March}, {Miquel}, {Nord}, {Roodman}, {Sanchez},
  {Santiago}, {Scarpine}, {Schindler}, {Schubnell}, {Sevilla-Noarbe}, {Smith},
  {Smith}, {Soares-Santos}, {Sobreira}, {Suchyta}, {Tarle}, {Thomas}, {Tucker},
  {Wechsler}, {Wolf}, \& {Yanny}}]{nagasawa18}
{Nagasawa}, D.~Q., {Marshall}, J.~L., {Li}, T.~S., {et~al.} 2018, \apj, 852,
  99, \dodoi{10.3847/1538-4357/aaa01d}

\bibitem[{{Navarro} {et~al.}(1996){Navarro}, {Frenk}, \& {White}}]{nfw96}
{Navarro}, J.~F., {Frenk}, C.~S., \& {White}, S.~D.~M. 1996, \apj, 462, 563,
  \dodoi{10.1086/177173}

\bibitem[{{Pace} \& {Li}(2019)}]{pace19}
{Pace}, A.~B., \& {Li}, T.~S. 2019, \apj, 875, 77,
  \dodoi{10.3847/1538-4357/ab0aee}

\bibitem[{{Pardy} {et~al.}(2019){Pardy}, {D'Onghia}, {Navarro}, {Grand },
  {Gomez}, {Marinacci}, {Pakmor}, {Simpson}, \& {Springel}}]{pardy19}
{Pardy}, S.~A., {D'Onghia}, E., {Navarro}, J., {et~al.} 2019, arXiv e-prints,
  arXiv:1904.01028.
\newblock \doarXiv{1904.01028}

\bibitem[{{Patel} {et~al.}(2017){Patel}, {Besla}, \& {Sohn}}]{patel17a}
{Patel}, E., {Besla}, G., \& {Sohn}, S.~T. 2017, \mnras, 464, 3825,
  \dodoi{10.1093/mnras/stw2616}

\bibitem[{{Patel} {et~al.}(2018){Patel}, {Carlin}, {Tollerud}, {Collins}, \&
  {Dooley}}]{patel18b}
{Patel}, E., {Carlin}, J.~L., {Tollerud}, E.~J., {Collins}, M. L.~M., \&
  {Dooley}, G.~A. 2018, \mnras, 480, 1883, \dodoi{10.1093/mnras/sty1946}

\bibitem[{{Pawlowski} \& {Kroupa}(2019)}]{pawlowski19}
{Pawlowski}, M.~S., \& {Kroupa}, P. 2019, \mnras, 2774,
  \dodoi{10.1093/mnras/stz3163}

\bibitem[{{Pawlowski} {et~al.}(2012){Pawlowski}, {Pflamm-Altenburg}, \&
  {Kroupa}}]{pawlowski12}
{Pawlowski}, M.~S., {Pflamm-Altenburg}, J., \& {Kroupa}, P. 2012, \mnras, 423,
  1109, \dodoi{10.1111/j.1365-2966.2012.20937.x}

\bibitem[{{Piatek} {et~al.}(2005){Piatek}, {Pryor}, {Bristow}, {Olszewski},
  {Harris}, {Mateo}, {Minniti}, \& {Tinney}}]{piatek05}
{Piatek}, S., {Pryor}, C., {Bristow}, P., {et~al.} 2005, \aj, 130, 95,
  \dodoi{10.1086/430532}

\bibitem[{{Piatek} {et~al.}(2007){Piatek}, {Pryor}, {Bristow}, {Olszewski},
  {Harris}, {Mateo}, {Minniti}, \& {Tinney}}]{piatek07}
---. 2007, \aj, 133, 818, \dodoi{10.1086/510456}

\bibitem[{{Piatek} {et~al.}(2003){Piatek}, {Pryor}, {Olszewski}, {Harris},
  {Mateo}, {Minniti}, \& {Tinney}}]{piatek03}
{Piatek}, S., {Pryor}, C., {Olszewski}, E.~W., {et~al.} 2003, \aj, 126, 2346,
  \dodoi{10.1086/378713}

\bibitem[{{Pietrzy{\'n}ski} {et~al.}(2008){Pietrzy{\'n}ski}, {Gieren},
  {Szewczyk}, {Walker}, {Rizzi}, {Bresolin}, {Kudritzki}, {Nalewajko}, {Storm},
  {Dall'Ora}, \& {Ivanov}}]{pietrzynski08}
{Pietrzy{\'n}ski}, G., {Gieren}, W., {Szewczyk}, O., {et~al.} 2008, \aj, 135,
  1993, \dodoi{10.1088/0004-6256/135/6/1993}

\bibitem[{{Plummer}(1911)}]{plummer11}
{Plummer}, H.~C. 1911, \mnras, 71, 460, \dodoi{10.1093/mnras/71.5.460}

\bibitem[{{Rizzi} {et~al.}(2007){Rizzi}, {Held}, {Saviane}, {Tully}, \&
  {Gullieuszik}}]{rizzi07}
{Rizzi}, L., {Held}, E.~V., {Saviane}, I., {Tully}, R.~B., \& {Gullieuszik}, M.
  2007, \mnras, 380, 1255, \dodoi{10.1111/j.1365-2966.2007.12196.x}

\bibitem[{{Sales} {et~al.}(2011){Sales}, {Navarro}, {Cooper}, {White}, {Frenk},
  \& {Helmi}}]{sales11}
{Sales}, L.~V., {Navarro}, J.~F., {Cooper}, A.~P., {et~al.} 2011, \mnras, 418,
  648, \dodoi{10.1111/j.1365-2966.2011.19514.x}

\bibitem[{{Sales} {et~al.}(2017){Sales}, {Navarro}, {Kallivayalil}, \&
  {Frenk}}]{sales17}
{Sales}, L.~V., {Navarro}, J.~F., {Kallivayalil}, N., \& {Frenk}, C.~S. 2017,
  \mnras, 465, 1879, \dodoi{10.1093/mnras/stw2816}

\bibitem[{{Sales} {et~al.}(2013){Sales}, {Wang}, {White}, \&
  {Navarro}}]{sales13}
{Sales}, L.~V., {Wang}, W., {White}, S.~D.~M., \& {Navarro}, J.~F. 2013,
  \mnras, 428, 573, \dodoi{10.1093/mnras/sts054}

\bibitem[{{Sanders} {et~al.}(2018){Sanders}, {Evans}, \& {Dehnen}}]{sanders18}
{Sanders}, J.~L., {Evans}, N.~W., \& {Dehnen}, W. 2018, \mnras, 478, 3879,
  \dodoi{10.1093/mnras/sty1278}

\bibitem[{{Santistevan} {et~al.}(2020){Santistevan}, {Wetzel}, {El-Badry},
  {Bland-Hawthorn}, {Boylan-Kolchin}, {Bailin}, {Faucher-Giguere}, \&
  {Benincasa}}]{santistevan20}
{Santistevan}, I.~B., {Wetzel}, A., {El-Badry}, K., {et~al.} 2020, arXiv
  e-prints, arXiv:2001.03178.
\newblock \doarXiv{2001.03178}

\bibitem[{{Sch{\"o}nrich} {et~al.}(2010){Sch{\"o}nrich}, {Binney}, \&
  {Dehnen}}]{schonrich10}
{Sch{\"o}nrich}, R., {Binney}, J., \& {Dehnen}, W. 2010, \mnras, 403, 1829,
  \dodoi{10.1111/j.1365-2966.2010.16253.x}

\bibitem[{{Simon}(2018)}]{simon18}
{Simon}, J.~D. 2018, \apj, 863, 89, \dodoi{10.3847/1538-4357/aacdfb}

\bibitem[{{Simon} \& {Geha}(2007)}]{simon07}
{Simon}, J.~D., \& {Geha}, M. 2007, \apj, 670, 313, \dodoi{10.1086/521816}

\bibitem[{{Simon} {et~al.}(2011){Simon}, {Geha}, {Minor}, {Martinez}, {Kirby},
  {Bullock}, {Kaplinghat}, {Strigari}, {Willman}, {Choi}, {Tollerud}, \&
  {Wolf}}]{simon11}
{Simon}, J.~D., {Geha}, M., {Minor}, Q.~E., {et~al.} 2011, \apj, 733, 46,
  \dodoi{10.1088/0004-637X/733/1/46}

\bibitem[{{Simon} {et~al.}(2015){Simon}, {Drlica-Wagner}, {Li}, {Nord}, {Geha},
  {Bechtol}, {Balbinot}, {Buckley-Geer}, {Lin}, {Marshall}, {Santiago},
  {Strigari}, {Wang}, {Wechsler}, {Yanny}, {Abbott}, {Bauer}, {Bernstein},
  {Bertin}, {Brooks}, {Burke}, {Capozzi}, {Carnero Rosell}, {Carrasco Kind},
  {D'Andrea}, {da Costa}, {DePoy}, {Desai}, {Diehl}, {Dodelson}, {Cunha},
  {Estrada}, {Evrard}, {Fausti Neto}, {Fernandez}, {Finley}, {Flaugher},
  {Frieman}, {Gaztanaga}, {Gerdes}, {Gruen}, {Gruendl}, {Honscheid}, {James},
  {Kent}, {Kuehn}, {Kuropatkin}, {Lahav}, {Maia}, {March}, {Martini}, {Miller},
  {Miquel}, {Ogando}, {Romer}, {Roodman}, {Rykoff}, {Sako}, {Sanchez},
  {Schubnell}, {Sevilla}, {Smith}, {Soares-Santos}, {Sobreira}, {Suchyta},
  {Swanson}, {Tarle}, {Thaler}, {Tucker}, {Vikram}, {Walker}, {Wester}, \& {DES
  Collaboration}}]{simon15}
{Simon}, J.~D., {Drlica-Wagner}, A., {Li}, T.~S., {et~al.} 2015, \apj, 808, 95,
  \dodoi{10.1088/0004-637X/808/1/95}

\bibitem[{{Simon} {et~al.}(2017){Simon}, {Li}, {Drlica-Wagner}, {Bechtol},
  {Marshall}, {James}, {Wang}, {Strigari}, {Balbinot}, {Kuehn}, {Walker},
  {Abbott}, {Allam}, {Annis}, {Benoit-L{\'e}vy}, {Brooks}, {Buckley-Geer},
  {Burke}, {Carnero Rosell}, {Carrasco Kind}, {Carretero}, {Cunha}, {D'Andrea},
  {da Costa}, {DePoy}, {Desai}, {Doel}, {Fernandez}, {Flaugher}, {Frieman},
  {Garc{\'\i}a-Bellido}, {Gaztanaga}, {Goldstein}, {Gruen}, {Gutierrez},
  {Kuropatkin}, {Maia}, {Martini}, {Menanteau}, {Miller}, {Miquel}, {Neilsen},
  {Nord}, {Ogando}, {Plazas}, {Romer}, {Rykoff}, {Sanchez}, {Santiago},
  {Scarpine}, {Schubnell}, {Sevilla-Noarbe}, {Smith}, {Sobreira}, {Suchyta},
  {Swanson}, {Tarle}, {Whiteway}, {Yanny}, \& {DES Collaboration}}]{simon17}
{Simon}, J.~D., {Li}, T.~S., {Drlica-Wagner}, A., {et~al.} 2017, \apj, 838, 11,
  \dodoi{10.3847/1538-4357/aa5be7}

\bibitem[{{Sohn} {et~al.}(2017){Sohn}, {Patel}, {Besla}, {van der Marel},
  {Bullock}, {Strigari}, {van de Ven}, {Walker}, \& {Bellini}}]{sohn17}
{Sohn}, S.~T., {Patel}, E., {Besla}, G., {et~al.} 2017, \apj, 849, 93,
  \dodoi{10.3847/1538-4357/aa917b}

\bibitem[{{Springel} {et~al.}(2001){Springel}, {Yoshida}, \& {White}}]{gadget}
{Springel}, V., {Yoshida}, N., \& {White}, S.~D.~M. 2001, \na, 6, 79,
  \dodoi{10.1016/S1384-1076(01)00042-2}

\bibitem[{{The Astropy Collaboration} {et~al.}(2018){The Astropy
  Collaboration}, {Price-Whelan}, {Sip{\H o}cz}, {G{\"u}nther}, {Lim},
  {Crawford}, {Conseil}, {Shupe}, {Craig}, {Dencheva}, {Ginsburg},
  {VanderPlas}, {Bradley}, {P{\'e}rez-Su{\'a}rez}, {de Val-Borro}, {Aldcroft},
  {Cruz}, {Robitaille}, {Tollerud}, {Ardelean}, {Babej}, {Bachetti}, {Bakanov},
  {Bamford}, {Barentsen}, {Barmby}, {Baumbach}, {Berry}, {Biscani}, {Boquien},
  {Bostroem}, {Bouma}, {Brammer}, {Bray}, {Breytenbach}, {Buddelmeijer},
  {Burke}, {Calderone}, {Cano Rodr{\'{\i}}guez}, {Cara}, {Cardoso},
  {Cheedella}, {Copin}, {Crichton}, {D{\'A}vella}, {Deil}, {Depagne},
  {Dietrich}, {Donath}, {Droettboom}, {Earl}, {Erben}, {Fabbro}, {Ferreira},
  {Finethy}, {Fox}, {Garrison}, {Gibbons}, {Goldstein}, {Gommers}, {Greco},
  {Greenfield}, {Groener}, {Grollier}, {Hagen}, {Hirst}, {Homeier}, {Horton},
  {Hosseinzadeh}, {Hu}, {Hunkeler}, {Ivezi{\'c}}, {Jain}, {Jenness}, {Kanarek},
  {Kendrew}, {Kern}, {Kerzendorf}, {Khvalko}, {King}, {Kirkby}, {Kulkarni},
  {Kumar}, {Lee}, {Lenz}, {Littlefair}, {Ma}, {Macleod}, {Mastropietro},
  {McCully}, {Montagnac}, {Morris}, {Mueller}, {Mumford}, {Muna}, {Murphy},
  {Nelson}, {Nguyen}, {Ninan}, {N{\"o}the}, {Ogaz}, {Oh}, {Parejko}, {Parley},
  {Pascual}, {Patil}, {Patil}, {Plunkett}, {Prochaska}, {Rastogi}, {Reddy
  Janga}, {Sabater}, {Sakurikar}, {Seifert}, {Sherbert}, {Sherwood-Taylor},
  {Shih}, {Sick}, {Silbiger}, {Singanamalla}, {Singer}, {Sladen}, {Sooley},
  {Sornarajah}, {Streicher}, {Teuben}, {Thomas}, {Tremblay}, {Turner},
  {Terr{\'o}n}, {van Kerkwijk}, {de la Vega}, {Watkins}, {Weaver}, {Whitmore},
  {Woillez}, \& {Zabalza}}]{astropy}
{The Astropy Collaboration}, {Price-Whelan}, A.~M., {Sip{\H o}cz}, B.~M.,
  {et~al.} 2018, ArXiv e-prints.
\newblock \doarXiv{1801.02634}

\bibitem[{{Torrealba} {et~al.}(2016{\natexlab{a}}){Torrealba}, {Koposov},
  {Belokurov}, \& {Irwin}}]{torrealba16a}
{Torrealba}, G., {Koposov}, S.~E., {Belokurov}, V., \& {Irwin}, M.
  2016{\natexlab{a}}, \mnras, 459, 2370, \dodoi{10.1093/mnras/stw733}

\bibitem[{{Torrealba} {et~al.}(2016{\natexlab{b}}){Torrealba}, {Koposov},
  {Belokurov}, {Irwin}, {Collins}, {Spencer}, {Ibata}, {Mateo}, {Bonaca}, \&
  {Jethwa}}]{torrealba16b}
{Torrealba}, G., {Koposov}, S.~E., {Belokurov}, V., {et~al.}
  2016{\natexlab{b}}, \mnras, 463, 712, \dodoi{10.1093/mnras/stw2051}

\bibitem[{{Torrealba} {et~al.}(2018){Torrealba}, {Belokurov}, {Koposov},
  {Bechtol}, {Drlica-Wagner}, {Olsen}, {Vivas}, {Yanny}, {Jethwa}, {Walker},
  {Li}, {Allam}, {Conn}, {Gallart}, {Gruendl}, {James}, {Johnson}, {Kuehn},
  {Kuropatkin}, {Martin}, {Martinez-Delgado}, {Nidever}, {No{\"e}l}, {Simon},
  {Stringfellow}, \& {Tucker}}]{torrealba18}
{Torrealba}, G., {Belokurov}, V., {Koposov}, S.~E., {et~al.} 2018, \mnras, 475,
  5085, \dodoi{10.1093/mnras/sty170}

\bibitem[{{van der Marel} {et~al.}(2002){van der Marel}, {Alves}, {Hardy}, \&
  {Suntzeff}}]{vdm02}
{van der Marel}, R.~P., {Alves}, D.~R., {Hardy}, E., \& {Suntzeff}, N.~B. 2002,
  \aj, 124, 2639, \dodoi{10.1086/343775}

\bibitem[{{van der Marel} {et~al.}(2012b){van der Marel}, {Besla}, {Cox},
  {Sohn}, \& {Anderson}}]{vdm12iii}
{van der Marel}, R.~P., {Besla}, G., {Cox}, T.~J., {Sohn}, S.~T., \&
  {Anderson}, J. 2012b, \apj, 753, 9, \dodoi{10.1088/0004-637X/753/1/9}

\bibitem[{{van der Marel} {et~al.}(2012a){van der Marel}, {Fardal}, {Besla},
  {Beaton}, {Sohn}, {Anderson}, {Brown}, \& {Guhathakurta}}]{vdm12ii}
{van der Marel}, R.~P., {Fardal}, M., {Besla}, G., {et~al.} 2012a, \apj, 753,
  8, \dodoi{10.1088/0004-637X/753/1/8}

\bibitem[{{van der Marel} \& {Kallivayalil}(2014)}]{vdmnk14}
{van der Marel}, R.~P., \& {Kallivayalil}, N. 2014, \apj, 781, 121,
  \dodoi{10.1088/0004-637X/781/2/121}

\bibitem[{van~der Walt {et~al.}(2011)van~der Walt, Colbert, \&
  Varoquaux}]{numpy}
van~der Walt, S., Colbert, S.~C., \& Varoquaux, G. 2011, Computing in Science
  Engineering, 13, 22, \dodoi{10.1109/MCSE.2011.37}

\bibitem[{{Vivas} \& {Mateo}(2013)}]{vivas13}
{Vivas}, A.~K., \& {Mateo}, M. 2013, \aj, 146, 141,
  \dodoi{10.1088/0004-6256/146/6/141}

\bibitem[{{Vivas} {et~al.}(2016){Vivas}, {Olsen}, {Blum}, {Nidever}, {Walker},
  {Martin}, {Besla}, {Gallart}, {van der Marel}, {Majewski}, {Kaleida},
  {Mu{\~n}oz}, {Saha}, {Conn}, \& {Jin}}]{vivas16}
{Vivas}, A.~K., {Olsen}, K., {Blum}, R., {et~al.} 2016, \aj, 151, 118,
  \dodoi{10.3847/0004-6256/151/5/118}

\bibitem[{{Walker} {et~al.}(2008){Walker}, {Mateo}, \& {Olszewski}}]{walker08}
{Walker}, M.~G., {Mateo}, M., \& {Olszewski}, E.~W. 2008, \apjl, 688, L75,
  \dodoi{10.1086/595586}

\bibitem[{{Walker} {et~al.}(2009{\natexlab{a}}){Walker}, {Mateo}, \&
  {Olszewski}}]{walker09b}
---. 2009{\natexlab{a}}, \aj, 137, 3100, \dodoi{10.1088/0004-6256/137/2/3100}

\bibitem[{{Walker} {et~al.}(2015){Walker}, {Mateo}, {Olszewski}, {Bailey},
  {Koposov}, {Belokurov}, \& {Evans}}]{walker15}
{Walker}, M.~G., {Mateo}, M., {Olszewski}, E.~W., {et~al.} 2015, \apj, 808,
  108, \dodoi{10.1088/0004-637X/808/2/108}

\bibitem[{{Walker} {et~al.}(2009{\natexlab{b}}){Walker}, {Mateo}, {Olszewski},
  {Pe{\~n}arrubia}, {Evans}, \& {Gilmore}}]{walker09c}
---. 2009{\natexlab{b}}, \apj, 704, 1274, \dodoi{10.1088/0004-637X/704/2/1274}

\bibitem[{{Weisz} {et~al.}(2014){Weisz}, {Dolphin}, {Skillman}, {Holtzman},
  {Gilbert}, {Dalcanton}, \& {Williams}}]{weisz14}
{Weisz}, D.~R., {Dolphin}, A.~E., {Skillman}, E.~D., {et~al.} 2014, \apj, 789,
  147, \dodoi{10.1088/0004-637X/789/2/147}

\bibitem[{{Zentner} \& {Bullock}(2003)}]{zentner03}
{Zentner}, A.~R., \& {Bullock}, J.~S. 2003, \apj, 598, 49,
  \dodoi{10.1086/378797}

\bibitem[{{Zhao}(1998)}]{zhao98}
{Zhao}, H. 1998, \apjl, 500, L149, \dodoi{10.1086/311413}

\bibitem[{{Zivick} {et~al.}(2018){Zivick}, {Kallivayalil}, {van der Marel},
  {Besla}, {Linden}, {Koz{\l}owski}, {Fritz}, {Kochanek}, {Anderson}, {Sohn},
  {Geha}, \& {Alcock}}]{zivick18}
{Zivick}, P., {Kallivayalil}, N., {van der Marel}, R.~P., {et~al.} 2018, \apj,
  864, 55, \dodoi{10.3847/1538-4357/aad4b0}

\end{thebibliography}
\bibliographystyle{aasjournal}




\appendix

\section{Results of Orbital Parameters for LMC1}
\label{sec:appendixA}

\begin{figure}[ht!]
    \centering
    \includegraphics[scale=0.75]{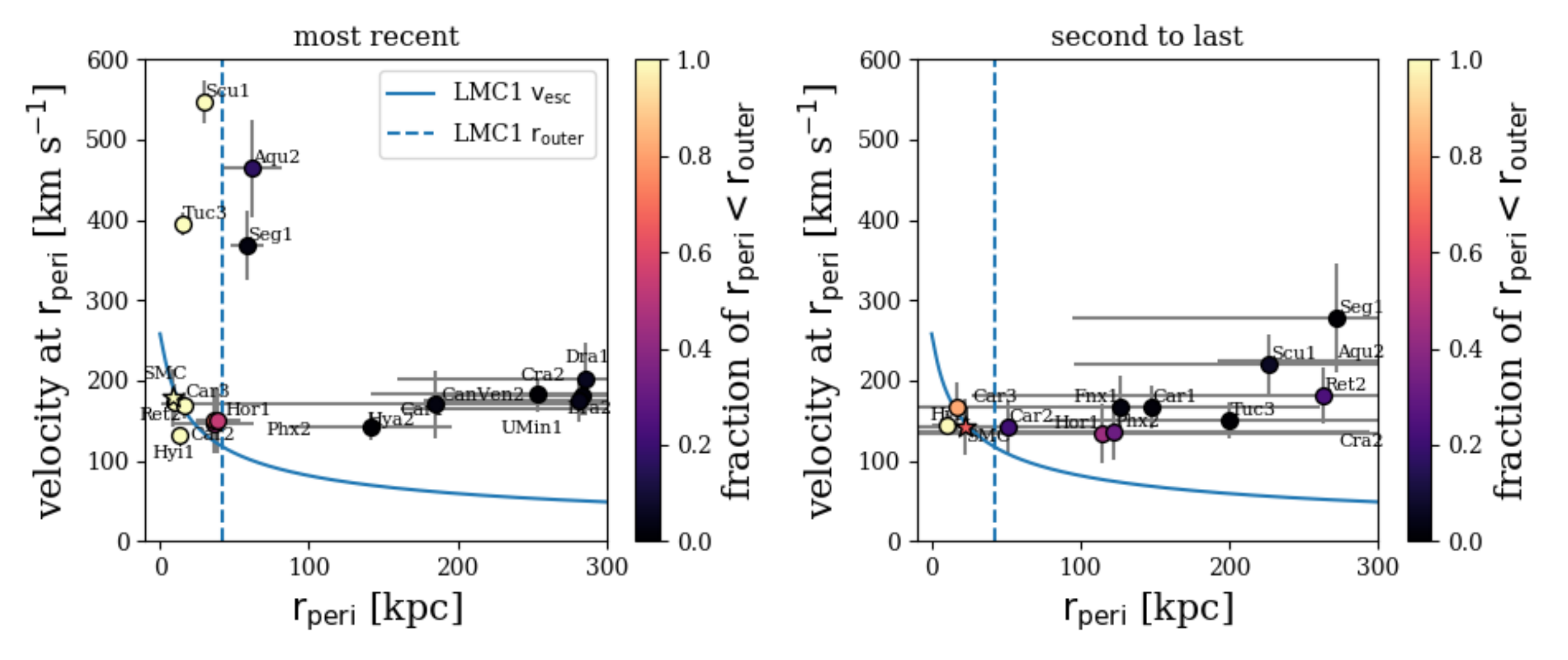}
    \caption{Same as Fig. \ref{fig:orbparams} except the orbital properties are calculated relative to LMC1 in MW1.}
\end{figure}

\begin{figure}[ht!]
    \centering
    \includegraphics[scale=0.75]{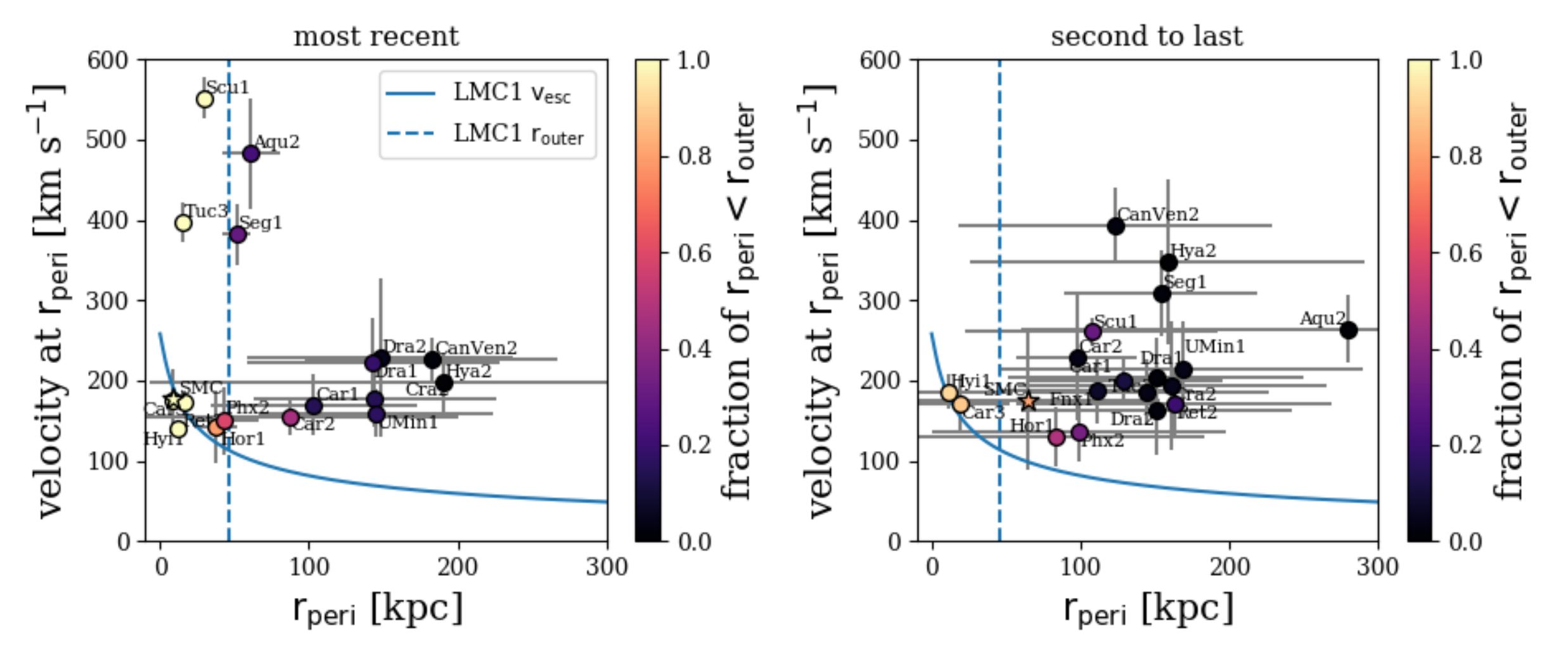}
    \caption{Same as Fig. \ref{fig:orbparams} except the orbital properties are calculated relative to LMC1 in MW2.}
\end{figure}

\begin{splitdeluxetable*}{lcccccccBlccccccc}
\tabletypesize{\footnotesize}
\label{tab:orbparams_MW1_LMC1}
\tablecaption{Orbital properties with respect to the LMC in MW1 using the LMC1 model. }
\tablehead{\colhead{Name} & \colhead{$\rm f_{peri,1}$ } & \colhead{$\rm f_{r_{outer},1}$} & \colhead{$\rm r_{peri,1}$ [kpc]} & \colhead{$\rm t_{peri,1}$ [Gyr]} & \colhead{$\rm f_{apo,1}$} & \colhead{$\rm r_{apo,1}$ [kpc]} & \colhead{$\rm t_{apo,1}$ [Gyr]} & \colhead{Name} & \colhead{$\rm f_{peri,2}$} &  \colhead{$\rm f_{r_{outer},2}$} & \colhead{$\rm r_{peri,2}$ [kpc]} & \colhead{$\rm t_{peri,2}$ [Gyr]} & \colhead{$\rm f_{apo,2}$} & \colhead{$\rm r_{apo,2}$ [kpc]} & \colhead{$\rm t_{apo,2}$ [Gyr]}\\ \hline
\multicolumn{8}{c}{most recent} & \multicolumn{8}{c}{second to last}}
\startdata
Aqu2 & 1.0 & 0.17 & 61.35$\pm$19.88 & 0.16$\pm$0.06 & 0.21 & 524.34$\pm$183.63 & 3.29$\pm$1.2 & Aqu2 & 1.0 &  0.0 & 384.45$\pm$192.93 & 4.22$\pm$1.15 & 0.07 & 426.56$\pm$167.58 & 4.83$\pm$0.82 \\ 
CanVen2 & 0.22 & 0.0 & 254.13$\pm$112.56 & 3.46$\pm$1.12 & 0.27 & 346.95$\pm$116.69 & 1.84$\pm$1.21 & CanVen2 & 0.22 &  0.0 & 214.71$\pm$160.25 & 4.76$\pm$0.64 & 0.05 & 369.94$\pm$124.55 & 4.81$\pm$0.83 \\ 
Car2 & 0.7 & 0.46 & 35.76$\pm$26.65 & 1.61$\pm$0.45 & 0.72 & 82.93$\pm$76.63 & 0.8$\pm$0.83 & Car2 & 0.7 &  0.21 & 51.12$\pm$58.67 & 4.63$\pm$0.88 & 0.54 & 141.03$\pm$107.07 & 3.92$\pm$1.04 \\ 
Car3 & 1.0 & 1.0 & 9.21$\pm$3.35 & 0.19$\pm$0.06 & 0.92 & 79.16$\pm$75.02 & 1.5$\pm$0.87 & Car3 & 1.0 &  0.82 & 16.4$\pm$30.76 & 2.53$\pm$1.13 & 0.69 & 66.18$\pm$32.89 & 3.34$\pm$1.1 \\ 
Cra2 & 1.0 & 0.02 & 283.37$\pm$96.07 & 2.38$\pm$0.68 & 1.0 & 358.54$\pm$72.79 & 1.51$\pm$0.38 & Cra2 & 1.0 &  0.0 & 497.01$\pm$188.07 & 5.05$\pm$0.63 & 0.92 & 485.06$\pm$177.36 & 4.48$\pm$0.6 \\ 
Dra2 & 0.43 & 0.0 & 343.9$\pm$181.82 & 4.05$\pm$1.21 & 0.69 & 549.2$\pm$165.28 & 3.07$\pm$1.16 & Dra2 & 0.43 &  0.0 & 421.15$\pm$178.1 & 5.06$\pm$0.78 & 0.1 & 436.93$\pm$139.38 & 4.78$\pm$0.8 \\ 
Hor1 & 0.96 & 0.68 & 35.74$\pm$11.27 & 0.2$\pm$0.31 & 0.84 & 211.45$\pm$228.29 & 2.01$\pm$1.66 & Hor1 & 0.96 &  0.44 & 113.94$\pm$185.17 & 2.49$\pm$1.61 & 0.51 & 101.38$\pm$140.55 & 3.07$\pm$1.21 \\ 
Hyi1 & 1.0 & 1.0 & 12.98$\pm$2.75 & 0.29$\pm$0.05 & 1.0 & 34.66$\pm$10.9 & 0.99$\pm$0.27 & Hyi1 & 1.0 &  0.99 & 10.2$\pm$9.79 & 1.64$\pm$0.36 & 0.99 & 35.57$\pm$17.28 & 2.28$\pm$0.53 \\ 
Hya2 & 0.25 & 0.0 & 141.1$\pm$54.5 & 0.81$\pm$1.03 & 0.13 & 207.28$\pm$95.15 & 0.84$\pm$1.01 & Hya2 & 0.25 &  0.0 & 182.77$\pm$116.04 & 4.94$\pm$1.09 & 0.04 & 422.79$\pm$178.87 & 4.32$\pm$0.98 \\ 
Phx2 & 0.96 & 0.53 & 38.7$\pm$14.94 & 0.35$\pm$0.3 & 0.75 & 235.18$\pm$217.07 & 2.56$\pm$1.42 & Phx2 & 0.96 &  0.32 & 122.39$\pm$172.24 & 3.35$\pm$1.37 & 0.37 & 146.0$\pm$184.07 & 4.04$\pm$1.06 \\ 
Ret2 & 1.0 & 1.0 & 16.23$\pm$2.88 & 0.12$\pm$0.02 & 0.86 & 339.18$\pm$232.4 & 2.8$\pm$1.41 & Ret2 & 1.0 &  0.24 & 262.55$\pm$235.88 & 3.31$\pm$1.36 & 0.54 & 325.57$\pm$259.65 & 4.01$\pm$1.14 \\ 
Seg1 & 0.99 & 0.03 & 58.43$\pm$11.07 & 0.31$\pm$0.1 & 0.99 & 70.39$\pm$7.3 & 0.12$\pm$0.05 & Seg1 & 0.99 &  0.0 & 271.96$\pm$177.62 & 2.01$\pm$1.25 & 0.97 & 300.59$\pm$189.41 & 1.71$\pm$1.23 \\ 
Tuc3 & 1.0 & 1.0 & 14.87$\pm$3.22 & 0.08$\pm$0.0 & 0.8 & 222.0$\pm$142.07 & 1.2$\pm$1.0 & Tuc3 & 1.0 &  0.01 & 199.72$\pm$121.01 & 1.33$\pm$1.12 & 0.61 & 325.56$\pm$139.88 & 2.15$\pm$1.18 \\ 
Car1 & 0.58 & 0.03 & 184.45$\pm$183.38 & 2.18$\pm$1.59 & 0.88 & 356.52$\pm$266.38 & 2.46$\pm$1.9 & Car1 & 0.58 &  0.02 & 147.17$\pm$61.4 & 5.01$\pm$0.65 & 0.43 & 307.5$\pm$125.57 & 4.25$\pm$0.84 \\ 
Dra1 & 0.95 & 0.06 & 285.94$\pm$126.55 & 3.7$\pm$0.92 & 1.0 & 393.89$\pm$79.58 & 2.11$\pm$0.38 & Dra1 & 0.95 &  0.0 & 272.23$\pm$297.87 & 5.92$\pm$0.03 & 0.27 & 466.2$\pm$195.48 & 5.54$\pm$0.34 \\ 
Fnx1 & 1.0 & 0.0 & 100.4$\pm$3.92 & 0.13$\pm$0.03 & 0.78 & 432.72$\pm$282.78 & 3.26$\pm$2.07 & Fnx1 & 1.0 &  0.03 & 127.04$\pm$133.85 & 2.01$\pm$1.07 & 0.27 & 297.93$\pm$102.75 & 4.47$\pm$0.8 \\ 
Scu1 & 1.0 & 0.99 & 29.13$\pm$4.9 & 0.11$\pm$0.01 & 0.76 & 351.74$\pm$79.49 & 2.1$\pm$0.66 & Scu1 & 1.0 &  0.07 & 226.53$\pm$131.15 & 4.14$\pm$0.78 & 0.19 & 286.11$\pm$118.52 & 5.34$\pm$0.49 \\ 
UMin1 & 0.97 & 0.04 & 281.14$\pm$103.66 & 3.18$\pm$0.89 & 0.99 & 361.69$\pm$58.79 & 1.86$\pm$0.29 & UMin1 & 0.97 &  0.0 & 470.22$\pm$293.96 & 5.74$\pm$0.27 & 0.58 & 515.69$\pm$186.78 & 5.36$\pm$0.37 \\ 
\enddata
\end{splitdeluxetable*}

\begin{splitdeluxetable*}{lcccccccBlccccccc}
\tabletypesize{\footnotesize}
\label{tab:orbparams_MW2_LMC1}
\tablecaption{Orbital properties with respect to the LMC in MW2 using the LMC1 model. }
\tablehead{\colhead{Name} & \colhead{$\rm f_{peri,1}$ } & \colhead{$\rm f_{r_{outer},1}$} & \colhead{$\rm r_{peri,1}$ [kpc]} & \colhead{$\rm t_{peri,1}$ [Gyr]} & \colhead{$\rm f_{apo,1}$} & \colhead{$\rm r_{apo,1}$ [kpc]} & \colhead{$\rm t_{apo,1}$ [Gyr]} & \colhead{Name} & \colhead{$\rm f_{peri,2}$} &  \colhead{$\rm f_{r_{outer},2}$} & \colhead{$\rm r_{peri,2}$ [kpc]} & \colhead{$\rm t_{peri,2}$ [Gyr]} & \colhead{$\rm f_{apo,2}$} & \colhead{$\rm r_{apo,2}$ [kpc]} & \colhead{$\rm t_{apo,2}$ [Gyr]}\\ \hline
\multicolumn{8}{c}{most recent} & \multicolumn{8}{c}{second to last}}
\startdata
Aqu2 & 1.0 & 0.22 & 61.23$\pm$19.75 & 0.16$\pm$0.06 & 0.37 & 438.95$\pm$209.94 & 2.26$\pm$0.96 & Aqu2 & 0.34 &  0.01 & 279.96$\pm$220.34 & 3.56$\pm$0.98 & 0.2 & 300.61$\pm$131.26 & 4.14$\pm$0.93 \\ 
CanVen2 & 0.46 & 0.02 & 182.51$\pm$84.93 & 2.43$\pm$1.05 & 0.5 & 298.37$\pm$95.93 & 1.1$\pm$1.11 & CanVen2 & 0.19 &  0.03 & 123.0$\pm$105.32 & 4.72$\pm$0.87 & 0.32 & 364.42$\pm$167.03 & 4.07$\pm$0.87 \\ 
Car2 & 0.96 & 0.46 & 87.01$\pm$113.4 & 2.1$\pm$1.77 & 0.99 & 156.82$\pm$160.13 & 1.31$\pm$1.39 & Car2 & 0.69 &  0.04 & 97.3$\pm$40.28 & 4.24$\pm$0.79 & 0.8 & 231.47$\pm$81.47 & 3.31$\pm$0.96 \\ 
Car3 & 1.0 & 1.0 & 8.48$\pm$3.37 & 0.19$\pm$0.05 & 0.99 & 70.29$\pm$69.51 & 1.19$\pm$0.64 & Car3 & 0.98 &  0.88 & 18.55$\pm$33.67 & 2.08$\pm$0.86 & 0.86 & 75.49$\pm$52.52 & 3.03$\pm$0.98 \\ 
Cra2 & 1.0 & 0.15 & 143.9$\pm$81.43 & 2.31$\pm$0.66 & 1.0 & 274.55$\pm$29.7 & 0.98$\pm$0.13 & Cra2 & 0.82 &  0.05 & 161.3$\pm$103.94 & 4.55$\pm$0.85 & 0.92 & 250.7$\pm$113.17 & 3.57$\pm$0.73 \\ 
Dra2 & 0.97 & 0.05 & 147.78$\pm$89.45 & 3.08$\pm$1.03 & 1.0 & 331.39$\pm$107.22 & 1.5$\pm$0.56 & Dra2 & 0.54 &  0.02 & 150.79$\pm$90.67 & 4.66$\pm$0.8 & 0.76 & 268.73$\pm$113.84 & 4.11$\pm$0.98 \\ 
Hor1 & 0.96 & 0.79 & 37.61$\pm$14.73 & 0.15$\pm$0.29 & 0.94 & 169.88$\pm$156.07 & 1.66$\pm$1.47 & Hor1 & 0.83 &  0.48 & 83.9$\pm$99.42 & 2.31$\pm$1.56 & 0.71 & 134.67$\pm$121.64 & 3.2$\pm$1.45 \\ 
Hyi1 & 1.0 & 1.0 & 12.1$\pm$2.88 & 0.29$\pm$0.04 & 0.99 & 55.09$\pm$78.57 & 1.21$\pm$0.84 & Hyi1 & 0.95 &  0.92 & 11.04$\pm$43.81 & 1.74$\pm$0.64 & 0.93 & 52.09$\pm$40.77 & 2.49$\pm$0.64 \\ 
Hya2 & 0.32 & 0.01 & 190.36$\pm$197.45 & 1.22$\pm$1.49 & 0.22 & 311.38$\pm$237.37 & 1.48$\pm$1.66 & Hya2 & 0.08 &  0.01 & 158.48$\pm$132.38 & 4.58$\pm$0.79 & 0.12 & 346.04$\pm$135.04 & 3.63$\pm$0.98 \\ 
Phx2 & 0.96 & 0.59 & 42.57$\pm$23.21 & 0.32$\pm$0.46 & 0.92 & 207.1$\pm$155.49 & 2.15$\pm$1.24 & Phx2 & 0.8 &  0.35 & 99.07$\pm$98.97 & 3.05$\pm$1.3 & 0.57 & 167.55$\pm$127.39 & 3.94$\pm$1.17 \\ 
Ret2 & 1.0 & 1.0 & 15.99$\pm$2.98 & 0.12$\pm$0.02 & 1.0 & 247.06$\pm$98.47 & 1.89$\pm$0.7 & Ret2 & 0.99 &  0.2 & 162.81$\pm$105.67 & 2.81$\pm$1.01 & 0.81 & 270.79$\pm$127.19 & 3.52$\pm$1.04 \\ 
Seg1 & 1.0 & 0.3 & 51.31$\pm$8.97 & 0.29$\pm$0.08 & 1.0 & 69.63$\pm$33.66 & 0.1$\pm$0.18 & Seg1 & 1.0 &  0.02 & 153.9$\pm$64.95 & 1.35$\pm$0.6 & 1.0 & 184.04$\pm$59.57 & 0.97$\pm$0.4 \\ 
Tuc3 & 1.0 & 1.0 & 15.23$\pm$3.24 & 0.07$\pm$0.01 & 0.92 & 178.01$\pm$117.92 & 0.99$\pm$0.83 & Tuc3 & 0.9 &  0.04 & 144.94$\pm$81.32 & 1.23$\pm$0.97 & 0.84 & 237.69$\pm$111.76 & 1.98$\pm$1.22 \\ 
Car1 & 0.95 & 0.15 & 103.28$\pm$69.42 & 1.6$\pm$1.14 & 1.0 & 164.79$\pm$121.2 & 0.92$\pm$0.9 & Car1 & 0.8 &  0.11 & 129.35$\pm$65.79 & 4.02$\pm$0.85 & 0.89 & 213.08$\pm$101.54 & 2.95$\pm$0.85 \\ 
Dra1 & 0.98 & 0.18 & 142.83$\pm$84.75 & 3.3$\pm$0.79 & 1.0 & 265.79$\pm$43.83 & 1.5$\pm$0.38 & Dra1 & 0.39 &  0.02 & 150.86$\pm$99.2 & 5.23$\pm$0.52 & 0.73 & 295.67$\pm$160.63 & 4.4$\pm$0.59 \\ 
Fnx1 & 1.0 & 0.0 & 100.91$\pm$3.83 & 0.12$\pm$0.02 & 1.0 & 232.67$\pm$174.02 & 1.57$\pm$1.34 & Fnx1 & 0.88 &  0.08 & 110.86$\pm$64.79 & 2.22$\pm$1.54 & 0.7 & 234.46$\pm$87.11 & 3.1$\pm$0.7 \\ 
Scu1 & 1.0 & 1.0 & 29.41$\pm$4.98 & 0.11$\pm$0.01 & 1.0 & 245.7$\pm$66.0 & 1.14$\pm$0.37 & Scu1 & 0.99 &  0.31 & 107.34$\pm$85.09 & 2.54$\pm$0.43 & 0.99 & 282.18$\pm$144.25 & 3.83$\pm$0.64 \\ 
UMin1 & 1.0 & 0.16 & 145.05$\pm$78.86 & 2.93$\pm$0.77 & 1.0 & 250.34$\pm$34.49 & 1.31$\pm$0.25 & UMin1 & 0.64 &  0.05 & 168.6$\pm$121.31 & 5.08$\pm$0.52 & 0.85 & 276.28$\pm$153.26 & 4.08$\pm$0.54 \\ 
\enddata
\end{splitdeluxetable*}
\clearpage

\section{Results of Orbital Parameters for LMC3}
\label{sec:appendixB}

\begin{figure}[ht!]
    \centering
    \includegraphics[scale=0.75]{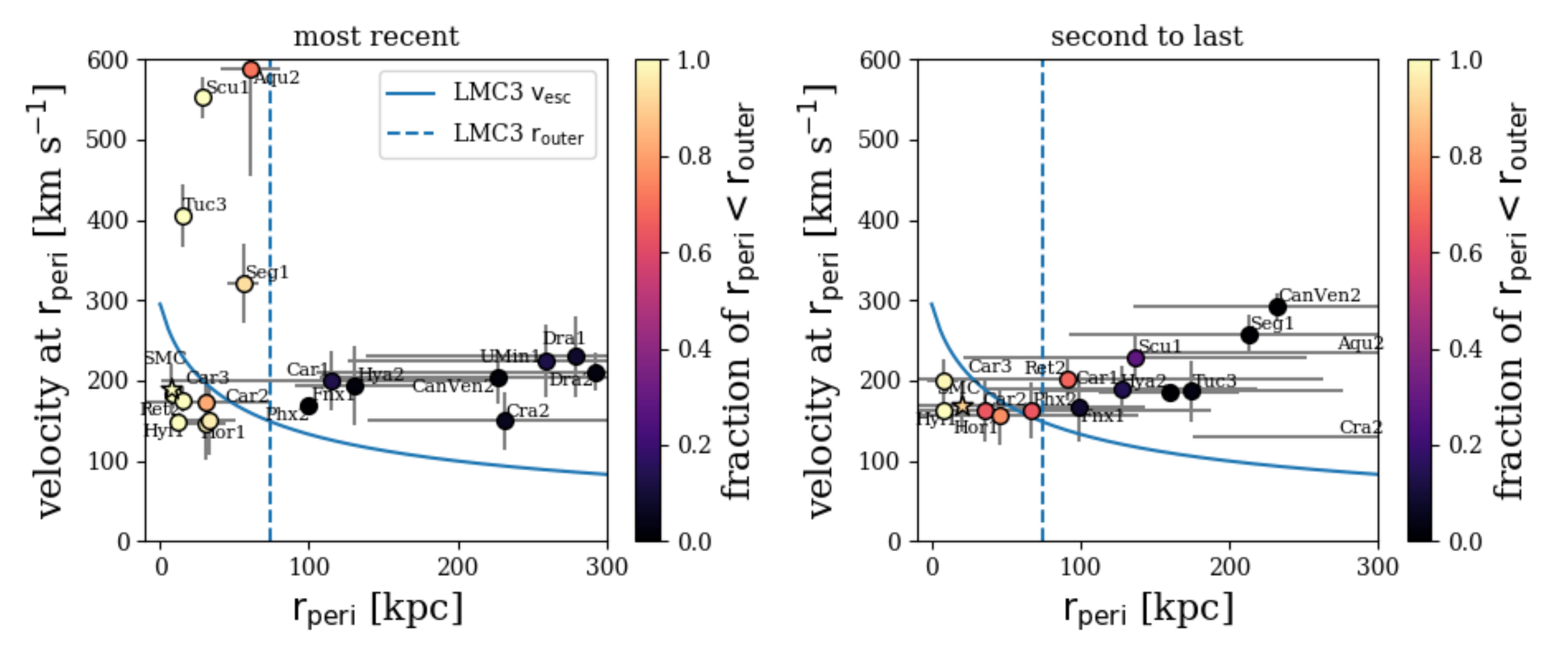}
    \caption{Same as Fig. \ref{fig:orbparams} except the orbital properties are calculated relative to LMC3 in MW1.}
\end{figure}

\begin{figure}[ht!]
    \centering
    \includegraphics[scale=0.75]{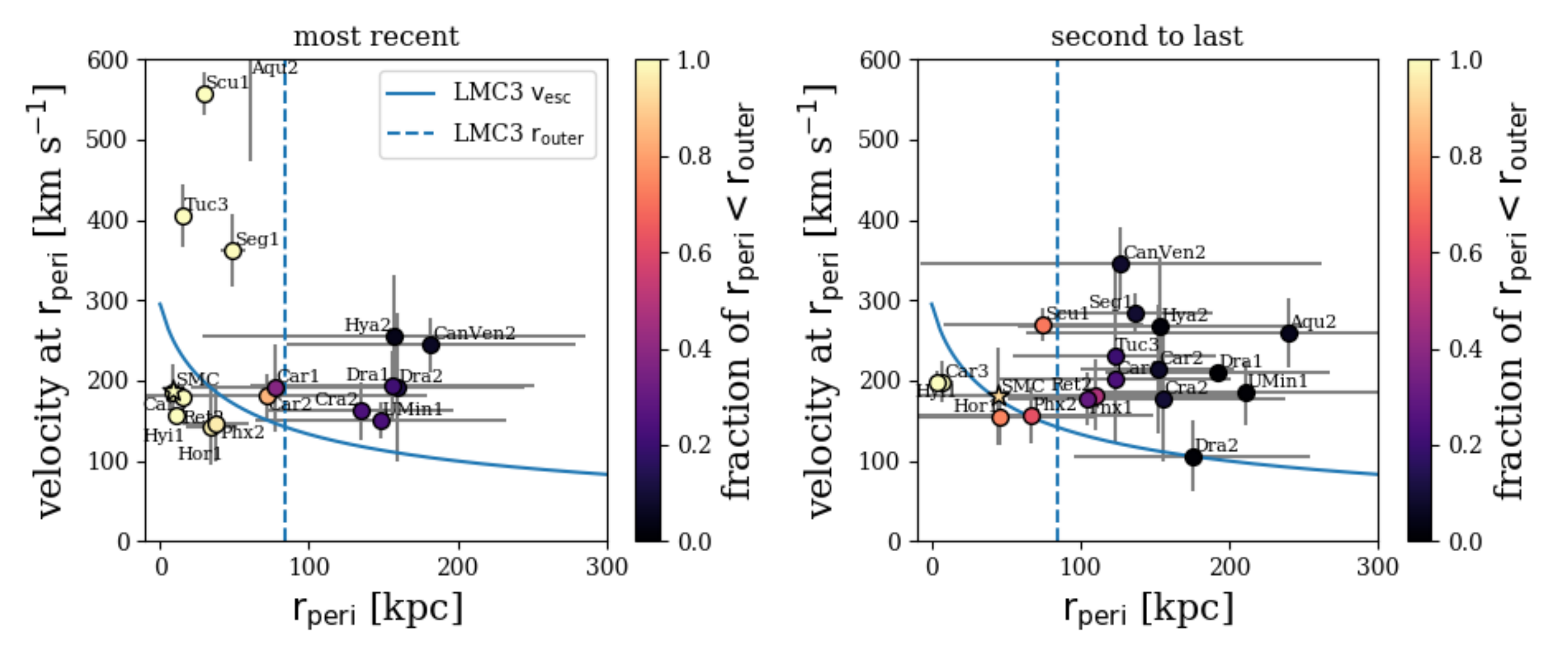}
    \caption{Same as Fig. \ref{fig:orbparams} except the orbital properties are calculated relative to LMC3 in MW2.}
\end{figure}

\begin{splitdeluxetable*}{lcccccccBlccccccc}
\tabletypesize{\footnotesize}
\label{tab:orbparams_MW1_LMC3}
\tablecaption{Orbital properties with respect to the LMC in MW1 using the LMC3 model. }
\tablehead{\colhead{Name} & \colhead{$\rm f_{peri,1}$ } & \colhead{$\rm f_{r_{outer},1}$} & \colhead{$\rm r_{peri,1}$ [kpc]} & \colhead{$\rm t_{peri,1}$ [Gyr]} & \colhead{$\rm f_{apo,1}$} & \colhead{$\rm r_{apo,1}$ [kpc]} & \colhead{$\rm t_{apo,1}$ [Gyr]} & \colhead{Name} & \colhead{$\rm f_{peri,2}$} &  \colhead{$\rm f_{r_{outer},2}$} & \colhead{$\rm r_{peri,2}$ [kpc]} & \colhead{$\rm t_{peri,2}$ [Gyr]} & \colhead{$\rm f_{apo,2}$} & \colhead{$\rm r_{apo,2}$ [kpc]} & \colhead{$\rm t_{apo,2}$ [Gyr]}\\ \hline
\multicolumn{8}{c}{most recent} & \multicolumn{8}{c}{second to last}}
\startdata
Aqu2 & 1.0 & 0.71 & 60.96$\pm$20.06 & 0.16$\pm$0.06 & 0.22 & 452.5$\pm$158.17 & 2.98$\pm$1.16 & Aqu2 & 1.0 &  0.01 & 305.78$\pm$163.31 & 4.13$\pm$1.04 & 0.08 & 358.57$\pm$125.89 & 4.74$\pm$0.82 \\ 
CanVen2 & 0.27 & 0.02 & 226.94$\pm$120.08 & 3.39$\pm$1.24 & 0.33 & 338.6$\pm$108.02 & 1.77$\pm$1.24 & CanVen2 & 0.27 &  0.0 & 231.59$\pm$95.9 & 4.75$\pm$0.87 & 0.1 & 326.49$\pm$111.18 & 4.78$\pm$1.0 \\ 
Car2 & 0.87 & 0.82 & 30.96$\pm$42.7 & 1.13$\pm$0.47 & 0.9 & 74.11$\pm$80.0 & 0.61$\pm$0.84 & Car2 & 0.87 &  0.64 & 36.12$\pm$42.96 & 3.75$\pm$0.97 & 0.82 & 119.57$\pm$91.3 & 2.76$\pm$0.97 \\ 
Car3 & 1.0 & 1.0 & 8.75$\pm$2.98 & 0.18$\pm$0.05 & 1.0 & 52.89$\pm$32.46 & 0.87$\pm$0.36 & Car3 & 1.0 &  0.98 & 7.51$\pm$10.96 & 1.54$\pm$0.69 & 0.95 & 52.58$\pm$21.17 & 2.15$\pm$0.73 \\ 
Cra2 & 1.0 & 0.06 & 231.45$\pm$92.57 & 2.78$\pm$0.89 & 1.0 & 338.84$\pm$46.2 & 1.43$\pm$0.21 & Cra2 & 1.0 &  0.01 & 366.44$\pm$191.27 & 4.87$\pm$0.7 & 0.84 & 389.8$\pm$156.15 & 4.59$\pm$0.73 \\ 
Dra2 & 0.33 & 0.03 & 292.17$\pm$186.9 & 4.53$\pm$1.13 & 0.63 & 529.19$\pm$155.54 & 3.16$\pm$1.12 & Dra2 & 0.33 &  0.0 & 356.7$\pm$176.3 & 5.11$\pm$0.71 & 0.07 & 391.59$\pm$143.6 & 4.93$\pm$0.75 \\ 
Hor1 & 0.98 & 0.98 & 30.92$\pm$13.45 & 0.31$\pm$0.38 & 0.94 & 114.81$\pm$150.67 & 1.38$\pm$1.27 & Hor1 & 0.98 &  0.76 & 46.11$\pm$92.67 & 1.92$\pm$1.21 & 0.78 & 73.33$\pm$81.39 & 2.51$\pm$1.19 \\ 
Hyi1 & 1.0 & 1.0 & 11.49$\pm$2.35 & 0.26$\pm$0.03 & 1.0 & 29.37$\pm$2.4 & 0.72$\pm$0.06 & Hyi1 & 1.0 &  1.0 & 7.83$\pm$2.1 & 1.15$\pm$0.11 & 1.0 & 27.76$\pm$5.06 & 1.55$\pm$0.17 \\ 
Hya2 & 0.27 & 0.03 & 130.97$\pm$40.22 & 0.99$\pm$1.28 & 0.17 & 241.56$\pm$152.22 & 1.24$\pm$1.56 & Hya2 & 0.27 &  0.0 & 159.45$\pm$47.16 & 4.63$\pm$1.11 & 0.07 & 365.61$\pm$151.89 & 4.06$\pm$1.1 \\ 
Phx2 & 0.98 & 0.97 & 32.99$\pm$17.76 & 0.49$\pm$0.49 & 0.94 & 157.54$\pm$164.04 & 2.04$\pm$1.26 & Phx2 & 0.98 &  0.64 & 67.03$\pm$121.01 & 2.74$\pm$1.2 & 0.66 & 96.48$\pm$107.71 & 3.44$\pm$1.06 \\ 
Ret2 & 1.0 & 1.0 & 15.57$\pm$2.93 & 0.13$\pm$0.02 & 0.95 & 154.62$\pm$189.67 & 1.62$\pm$1.37 & Ret2 & 1.0 &  0.68 & 91.36$\pm$171.87 & 2.09$\pm$1.33 & 0.81 & 120.01$\pm$166.05 & 2.59$\pm$1.24 \\ 
Seg1 & 1.0 & 0.93 & 55.74$\pm$10.63 & 0.32$\pm$0.11 & 1.0 & 70.02$\pm$7.18 & 0.11$\pm$0.04 & Seg1 & 1.0 &  0.01 & 212.87$\pm$120.51 & 1.77$\pm$1.01 & 0.98 & 240.9$\pm$129.01 & 1.39$\pm$0.91 \\ 
Tuc3 & 1.0 & 1.0 & 14.76$\pm$3.25 & 0.08$\pm$0.01 & 0.72 & 210.6$\pm$140.48 & 1.22$\pm$1.01 & Tuc3 & 1.0 &  0.04 & 174.36$\pm$102.45 & 1.4$\pm$1.18 & 0.54 & 267.85$\pm$135.58 & 2.09$\pm$1.42 \\ 
Car1 & 0.74 & 0.13 & 115.56$\pm$115.01 & 1.58$\pm$1.13 & 0.95 & 235.88$\pm$207.05 & 1.62$\pm$1.7 & Car1 & 0.74 &  0.14 & 127.28$\pm$91.4 & 4.31$\pm$1.03 & 0.61 & 229.62$\pm$114.04 & 3.64$\pm$1.13 \\ 
Dra1 & 0.67 & 0.08 & 279.59$\pm$140.87 & 4.51$\pm$0.97 & 0.97 & 424.29$\pm$92.17 & 2.64$\pm$0.54 & Dra1 & 0.67 &  0.0 & 326.4$\pm$14.86 & 4.23$\pm$0.51 & 0.04 & 432.25$\pm$211.38 & 5.38$\pm$0.64 \\ 
Fnx1 & 1.0 & 0.0 & 99.93$\pm$4.73 & 0.15$\pm$0.08 & 0.88 & 318.34$\pm$240.72 & 2.53$\pm$2.04 & Fnx1 & 1.0 &  0.09 & 98.94$\pm$44.37 & 1.54$\pm$0.78 & 0.46 & 232.8$\pm$78.98 & 3.78$\pm$0.82 \\ 
Scu1 & 1.0 & 1.0 & 28.81$\pm$4.87 & 0.11$\pm$0.01 & 0.89 & 330.82$\pm$91.61 & 2.06$\pm$0.84 & Scu1 & 1.0 &  0.28 & 136.77$\pm$115.49 & 4.25$\pm$0.92 & 0.26 & 257.26$\pm$52.52 & 5.15$\pm$0.65 \\ 
UMin1 & 0.79 & 0.13 & 258.84$\pm$132.6 & 4.13$\pm$1.06 & 0.98 & 392.29$\pm$67.75 & 2.31$\pm$0.42 & UMin1 & 0.79 &  0.0 & 277.26$\pm$52.98 & 4.8$\pm$0.93 & 0.18 & 495.87$\pm$190.92 & 5.5$\pm$0.54 \\ 
\enddata
\end{splitdeluxetable*}

\begin{splitdeluxetable*}{lcccccccBlccccccc}
\tabletypesize{\footnotesize}
\label{tab:orbparams_MW2_LMC3}
\tablecaption{Orbital properties with respect to the LMC in MW2 using the LMC3 model. }
\tablehead{\colhead{Name} & \colhead{$\rm f_{peri,1}$ } & \colhead{$\rm f_{r_{outer},1}$} & \colhead{$\rm r_{peri,1}$ [kpc]} & \colhead{$\rm t_{peri,1}$ [Gyr]} & \colhead{$\rm f_{apo,1}$} & \colhead{$\rm r_{apo,1}$ [kpc]} & \colhead{$\rm t_{apo,1}$ [Gyr]} & \colhead{Name} & \colhead{$\rm f_{peri,2}$} &  \colhead{$\rm f_{r_{outer},2}$} & \colhead{$\rm r_{peri,2}$ [kpc]} & \colhead{$\rm t_{peri,2}$ [Gyr]} & \colhead{$\rm f_{apo,2}$} & \colhead{$\rm r_{apo,2}$ [kpc]} & \colhead{$\rm t_{apo,2}$ [Gyr]}\\ \hline
\multicolumn{8}{c}{most recent} & \multicolumn{8}{c}{second to last}}
\startdata
Aqu2 & 1.0 & 0.9 & 60.86$\pm$19.92 & 0.16$\pm$0.06 & 0.36 & 369.57$\pm$159.98 & 1.98$\pm$0.8 & Aqu2 & 0.34 &  0.05 & 239.75$\pm$175.98 & 3.41$\pm$0.99 & 0.2 & 293.09$\pm$121.33 & 4.21$\pm$0.93 \\ 
CanVen2 & 0.5 & 0.07 & 181.89$\pm$96.91 & 2.24$\pm$1.03 & 0.54 & 292.48$\pm$97.44 & 1.04$\pm$1.14 & CanVen2 & 0.22 &  0.08 & 127.12$\pm$134.84 & 4.72$\pm$0.78 & 0.37 & 380.28$\pm$185.18 & 4.0$\pm$0.84 \\ 
Car2 & 1.0 & 0.84 & 72.08$\pm$107.47 & 1.3$\pm$1.2 & 1.0 & 103.24$\pm$121.29 & 0.76$\pm$0.98 & Car2 & 0.86 &  0.08 & 151.68$\pm$52.0 & 3.88$\pm$0.89 & 0.97 & 258.48$\pm$94.63 & 3.07$\pm$0.98 \\ 
Car3 & 1.0 & 1.0 & 8.1$\pm$3.0 & 0.18$\pm$0.05 & 1.0 & 47.55$\pm$27.41 & 0.77$\pm$0.26 & Car3 & 1.0 &  1.0 & 6.74$\pm$8.18 & 1.36$\pm$0.52 & 0.98 & 55.35$\pm$37.18 & 2.07$\pm$0.75 \\ 
Cra2 & 1.0 & 0.24 & 134.93$\pm$62.04 & 2.28$\pm$0.58 & 1.0 & 265.35$\pm$30.15 & 0.94$\pm$0.14 & Cra2 & 0.8 &  0.1 & 155.33$\pm$82.4 & 4.38$\pm$0.94 & 0.93 & 239.43$\pm$105.62 & 3.56$\pm$0.78 \\ 
Dra2 & 0.95 & 0.14 & 159.49$\pm$85.59 & 3.24$\pm$1.01 & 0.99 & 326.01$\pm$106.52 & 1.54$\pm$0.56 & Dra2 & 0.35 &  0.01 & 174.94$\pm$79.61 & 4.72$\pm$0.81 & 0.59 & 293.11$\pm$105.52 & 4.33$\pm$0.96 \\ 
Hor1 & 0.98 & 0.98 & 34.25$\pm$17.33 & 0.26$\pm$0.44 & 0.98 & 106.46$\pm$119.89 & 1.2$\pm$1.19 & Hor1 & 0.89 &  0.74 & 46.15$\pm$64.79 & 1.69$\pm$1.27 & 0.8 & 90.62$\pm$92.33 & 2.31$\pm$1.2 \\ 
Hyi1 & 1.0 & 1.0 & 10.79$\pm$2.4 & 0.26$\pm$0.03 & 1.0 & 33.17$\pm$4.76 & 0.75$\pm$0.1 & Hyi1 & 1.0 &  1.0 & 3.76$\pm$2.26 & 1.2$\pm$0.18 & 1.0 & 35.79$\pm$10.9 & 1.67$\pm$0.28 \\ 
Hya2 & 0.33 & 0.05 & 156.7$\pm$128.68 & 1.25$\pm$1.59 & 0.24 & 293.93$\pm$200.07 & 1.52$\pm$1.66 & Hya2 & 0.11 &  0.02 & 153.15$\pm$95.69 & 4.38$\pm$0.96 & 0.14 & 327.82$\pm$145.06 & 3.3$\pm$1.02 \\ 
Phx2 & 0.97 & 0.96 & 37.34$\pm$22.2 & 0.42$\pm$0.53 & 0.98 & 146.69$\pm$133.48 & 1.78$\pm$1.25 & Phx2 & 0.85 &  0.62 & 67.32$\pm$81.05 & 2.37$\pm$1.17 & 0.72 & 130.55$\pm$108.4 & 3.27$\pm$1.13 \\ 
Ret2 & 1.0 & 1.0 & 15.32$\pm$3.03 & 0.13$\pm$0.02 & 1.0 & 181.5$\pm$128.26 & 1.62$\pm$0.93 & Ret2 & 0.98 &  0.48 & 109.54$\pm$106.68 & 2.41$\pm$1.27 & 0.85 & 185.42$\pm$140.22 & 3.0$\pm$1.23 \\ 
Seg1 & 1.0 & 1.0 & 48.96$\pm$8.77 & 0.29$\pm$0.08 & 1.0 & 68.29$\pm$6.27 & 0.09$\pm$0.03 & Seg1 & 1.0 &  0.09 & 136.35$\pm$52.78 & 1.29$\pm$0.5 & 1.0 & 170.25$\pm$54.15 & 0.92$\pm$0.35 \\ 
Tuc3 & 1.0 & 1.0 & 15.08$\pm$3.24 & 0.07$\pm$0.01 & 0.84 & 164.48$\pm$117.1 & 0.95$\pm$0.85 & Tuc3 & 0.82 &  0.2 & 122.9$\pm$67.91 & 1.19$\pm$0.88 & 0.72 & 219.52$\pm$118.4 & 1.96$\pm$1.24 \\ 
Car1 & 0.97 & 0.38 & 77.53$\pm$56.63 & 1.35$\pm$0.85 & 1.0 & 134.41$\pm$84.32 & 0.68$\pm$0.67 & Car1 & 0.72 &  0.24 & 123.55$\pm$77.41 & 3.43$\pm$1.1 & 0.84 & 183.86$\pm$93.43 & 2.79$\pm$1.13 \\ 
Dra1 & 0.78 & 0.21 & 156.11$\pm$95.25 & 3.87$\pm$1.14 & 0.9 & 290.96$\pm$53.55 & 1.98$\pm$0.89 & Dra1 & 0.12 &  0.0 & 192.53$\pm$75.39 & 5.09$\pm$0.9 & 0.35 & 316.99$\pm$143.2 & 4.66$\pm$0.77 \\ 
Fnx1 & 1.0 & 0.0 & 100.64$\pm$3.91 & 0.13$\pm$0.03 & 1.0 & 188.88$\pm$142.66 & 1.22$\pm$1.18 & Fnx1 & 0.93 &  0.3 & 105.01$\pm$53.94 & 1.85$\pm$1.42 & 0.78 & 202.39$\pm$80.26 & 2.77$\pm$0.68 \\ 
Scu1 & 1.0 & 1.0 & 29.1$\pm$4.95 & 0.11$\pm$0.01 & 1.0 & 221.94$\pm$47.15 & 1.04$\pm$0.25 & Scu1 & 1.0 &  0.72 & 75.04$\pm$66.9 & 2.45$\pm$0.54 & 0.99 & 261.89$\pm$96.32 & 3.81$\pm$0.75 \\ 
UMin1 & 0.89 & 0.24 & 148.29$\pm$84.61 & 3.64$\pm$1.06 & 0.96 & 273.14$\pm$43.12 & 1.74$\pm$0.77 & UMin1 & 0.21 &  0.01 & 210.5$\pm$95.03 & 5.22$\pm$0.63 & 0.48 & 297.87$\pm$145.48 & 4.53$\pm$0.66 \\ 
\enddata
\end{splitdeluxetable*}
\clearpage

\section{Orbital Properties with respect to the Milky Way}
\label{sec:appendixC}

\begin{splitdeluxetable*}{lccccccBlcccccc}
\tabletypesize{\footnotesize}
\label{tab:orbparams_wrtMW}
\tablecaption{Orbital properties with respect to the MW for the fiducial LMC model.}
\tablehead{\colhead{Name} & \colhead{$\rm f_{peri,1}$} & \colhead{$\rm r_{peri,1}$ [kpc]} & \colhead{$\rm t_{peri,1}$ [Gyr]} & \colhead{$\rm f_{apo,1}$} & \colhead{$\rm r_{apo,1}$ [kpc]} & \colhead{$\rm t_{apo,1}$ [Gyr]} & \colhead{Name} & \colhead{$\rm f_{peri,1}$} & \colhead{$\rm r_{peri,1}$ [kpc]} & \colhead{$\rm t_{peri,1}$ [Gyr]} & \colhead{$\rm f_{apo,1}$} & \colhead{$\rm r_{apo,1}$ [kpc]} & \colhead{$\rm t_{apo,1}$ [Gyr]}\\ \hline
\multicolumn{7}{c}{MW1} & \multicolumn{7}{c}{MW2}}
\startdata
Aqu2 & 0.91 & 93.31$\pm$24.3 & 0.17$\pm$0.23 & 0.19 & 208.13$\pm$87.52 & 2.86$\pm$1.13 & Aqu2 &  0.91 & 91.9$\pm$25.56 & 0.18$\pm$0.22 & 0.29 & 230.05$\pm$107.51 & 2.62$\pm$1.04 \\ 
CanVen2 & 0.13 & 59.29$\pm$83.27 & 4.64$\pm$0.79 & 0.25 & 349.77$\pm$96.71 & 3.01$\pm$1.32 & CanVen2 &  0.3 & 81.13$\pm$79.04 & 3.88$\pm$1.0 & 0.4 & 306.97$\pm$82.59 & 2.1$\pm$1.03 \\ 
Car2 & 1.0 & 28.05$\pm$1.25 & 0.08$\pm$0.01 & 0.93 & 318.46$\pm$120.33 & 3.24$\pm$1.19 & Car2 &  1.0 & 27.55$\pm$1.3 & 0.08$\pm$0.01 & 1.0 & 145.95$\pm$30.25 & 1.28$\pm$0.36 \\ 
Car3 & 1.0 & 28.81$\pm$1.26 & 0.01$\pm$0.0 & 0.88 & 395.63$\pm$121.82 & 3.46$\pm$1.01 & Car3 &  1.0 & 29.09$\pm$9.2 & 0.02$\pm$0.15 & 1.0 & 251.66$\pm$83.71 & 2.19$\pm$0.73 \\ 
Cra2 & 1.0 & 18.95$\pm$9.65 & 2.06$\pm$0.34 & 1.0 & 143.23$\pm$11.6 & 0.76$\pm$0.17 & Cra2 &  1.0 & 17.93$\pm$8.75 & 1.55$\pm$0.19 & 1.0 & 132.0$\pm$8.09 & 0.51$\pm$0.09 \\ 
Dra2 & 0.39 & 31.86$\pm$21.61 & 3.75$\pm$1.2 & 0.55 & 246.86$\pm$97.34 & 2.53$\pm$1.25 & Dra2 &  0.92 & 31.81$\pm$13.0 & 2.61$\pm$1.15 & 0.98 & 177.78$\pm$91.76 & 1.45$\pm$0.87 \\ 
Hor1 & 0.87 & 181.39$\pm$138.95 & 1.58$\pm$1.43 & 0.92 & 205.97$\pm$144.31 & 1.28$\pm$1.39 & Hor1 &  0.93 & 112.34$\pm$75.01 & 1.36$\pm$1.32 & 0.98 & 153.1$\pm$96.13 & 0.9$\pm$1.13 \\ 
Hyi1 & 0.95 & 344.25$\pm$106.06 & 3.06$\pm$0.47 & 0.95 & 361.46$\pm$100.55 & 2.67$\pm$0.52 & Hyi1 &  0.94 & 184.14$\pm$86.37 & 3.12$\pm$1.26 & 1.0 & 251.17$\pm$94.39 & 2.19$\pm$0.75 \\ 
Hya2 & 1.0 & 133.21$\pm$24.97 & 0.18$\pm$0.21 & 0.08 & 275.88$\pm$69.99 & 3.79$\pm$0.94 & Hya2 &  1.0 & 131.85$\pm$26.96 & 0.19$\pm$0.21 & 0.15 & 252.34$\pm$71.86 & 3.22$\pm$0.93 \\ 
Phx2 & 0.7 & 198.82$\pm$143.23 & 2.55$\pm$1.67 & 0.87 & 263.21$\pm$141.79 & 2.15$\pm$1.47 & Phx2 &  0.92 & 106.58$\pm$63.79 & 2.13$\pm$1.65 & 0.97 & 176.07$\pm$95.04 & 1.29$\pm$1.12 \\ 
Ret2 & 0.81 & 165.96$\pm$160.3 & 3.35$\pm$1.26 & 0.96 & 259.06$\pm$147.94 & 2.47$\pm$1.18 & Ret2 &  1.0 & 59.68$\pm$44.63 & 2.09$\pm$0.98 & 1.0 & 116.76$\pm$68.53 & 1.19$\pm$0.64 \\ 
Seg1 & 1.0 & 19.62$\pm$5.12 & 0.1$\pm$0.02 & 1.0 & 63.89$\pm$34.84 & 0.67$\pm$0.32 & Seg1 &  1.0 & 18.82$\pm$5.23 & 0.1$\pm$0.02 & 1.0 & 50.71$\pm$18.8 & 0.51$\pm$0.14 \\ 
Tuc3 & 0.99 & 2.01$\pm$1.8 & 0.64$\pm$0.12 & 0.99 & 52.11$\pm$13.76 & 0.3$\pm$0.15 & Tuc3 &  0.99 & 2.09$\pm$1.81 & 0.49$\pm$0.12 & 1.0 & 45.57$\pm$21.66 & 0.23$\pm$0.21 \\ 
Car1 & 1.0 & 80.14$\pm$18.25 & 0.78$\pm$0.19 & 0.99 & 146.66$\pm$54.19 & 1.52$\pm$1.53 & Car1 &  1.0 & 67.04$\pm$17.79 & 0.82$\pm$0.1 & 1.0 & 133.58$\pm$29.83 & 1.35$\pm$1.1 \\ 
Dra1 & 1.0 & 84.94$\pm$19.16 & 2.8$\pm$0.78 & 1.0 & 137.06$\pm$26.52 & 1.21$\pm$0.42 & Dra1 &  1.0 & 67.93$\pm$14.3 & 1.67$\pm$0.34 & 1.0 & 108.96$\pm$13.31 & 0.64$\pm$0.18 \\ 
Fnx1 & 0.91 & 108.09$\pm$25.51 & 1.37$\pm$0.26 & 0.99 & 160.39$\pm$51.56 & 0.59$\pm$1.19 & Fnx1 &  0.99 & 87.33$\pm$24.96 & 1.32$\pm$0.31 & 1.0 & 146.22$\pm$10.88 & 0.22$\pm$0.33 \\ 
Scu1 & 1.0 & 57.21$\pm$6.24 & 0.37$\pm$0.04 & 0.96 & 296.98$\pm$55.37 & 3.49$\pm$0.71 & Scu1 &  1.0 & 53.12$\pm$6.27 & 0.37$\pm$0.03 & 1.0 & 198.38$\pm$23.77 & 2.06$\pm$0.24 \\ 
UMin1 & 1.0 & 77.05$\pm$16.17 & 2.52$\pm$0.66 & 1.0 & 124.18$\pm$22.28 & 1.08$\pm$0.39 & UMin1 &  1.0 & 63.45$\pm$13.0 & 1.54$\pm$0.29 & 1.0 & 100.67$\pm$9.79 & 0.57$\pm$0.17 \\ 
\enddata
\tablecomments{Orbital parameters calculated with respect to the MW. All values are still calculated for the fiducial LMC model. Columns 1-8 provide the results in the MW1 potential and Columns 9-16 list results for MW2. }
\end{splitdeluxetable*}

\end{document}